\documentclass[%
 reprint,
superscriptaddress,
amsmath,amssymb,
aps,
]{revtex4-2}

\usepackage{graphicx}
\usepackage{dcolumn}
\usepackage{bm}
\usepackage{hyperref}
\usepackage{verbatim} 
\hypersetup{
colorlinks,
citecolor=blue,
filecolor=blue, 
linkcolor=blue, 
urlcolor=blue
}

\usepackage{color}
\usepackage{xcolor}
\usepackage{ulem}
\usepackage{physics}

\newcommand{\yambo}{\textsc{yambo}}
\newcommand{\qe}{\textsc{Quantum Espresso}}
\newcommand{\gpaw}{\textsc{gpaw}}

\newcommand{\q}{{\mathbf q}}
\newcommand{\kk}{{\mathbf k}}
\newcommand{\G}{{\mathbf G}}

\renewcommand{\Im}{\mathrm{Im}} 
\renewcommand{\Re}{\mathrm{Re}}

\newcommand{\editor}[2]{%
  \expandafter\newcommand\csname #1note\endcsname[1]{%
    \textcolor{#2}{(\textbf{#1:} \textit{##1})}}%
  \expandafter\newcommand\csname #1\endcsname[1]{%
    \textcolor{#2}{##1}}%
  \expandafter\newcommand\csname #1cancel\endcsname[1]{%
    \textcolor{#2}{\sout{##1}}}%
\expandafter\newcommand\csname #1can\endcsname[1]{%
    \textcolor{#2}{\sout{##1}}}%
  \expandafter\newcommand\csname #1change\endcsname[2]{%
    \textcolor{#2}{\sout{##1} ##2}}%
      \expandafter\newcommand\csname #1ch\endcsname[2]{%
    \textcolor{#2}{\sout{##1} ##2}}%
  \newenvironment{#1text}{\color{#2}}{\color{black}}
}
\definecolor{Blu}{rgb}{0.00,0.00,1.00}
\definecolor{Red}{rgb}{1.00,0.00,0.00}
\definecolor{Cyan}{rgb}{0.00,0.50,0.50}
\definecolor{Green}{rgb}{0.00,0.70,0.00}

\editor{DAL}{Green}
\editor{CC}{Cyan}
\editor{kb}{orange}
\editor{KB}{orange}
\editor{R}{Red}

\renewcommand{\emph}{\textit}
\newcommand{\suppinfo}{Ref.~\cite{supp-info}}

\begin{document}

\title{
Spectral properties from an efficient analytical representation of the $GW$ self-energy within a multipole approximation
}

\author{Dario A. Leon}
\email{dario.alejandro.leon.valido@nmbu.no}
\affiliation{
 Department of Mechanical Engineering and Technology Management, \\ Norwegian University of Life Sciences, NO-1432 Ås, Norway
}%

\author{Kristian Berland}
\affiliation{
 Department of Mechanical Engineering and Technology Management, \\ Norwegian University of Life Sciences, NO-1432 Ås, Norway
}

\author{Claudia Cardoso}
\affiliation{
 S3 Centre, Istituto Nanoscienze, CNR, 41125 Modena, Italy
}

\date{\today}

\begin{abstract}
We propose an efficient analytical representation of the frequency-dependent $GW$ self-energy $\Sigma$ via a multipole approximation (MPA-$\Sigma$). The MPA self-energy model is interpolated from a small set of numerical evaluations of $\Sigma$ in the complex frequency plane, similar to the MPA interpolation developed for the screened Coulomb interaction (MPA-$W$) [D. A. Leon {\it et al.}, Phys. Rev. B \textbf{104}, 115157 (2021)]. 
Crucially, MPA-$\Sigma$ enables a multipole representation for the interacting Green's function $G$ (MPA-$G$), and in turn, access to all the spectral properties, including 
quasiparticle energies (QP) and renormalization factors beyond the linearized QP equation. 
We validate the MPA-$\Sigma$ and MPA-$G$ approaches for a diverse set of systems:
bulk Si, Na and Cu, monolayer MoS$_2$, the NaCl ion-pair, and the F$_2$ molecule.
We show that, just as for MPA-$W$, an appropriate choice of frequency sampling in MPA-$\Sigma$ is critical to guarantee computational efficiency and high accuracy. 
Moreover, the combined MPA-$W$ and MPA-$\Sigma$ scheme considerably reduces the cost of full-frequency self-energy calculations, especially for spectral band structures over a wide energy range.  
\end{abstract}

\maketitle


\section{Introduction}
In condensed matter physics, first principle methods such as 
density functional theory (DFT) in the Kohn-Sham (KS) approximation 
provide accurate ground-state properties and have been immensely useful for understanding the electronic structure of materials. 
However, they 
fail to reliably provide accurate band structures, which requires including many-body effects beyond 
the mean field DFT level. The description of electron addition or removal energies and related excited-state properties is usually treated with methods such as the $GW$ approximation, based on the Green's function formalism~\cite{Aryasetiawan1998RPP,Hedin1999JPCM,Onida2002RMP,martin2016book,Reining2018wcms,Marzari2021NatureMat}. 

In common $GW$ implementations, the Green's function $G$ and the screened Coulomb potential $W$ are constructed perturbatively. Starting from DFT, the KS quasiparticle (QP) energies are corrected by an exchange-correlation self-energy $\Sigma$. 
This correction can be 
obtained iteratively within a self-consistent $GW$ approach, or done in a computationally cheaper one-shot $G_0W_0$ procedure. 
Since the imaginary part of $G$ is closely related to the spectral function obtained from photoemission experiments~\cite{Damascelli2004PS,hufner2013photoelectron}, a dynamical self-energy $\Sigma (\omega)$ can account for many-body features, such as finite QP lifetimes and satellite structures~\cite{Strinati1980PRL,Strinati1982PRB,Hedin1999JPCM,Dolado2001PRB,Marini2002PRB,Kheifets2003PRB,Arnaud2005PRB,Cazzaniga2012PRB,Reining2018wcms,Zhou2020pnas}.

$G_0W_0$ is the state-of-the-art {\it ab initio} method for the description of angle-resolved photoemission and inverse photoemission spectroscopy measurements, giving generally a very accurate agreement with experiment (see, e.g., Refs.~\cite{Hybertsen1985PRL,vanSchilfgaarde2006PRL,Huser2013PRB,Golze2019FrontChem}). More accurate spectral functions can be obtained with self-consistent approaches, including cumulant expansions of $\Sigma$ and vertex corrections (see, e.g., Refs.~\cite{Kheifets2003PRB,martin2016book,Gumhalter2016PRB,Zhou2018PRB,Nery2018PRB,Reining2018wcms,Zhou2020pnas}).

The $GW$ self-energy $\Sigma(\omega)$ is given by a frequency convolution of $G(\omega)$ and $W(\omega)$. 
This convolution can be evaluated with different full-frequency (FF) methods, based on numerical integrations along the frequency real axis~\cite{Marini2002PRL, Shishkin2006PRB, Huser2013PRB, Liu2015JComputPhys}, or through an integration in the complex frequency plane using contour deformation and analytic continuation techniques~\cite{Godby1988PRB, book_Anisimov2000, Kotani2007PRB,DalingPRB1991,Engel1991PRB,Duchemin2020JCTC}. 
Such numerical FF evaluations tend to be computationally expensive. 
A less costly alternative 
is to represent $W$ (or the dielectric function) with a simple model, such as in the plasmon pole approximation (PPA), that allows for an analytical integration of the frequency convolution in $\Sigma$~\cite{Hybertsen1986PRB, Zhang1989PRB, Godby1989PRL, vonderLinden1988PRB, Engel1993PRB}, 
but in many cases this has limited accuracy. 
Higher accuracy can be obtained with multipole models and Pad\'e approximants~\cite{FaridPRB1991, Engel1991PRB, Lee1994PRB, Soininen2003JPCM}, including the recently developed MPA-$W$ method~\cite{Leon2021PRB,Leon2023PRB,Guandalini2024PRB}.

Analytical models of the dielectric response are also widely used in the study of optical and electronic properties of materials~\cite{book_Raether1980,Giuliani_Vignale_2005,Pines2018book}. They have been used in the study of, e.g., optical excitations~\cite{Allen1977PRB,Smith1986PRB}, electron energy loss~\cite{Lee1996PRB,Jin1999PRB}, and x-ray absorption spectra~\cite{Kas2007PRB,Kas2009JPCS}. Simple models have also been used to account for dynamical effects arising from electron-hole interactions in doped systems~\cite{Liang2015PRL,Champagne2023nanolett}, or in the {\it ab initio} description of plasmon-phonon hybridization in doped semiconductors~\cite{Lihm2024PRL}. 
Less common is the use of models for $\Sigma$~\cite{Riegera1999CPC, Soininen2005PS, vanSetten2015JCTC, Chiarotti2022PRR}, the interacting $G$~\cite{Chiarotti2022PRR, Chiarotti2024PRR, Ferretti2024PRB,Quinzi2025PRB}, or for total energy calculations~\cite{Ismail-Beigi2010PRB,Chiarotti2022PRR,Chiarotti2024PRR,Ferretti2024PRB,Quinzi2025PRB,Guo2024Arxiv}, and usually these approaches mainly aim at improving the computational efficiency of such calculations.

In this work, we present an efficient multipole approximation for the self-energy (MPA-$\Sigma$). This method yields simple analytical representations of all $GW$ operators, including a multipole-Pad\'e representation of the Green's function (MPA-$G$). 
Moreover, the combination of MPA-$\Sigma$ with the previous MPA-$W$ method considerably reduces the cost of evaluating $\Sigma$ and $G$ in its full-frequency domain. 
Although this study is limited to the $G_0W_0$ approximation, 
extending it to higher levels of theory, e.g., self-consistent $GW$, and the inclusion of vertex corrections or cumulant expansions, is straightforward. 

The paper is organized as follows: In the methods section (Sec.~\ref{sec:methods}) we summarize the general $GW$ equations (Sec.~\ref{sec:qp_eqs}), and the previously introduced MPA-$W$ method (Sec.~\ref{sec:MPA-W}).
Thereafter, we present 
the new MPA-$\Sigma$ (Sec.~\ref{sec:MPA-S}) and MPA-$G$ (Sec.~\ref{sec:MPA-G}) approaches, and provide computational details of the $GW$ calculations (Sec.~\ref{sec:comp_details}). In the results section (Sec.~\ref{sec:results}) we benchmark MPA-$\Sigma$ and MPA-$G$ on different prototypical materials (Sec.~\ref{sec:benchmarks}), and build spectral band structures (Sec.~\ref{sec:spectral_bs}). 
Last Sec.~\ref{sec:comclusions} holds our conclusions. 
In addition, Appendix~\ref{sec:toy_models} provides an analysis of the QP particle and the renormalization factor in terms of two MPA toy models, while Appendix~\ref{sec:spline_interpolation} 
presents numerical details for interpolating spectral functions in momentum space.

\section{Methods}
\label{sec:methods}

\subsection{Quasiparticle $GW$ equations}
\label{sec:qp_eqs}
In terms of KS states, the non-interacting time-ordered Green's function can be written in the Lehmann representation~\cite{Lehmann1972,martin2016book}, analytically continued to the complex frequency plane, as
\begin{multline}
       G_0(z) = \sum_{m}  \rho_m^{\text{KS}} 
       \left[ \frac{f_m^{\text{KS}}}{z -\varepsilon_{m}^{\text{KS}} -i 0^+}
        +\frac{1-f_m^{\text{KS}}}{z - \varepsilon_{m}^{\text{KS}} +i 0^+}
        \right],
    \label{eq:G0}
\end{multline}
where the sum runs over the KS states, $m$, with the projectors $\rho_m^{\text{KS}} = |\psi_m ^{\text{KS}} \rangle \langle \psi_m ^{\text{KS}}|$, KS energies $\varepsilon_{m}^{\text{KS}}$, and occupation numbers $f_m^{\text{KS}} \in [0,1]$. The complex frequency is given by $z \equiv \omega + i \varpi$, which is evaluated in the first and third quadrants ($\omega \varpi > 0$), opposite to the pole position according to the time ordering (see notation in Table~\ref{table:notation}). 

The projection of $G_0$ onto the KS states ($n \kk$) is given by
\begin{equation}
\begin{aligned}
G_{0 n \kk } (z) & \equiv  \mel{\psi_{n \kk}^{\text{KS}}}{G_0(z)}{\psi_{n \kk} ^{\text{KS}}} & \\ 
    & = \frac{f_{n \kk}^{\text{KS}}}{z - \varepsilon^{\text{KS}}_{n \kk} -i 0^+} +\frac{1-f_{n \kk}^{\text{KS}}}{z - \varepsilon^{\text{KS}}_{n \kk} +i 0^+}, &
   \label{eq:G0_nk}
\end{aligned}
\end{equation}
where the spectral function $\Im [G_{0 n \kk }]$ is 
a Dirac delta function centered on $\epsilon^{\text{KS}}_{n \kk}$.
The interacting Green's function is given by the Dyson equation for this operator, in which the DFT exchange and correlation potential is subtracted from the self-energy: 
\begin{equation}
   G_{n \kk}^{-1} (z) = G_{0 n \kk}^{-1} (z) - \Sigma_{n \kk} (z) +v_{xc}^{\text{KS}},
   \label{eq:G1_nk}
\end{equation}
where, as commonly done, the off-diagonal elements ($n\kk\neq n'\mathbf{k'}$) have been neglected. 
At the $G_0W_0$ level, $\Sigma$ is given by the convolution of $G_0$ and $W_0$:
\begin{equation}
    \Sigma (z) = 
     \frac{i}{2 \pi}\int_{-\infty}^{+\infty} 
     d\omega' e^{-i \omega' 0^+} G_0(z-\omega') W_0(\omega').
    \label{eq:GW}
\end{equation}
The QP energies correspond to the poles of $G$, 
which are determined by solving the QP equation:
\begin{equation}
  \epsilon_{n \kk} =   \epsilon_{n \kk}^{\text{KS}} +  \langle \psi_{n \kk}^{\text{KS}}|\Sigma (\epsilon_{n \kk})-v_{xc}^{\text{KS}}|\psi_{n \kk}^{\text{KS}} \rangle
   \label{eq:QPeq}.
\end{equation}

The frequency dependence of $G$ has a structure typically dominated by a well-defined main peak, the QP pole~\cite{hufner2013photoelectron}, and satellite structures at larger energies~\cite{Farid2002PMB,martin2016book,Reining2018wcms}.
Like the QP pole, the satellites are also formal solutions of Eq.~\eqref{eq:QPeq}. They arise from many-body excitations accounted for in $\Sigma$, such as the plasmonic structures in $GW$, and give rise to replicas of the QP band structure~\cite{martin2016book,Reining2018wcms}. In the so-called QP picture, satellites are disregarded and only energies around the QP pole are considered. As such, the QP picture resembles the independent particle picture, but with the KS energies corrected by the real part of $\Sigma$, while the finite imaginary part accounts for the broadening of the QP pole, according to its lifetime. 

\begin{table}
\begin{ruledtabular}
\begin{tabular}{lcc} 
\\[-8pt]
 {\bf Complex quantity}        & {\bf Energy/Poles} & {\bf Residues} \\
  \hline\\[-8pt] 
 Energy/frequency      & $z = \omega + i \varpi$ & - \\
 $G_0 (z)$       & $\epsilon^{\text{KS}} = \varepsilon^{\text{KS}} \pm i 0^{+}$ & $1$ \\
  $G (z)$       & $\epsilon_p = \varepsilon_p + i \eta_p$ & $Z_p$ \\
 $W (z)$       & $\Omega_p = \omega_p + i \varpi_p$ & $R_p$ \\
 $\Sigma (z)$ & $\xi_p = \zeta_p + i \varsigma_p$ & $S_p$ \\
\end{tabular}
\end{ruledtabular}
 \caption{List of complex quantities relevant for this work, and definition of the used notation. In the case of $G_0$, each state is represented by a single pole with a vanishing imaginary part whose sign follows the time ordering. The residue of such pole carries all the spectral weight.}
 \label{table:notation}
\end{table}

Due to the non-linearity of Eq.~\eqref{eq:QPeq}, its numerical evaluation requires a recursive procedure, such as the secant method. Alternatively, it can be approximated by a linearized equation:
\begin{equation}
   \epsilon_{n \kk} \approx \epsilon^{\text{lin}}_{n \kk} \equiv \epsilon_{n \kk}^{\text{KS}} + Z^{\text{lin}}_{n \kk} \langle \psi_{n \kk}^{\text{KS}}|\Sigma (\epsilon_{n \kk}^{\text{KS}})-v_{xc}^{\text{KS}}|\psi_{n \kk}^{\text{KS}} \rangle,
   \label{eq:lQPeq}
\end{equation}
with the corresponding linearized renormalization factor, $Z^{\text{lin}}_{n \kk}$, given by
\begin{equation}
   Z^{\text{lin}}_{n \kk} = \left[ 1-\langle \psi_{n \kk}^{\text{KS}}|\frac{\partial\Sigma (z)}{\partial z}\bigg|_{z=\epsilon_{n \kk}^{\text{KS}}}|\psi_{n \kk}^{\text{KS}} \rangle \right]^{-1},
   \label{eq:qp_z}
\end{equation}
which approximates the spectral weight of the QP pole, $Z_{n \kk}$, based on the assumption that the QP correction ($ \epsilon_{n \kk} - \epsilon^{\text{KS}}_{n \kk}$) is small. 
Since the satellite structures have a non-vanishing weight, $Z_{n \kk}$ usually has a very small imaginary part and a real part ranging from 0.5 to 1. This interval is taken as a typical validity range of the QP picture~\cite{Farid2002PMB,Reining2018wcms,Rasmussen2021NPJComputMater}, while strong correlation effects can lead to situations where the spectral weight is concentrated in the satellites, like in Mott insulators (see, e.g., Refs.~\cite{Georges1996RevModPhys,martin2016book}). 
When computed in a consistent way, the spectral weights of the QP pole and the satellites sum exactly to one, since they comply with the sum rule for the number of particles and holes~\cite{vonBarth1996PRB}:
\begin{equation}
   \frac{1}{\pi} \int_{-\infty}^{\infty} \Im ~G_{n \kk} (\omega) d \omega  
    = 1 .
    \label{eq:sumrule_integralG}
\end{equation}
%

\subsection{MPA for the screening interaction}
\label{sec:MPA-W}
The screened Coulomb potential can be separated in a static bare Coulomb and a correlation term: $W (\omega) = v+ W_{\text c} (\omega)$. As detailed in Ref.~\cite{Leon2021PRB}, the frequency dependence of each matrix element $W^{\text c}_{\q \G \G'}$
can be described by a multipole model with a small number of complex poles for each transferred momentum, $\q$, and reciprocal lattice vectors, $\G \G'$. 
$W$ is then given by: 
\begin{multline}
    W^{\text{MPA}}_{\G\G'} (\q, z) = v_{\G\G'} (\q) + \sum_{p=1}^{n_W} R_{p \q\G\G'} \\
   \times  \left[ \frac{ 1}{z-\Omega_{p \q\G\G'}} - \frac{ 1}{z + \Omega_{p \q\G\G'}} \right] ,
    \label{eq:Xmp}
\end{multline}
where $\Omega_{p \q\G\G'}$ are the MPA poles and $R_{p \q\G\G'}$ their residues,  
and $n_W$, the number of poles. 
The time ordering of $W$ implies that $\Re [\Omega_{p \q\G\G'}] \times \Im [\Omega_{p \q\G\G'}] < 0$. 
Such poles represent effective plasmon-like quasiparticles emerging from a large set of single-particle transitions from valence to conduction states~\cite{FaridPRB1991,Leon2023PRB}. 

For each $W^{\text c}_{\q \G \G'}$ matrix element, all poles and residues are obtained through a non-linear interpolation of values numerically evaluated in a conveniently selected set of complex frequencies $\{ z_i, i=1,\ldots,2 n_W\}$. 
We use a frequency sampling along two lines parallel to the real axis (double-parallel sampling), typically along $\Im ~z=0.1$ and $\Im ~z=1$~Ha, respectively, with $\Re ~z_i$ distributed inhomogeneously. The double-parallel sampling, in particular the line of points with the largest imaginary part, reduces the noise resulting from the coarse Brillouin zone sampling of $W$. The inhomogeneous sampling distributions 
along the real axis are denser closer to the origin, following 
Eq.~(10) of Ref.~\cite{Leon2023PRB}, which limits the number of poles needed~\cite{Leon2021PRB,Leon2023PRB}.
We use two types of distributions, a linear and a quadratic semi-homogeneous partition, depending on the given system.
The frequency range is also specific for each system, since it must encompass the main structures of $W$. The classical plasmon energy, or the maximum single-particle transition from the valence to the conduction bands can be used as a reference energy scale in setting the sampling. More practical details, including a measure of the representability error, can be found in Refs.~\cite{Leon2021PRB,Leon2023PRB}.

With such an MPA representation, the frequency integral in the $G_0W_0$ self-energy of Eq.~\eqref{eq:GW} can be solved analytically. In a plane-wave basis set, it results in 
\begin{multline}
\Sigma_{n \kk}^{\text{MPA-}W} (z) = \Sigma_{n \kk}^{\text x}  + 
\sum\limits_m \sum\limits_{\G\G'}\sum\limits_{p=1}^{n_W}
     \int\frac{d\q}{(2\pi)^3}
     S^{nm}_{p\G\G'}(\kk,\q)
      \\ \times
    \left [ \frac{f_{m\kk-\q}^{\text{KS}}}
     {z-\epsilon_{m\kk-\q}^{\text{KS}}+\Omega_{p\q\G\G'}} + \frac{1-f_{m\kk-\q}^{\text{KS}}}
     {z-\epsilon_{m\kk-\q}^{\text{KS}}-\Omega_{p\q\G\G'}} \right ],
    \label{eq:Sc}
\end{multline}
where
\begin{equation}
\begin{aligned}
    S^{nm}_{p\G\G'}(\kk,\q) & \equiv -2 \rho_{nm}^{\text{KS}}(\kk,\q,\G)R_{p\q\G\G'}{\rho_{nm}^{\text{KS}}}^*(\kk,\q,\G') \\
    \rho_{nm}^{\text{KS}}(\kk,\q,\G) & \equiv \mel{n\mathbf{k}}{e^{i(\mathbf{q}+\mathbf{G}) \cdot \mathbf{r}}}{m\mathbf{k-q}}.
\end{aligned}
\end{equation}
Note that the time ordering of $W$ carries over to the time ordering of $\Sigma$, and since $\Im [\Omega_{p\q\G\G'}]$ is finite the vanishing imaginary part of the  $\epsilon_{m\kk-\q}^{\text{KS}}$ poles of $G_0$ can be disregarded. 
The derivative of $\Sigma^{\text{MPA-}W} (z)$ and therefore the linearized renormalization factor [see Eq.~\eqref{eq:qp_z}] can also be computed analytically as
\begin{multline}
\frac{\partial\Sigma_{n \kk}^{\text{MPA-}W} (z)}{\partial z} = 
-\sum\limits_m \sum\limits_{\G\G'}\sum\limits_{p=1}^{n_W}
     \int\frac{d\q}{(2\pi)^3}
     S^{nm}_{p\G\G'}(\kk,\q)
      \\ \times
    \left [ \frac{f_{m\kk-\q}^{\text{KS}}}
     {(z-\epsilon_{m\kk-\q}^{\text{KS}}+\Omega_{p\q\G\G'})^2} + \frac{1-f_{m\kk-\q}^{\text{KS}}}
     {(z-\epsilon_{m\kk-\q}^{\text{KS}}-\Omega_{p\q\G\G'})^2} \right ].
    \label{eq:ZmpaW}
\end{multline}
The case $n_W=1$ is analogous to the 
PPA approach, which only uses one or two frequency evaluations of $W$.  
On the other hand,
FF approaches on the real-axis can require as much as 1000 frequency points. By increasing $n_W$, MPA-$W$ typically provides an accuracy similar to FF with about 10 poles, interpolated from 20 frequency points. Thus, MPA-$W$ can be viewed as an effective FF approach  
requiring around 50 times fewer $W$ evaluations, with the corresponding savings in memory allocation~\cite{Leon2021PRB}.
The MPA-$W$ method is currently implemented in \yambo~\cite{Marini2009CPC,Sangalli2019JPCM} and \gpaw~\cite{gpaw2024JCP}.

\subsection{MPA for the self-energy}
\label{sec:MPA-S}
The MPA-$W$ representation of Eq.~\eqref{eq:Sc} shows that $\Sigma_{\text c}$, the correlation part of $\Sigma$, can be written as a sum of poles.
However, the evaluation of Eq.~\eqref{eq:Sc} for each frequency point still requires a large number of matrix multiplications due to the dependence of $\rho_{nm}^{\text{KS}}$, $\Omega_p$ and $R_p$ on the $\q \G \G'$ indices. 
To solve the linearized QP equation in Eq.~\eqref{eq:lQPeq}, 
$\Sigma$ only needs to be evaluated at two frequencies, or one if 
Eq.~\eqref{eq:ZmpaW} is used to compute the renormalization factor. 
However, to obtain spectral properties beyond the QP pole and the renormalization factor,  
a wide frequency range is needed.

To avoid the direct evaluation of the self-energy projection for each KS state on a dense frequency grid, $\Sigma_{n \kk}$ can be modeled as a simple multipole-Pad\'e approximant with a small number of $n_{\Sigma}$ poles that do not depend explicitly on $\q \G \G'$, but are consistent with Eq.~\eqref{eq:Sc}:
\begin{equation}
    \Sigma^{\text{MPA-}\Sigma}_{n \kk} (z) = \Sigma^{\text{x}}_{n \kk} + \sum_{p=1}^{n_{\Sigma}} \frac{S_{n \kk p}}{z-\xi_{n \kk p}}.
    \label{eq:SEmpa}
\end{equation}
The corresponding derivative is given by:
\begin{equation}
  \frac{\partial\Sigma^{\text{MPA-}\Sigma}_{n \kk} (z)}{\partial z} = -
  \sum_{p=1}^{n_{\Sigma}} \frac{S_{n \kk p}}{(z-\xi_{n \kk p})^2}.
    \label{eq:ZmpaS}
\end{equation}
Therefore, analogously to MPA-$W$, $\Sigma$ can be explicitly computed for a small number of frequency points used to interpolate the $\Sigma^{\text{MPA-}\Sigma}$ model. For each $n \kk$ (omitted for simplicity), the poles, $\xi_{p}$, and residues, $S_p$, are obtained by solving the following system of $2 n_{\Sigma}$ equations and variables:
\begin{equation}
   \Sigma_{\text c}^{\text{MPA-}\Sigma} (z_i) \equiv \sum_p^{n_{\Sigma}} \frac{S_p}{z_i-\xi_p} = 
     \Sigma_c (z_i)\text{, } i = 1, \ldots, 2 n_{\Sigma}.
    \label{eq:SmpaEqs}
\end{equation}

\begin{figure}
    \centering
    \includegraphics[width=0.49\textwidth]{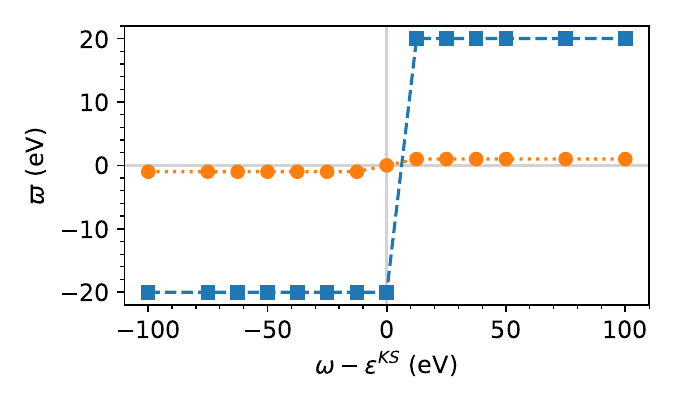}
    \caption{Example of an asymmetric MPA-$\Sigma$ sampling in the complex frequency plane with two branches close and far from the real axis, with imaginary part $\varpi=\pm 1$~eV (orange circles) and $\varpi=\pm 20$~eV (blue squares), each having six points in the positive side and eight in the negative one distributed according to the linear semi-homogeneous partition of Refs.~\cite{Leon2021PRB,Leon2023PRB}. 
    }
    \label{fig:sampling}
\end{figure}

The solution of Eq.~\eqref{eq:SmpaEqs} is obtained with a procedure similar to the one used in MPA-$W$, in which the correlation part of the MPA-$\Sigma$ model is rewritten in its Pad\'e form, i.e., as a fraction of two polynomials:
\begin{equation}
    \sum_p^{n_{\Sigma}} \frac{S_p}{z-\xi_{p}}
    = \frac{A_{n_{\Sigma} -1} (z)}{B_{n_{\Sigma}} (z)}.
    \label{eq:SEcMPAPade}
\end{equation}
The coefficients of the polynomial $B_{n_{\Sigma}} (z)$ can be evaluated from the numerical reference data, $\{z_i, \Sigma_c (z_i) \}$, using one of the two methods developed in Ref.~\cite{Leon2021PRB}, based on linear algebra and Thiele’s Pad\'e interpolation (see details in Sec.~I of \suppinfo). Moreover, its factorization can be performed using the companion matrix method~\cite{Leon2021PRB} and is given by
\begin{equation}
    B_{n_{\Sigma}} (z) = \prod_p^{n_{\Sigma}} (z - \xi_p).
    \label{eq:Spoles}
\end{equation}

In both methods the sampling points are divided into two sets, used to separate the problem of finding the poles, $\xi_p$, from the much simpler problem of finding the residues, $S_p$, once the poles are known. Such separation is computationally advantageous since the nonlinear problem of $2 n_{\Sigma}$ variables in Eq.~\eqref{eq:SmpaEqs}, is reduced to two problems of size $n_{\Sigma}$, one nonlinear for the poles and the other linear for the residues. Moreover, by first obtaining the poles, it is then possible to apply physical constraints. We impose the time ordering to the complex poles $\xi_p = \zeta_p + i \varsigma_p$, and that they lay in the vicinity of the real frequency axis, as done for MPA-$W$~\cite{Leon2021PRB}, which results in $\zeta_p/\varsigma_p < -1$. The residues $S_p$ can then be found by solving a simple linear least-squares problem (see details in Sec.~I of \suppinfo).   

As for MPA-$W$, an adequate frequency sampling of $\Sigma$ in the complex plane is essential to obtain an effective MPA-$\Sigma$ representation. We adopted the same type of inhomogeneous samplings parallel to the real axis used for MPA-$W$. 
Unlike $W$ [Eq.~\eqref{eq:Xmp}], $\Sigma$ is not symmetric in $\omega$ and therefore consists of single poles rather than pairs at $z= \pm \Omega_{p}$.  
For this reason, $\Sigma$ requires sampling along both the positive and negative axes, with a denser sampling in the region with the maximum variability. This corresponds to negative frequencies for the valence states, and positive for the conduction. 
A illustrated in Fig.~\ref{fig:sampling}, the sampling is chosen so that it complies with time ordering, having a small positive (negative) imaginary part for energies larger (smaller) than the KS energies, typically of $\varpi=\pm 0.1$~eV. The parallel sampling is done along the orange line, while the double parallel would use both the orange and the blue points.
Since $\Sigma$ has a smoother structure than $W$, it is sufficient to sample it along a single line parallel to the real frequency axis. 

The pole structure of $\Sigma$ is expected to resemble that of $W$, therefore the same sampling distribution can be used for $\Sigma$, as long as it is replicated on the negative side of the imaginary axis, as illustrated in the example of Fig.~\ref{fig:sampling}. The use of an even number of points including the origin results in an asymmetric distribution. 
Despite the need to sample $\Sigma$ along both the positive and negative parts of the real axis, using a single line allows us to use around the same number of sampling points as in the double-parallel sampling of $W$, typically about 20, for both MPA-$W$ and MPA-$\Sigma$.
More details on the MPA-$\Sigma$ sampling and the convergence with the number of poles $n_{\Sigma}$ can be found in Sec.~II of \suppinfo.

MPA-$\Sigma$ requires a particularly accurate interpolation around $z = 0$ for obtaining accurate QP energies, which sometimes  requires a more precise sampling. This can be done by benchmarking the sampling for one or a few selected QP states against the corresponding FF calculations. The sampling can then be replicated for the remaining QPs, without extending the FF calculations.
As in the case of MPA-$W$, the computational cost of the $\Sigma$ evaluation can be compared in terms of the number of frequency points for which $\Sigma$ is explicitly evaluated. Therefore, given a FF grid spacing of $\Delta \omega = 0.1$~eV in a frequency interval of $100$~eV,  MPA-$\Sigma$ is typically around 50 times more efficient than the FF $\Sigma$ evaluation.

\subsection{MPA for the Green's function}
\label{sec:MPA-G}
As in MPA-$W$ and MPA-$\Sigma$, one could construct a multipole-Pad\'e representation of the Green's function from the interpolation of the numerical data, $\{z_i, G(z_i)\}$.  
However, with MPA-$\Sigma$ in place, it is more convenient to obtain an MPA-$G$ representation 
from the Dyson equation in Eq.~\eqref{eq:G1_nk} (see also Refs.~\cite{Engel1991PRB, Gesenhue2017PRB} and the algorithmic-inversion-method in a sum-over-poles (AIM-SOP) 
representation of Refs.~\cite{Chiarotti2022PRR, Chiarotti2024PRR, Quinzi2025PRB}).

Given Eq.~\eqref{eq:SEcMPAPade}, the total MPA-$\Sigma$ can be written in its Pad\'e representation as 
\begin{equation}
    \Sigma^{\text{MPA-}\Sigma} (z) 
    = \frac{\Sigma_x B_{n_{\Sigma}} (z) + A_{n_{\Sigma} -1} (z)}{B_{n_{\Sigma}} (z)}.
    \label{eq:SEmpaPade}
\end{equation}
Notice that if we apply physical constraints after the factorization in Eq.~\eqref{eq:Spoles}, and fit the residues thereafter, we will need to reconstruct both the $A_{n_{\Sigma} -1} (z)$ and $B_{n_{\Sigma}} (z)$ polynomials from the new poles and residues using Eq.~\eqref{eq:SEmpaPade}, which is straightforward.
We can then obtain MPA-$G$ as
\begin{equation}
\begin{aligned}
    G^{\text{MPA-}\Sigma} (z) &\equiv \frac{B_{n_{\Sigma}} (z)}{C_{n_{\Sigma}+1} (z)}, \\
   C_{n_{\Sigma}+1} (z) &= z B_{n_{\Sigma}} (z) -\Sigma_x B_{n_{\Sigma}} (z) - A_{n_{\Sigma} -1} (z).
\end{aligned}
    \label{eq:GmpaPade}
\end{equation}
The poles of $G^{\text{MPA-}\Sigma}$ will then correspond to the zeros of $C_{n_{\Sigma}+1} (z)$. Analogous to $B_{n_{\Sigma}} (z)$, the $C_{n_{\Sigma}+1} (z)$ polynomial can be factorized, e.g., using the companion matrix method~\cite{Leon2021PRB}:
\begin{equation}
    C_{n_{\Sigma}+1} (z) = \prod_p^{n_{\Sigma}+1} (z - \epsilon_p),
\end{equation}
while the residues can be computed using the residue theorem. 
Constructed in this fashion, the poles of $G^{\text{MPA-}\Sigma}$ 
do not necessarily respect  
the time ordering, although this can be imposed in a second step.

The resulting multipole-Pad\'e representation of $G$ for each state $n \kk$ is then given by
\begin{equation}
    G^{\text{MPA-}\Sigma}_{n \kk} (z) = \sum_{p=1}^{n_{\Sigma} + 1} \frac{Z_{n \kk p}}{z-\epsilon_{n \kk p}} ,
    \label{eq:Gmpa}
\end{equation}
where the poles are obtained (see also Ref.~\cite{Chiarotti2022PRR}) as
\begin{equation}
    Z_{n \kk p} = \frac{\prod_i^{n_{\Sigma}} (\epsilon_{n \kk p} - \xi_{n \kk i})}{\prod_{i \neq p}^{n_{\Sigma} + 1} (\epsilon_{n \kk p} - \epsilon_{n \kk i})}.
    \label{eq:GmpaResidues}
\end{equation}
Notice that $G^{\text{MPA-}\Sigma}$ has one pole more than $\Sigma^{\text{MPA-}\Sigma}$, corresponding to the QP pole. As mentioned in Sec.~\ref{sec:qp_eqs} and as will be illustrated in Appendix.~\ref{sec:toy_models}, the remaining poles correspond to satellites that emerge from the poles of $\Sigma^{\text{MPA-}\Sigma}$. 
All the QP and satellites $\epsilon_{n \kk p}$ poles are solutions of the QP equation in Eq.~\eqref{eq:QPeq}.

The MPA-$G$ representation overcomes the limitations of the approximation introduced by the linearized QP equation in Eq.~\eqref{eq:lQPeq}, for both the positions and the residues of the QP poles. It also preserves important analytical properties of the solutions of non-linear eigenvalue equations with rational self-energy potentials~\cite{Guttel2017ActNum}.
From Eq.~\eqref{eq:GmpaResidues}, it follows that
\begin{equation}
    Z_{n \kk p} = \left[ 1-\frac{\partial\Sigma_{n \kk}^{\text{MPA-}\Sigma} (z)}{\partial z}\bigg|_{z=\epsilon_{n \kk p}} \right]^{-1},
   \label{eq:ZmpaG}
\end{equation}
while the sum rule of Eq.~\eqref{eq:sumrule_integralG} is obeyed and simplifies to 
\begin{equation}
    \sum_p Z_{n \kk p} = 1.
    \label{eq:sumrule_G}
\end{equation}
Therefore, the numerical accuracy of the QP and satellites spectral weights of Eq.~\eqref{eq:GmpaResidues} depends only on the quality of the MPA-$\Sigma$ interpolation, while always comply exactly with the sum rule for the number of particles and holes.

\subsection{Computational details}
\label{sec:comp_details}
DFT calculations were performed using the plane-wave {\qe} package~\cite{QE1,QE2} with the Perdew-Burke-Ernzerhof (PBE) variant of the generalized gradient approximation (GGA)~\cite{Perdew1996PRL}. 
We adopted the norm-conserving optimized Vanderbilt pseudopotentials of Ref.~\cite{Hamann2013PRB}, with a kinetic energy cutoff for the wave-functions of $70$, $30$, $100$, and $60$~Ry respectively for Na, Si, Cu, and monolayer MoS$_2$, and $85$~Ry for both the NaCl ion-pair and the F$_2$ molecule. The Brillouin zone was sampled with a $16\times16\times16$ Monkhorst-Pack grid for Si, Na, and Cu, $18\times18\times1$ for the monolayer MoS$_2$ and $\Gamma$-only for NaCl and F$_2$.

The $G_0W_0$ calculations were performed with {\yambo}~\cite{Marini2009CPC,Sangalli2019JPCM}.
In all the cases, the screened Coulomb potential was computed within MPA-$W$, using Eq.~\eqref{eq:Xmp} with $n_{W}=8$ for Si and Na, and $n_{W}=12$ for Cu, the same sampling as in Refs.~\cite{Leon2021PRB,Leon2023PRB}. Similarly, for MoS$_2$, NaCl and F$_2$ we used a sampling with 8 poles and a linear distribution. The method to evaluate the self-energy $\Sigma$, using full-frequency or the new MPA-$\Sigma$ method, is specified in each case.
Since we are considering a MoS$_2$ monolayer, we used the Monte-Carlo based averaging method ($W$-av) for 2D semiconductors, first developed in Ref.~\cite{Guandalini2023npjCM} and then merged with MPA-$W$ in Ref.~\cite{Guandalini2024PRB}.
For metals, we used the constant approximation (CA) method~\cite{Leon2023PRB} to treat the long-wavelength limit of the intraband contributions. Both, $W$-av and CA are methods that can greatly accelerate the $\kk$-point convergence of $GW$. 

 \begin{figure*}
    \centering
\includegraphics[width=0.995\textwidth]{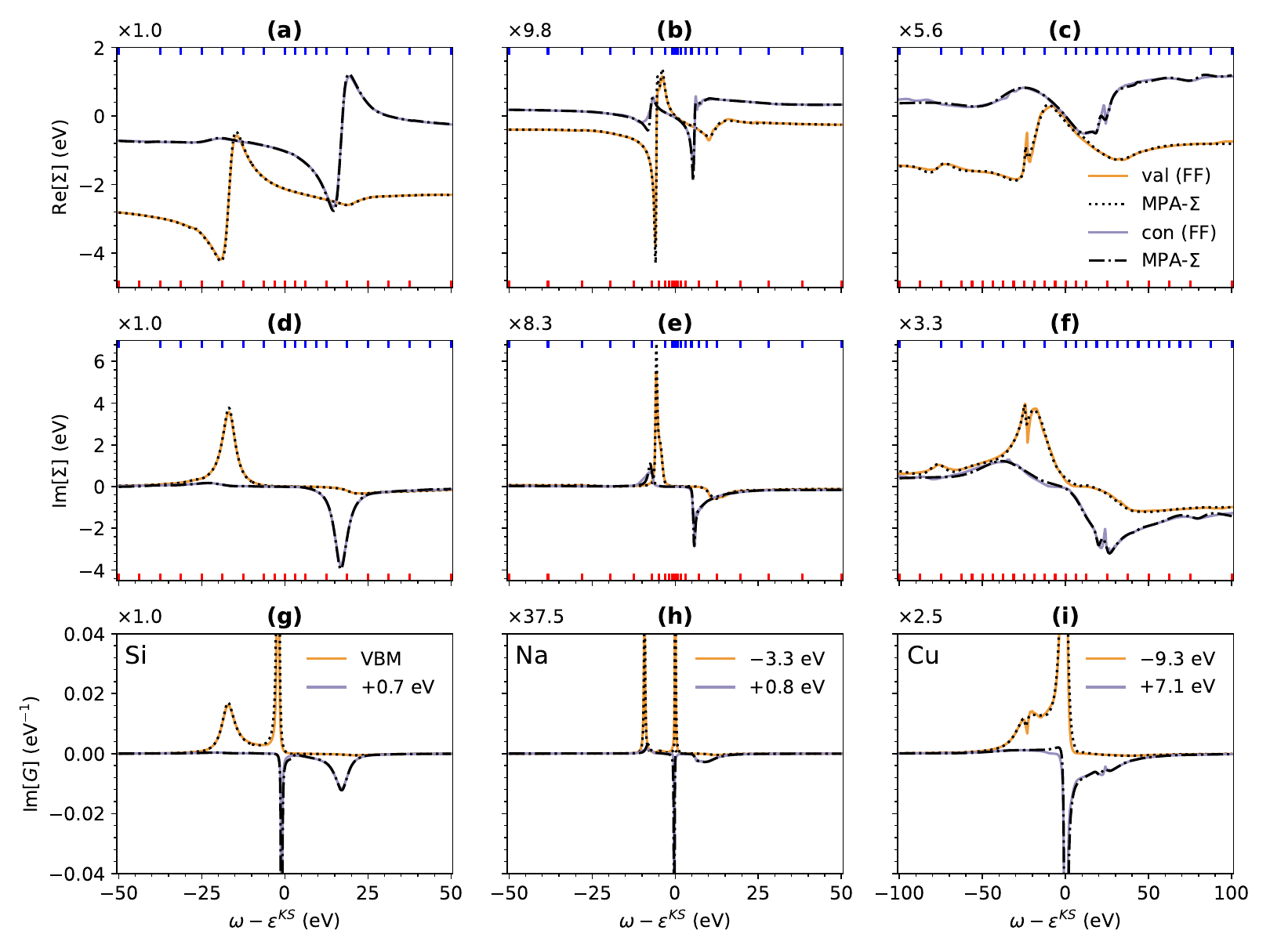}
    \caption{MPA-$\Sigma$ and MPA-$G$ results compared with the corresponding full-frequency (FF) $\Sigma$ and interacting $G$, for semiconducting and metallic materials. The plots show the spectra of a selected valence (yellow) and a conduction (purple) state of Si (left), Na (middle), and Cu (right), with KS energies relative to the valence band minimum (VBM), as specified in their respective panels. 
    (a)-(c) and (d)-(f) Compares the real and imaginary parts of $\Sigma$, and (g)-(i) the imaginary part of $G$, for the respective materials.
The solid orange and purple curves indicate the FF results, while the corresponding MPA-$\Sigma$ and MPA-$G$ results are plotted as black dotted (valence) and dashed-dotted (conduction) lines. 
In the top and center panels, the red (blue) ticks along the horizontal panel borders indicate the distribution of sampling points along the real frequency axis used as input in the MPA-$\Sigma$ interpolation of the given valence (conduction) state.}
\label{fig:Si-Na-Cu_mpaS}
\end{figure*}

 \begin{figure*}
    \centering
\includegraphics[width=0.995\textwidth]{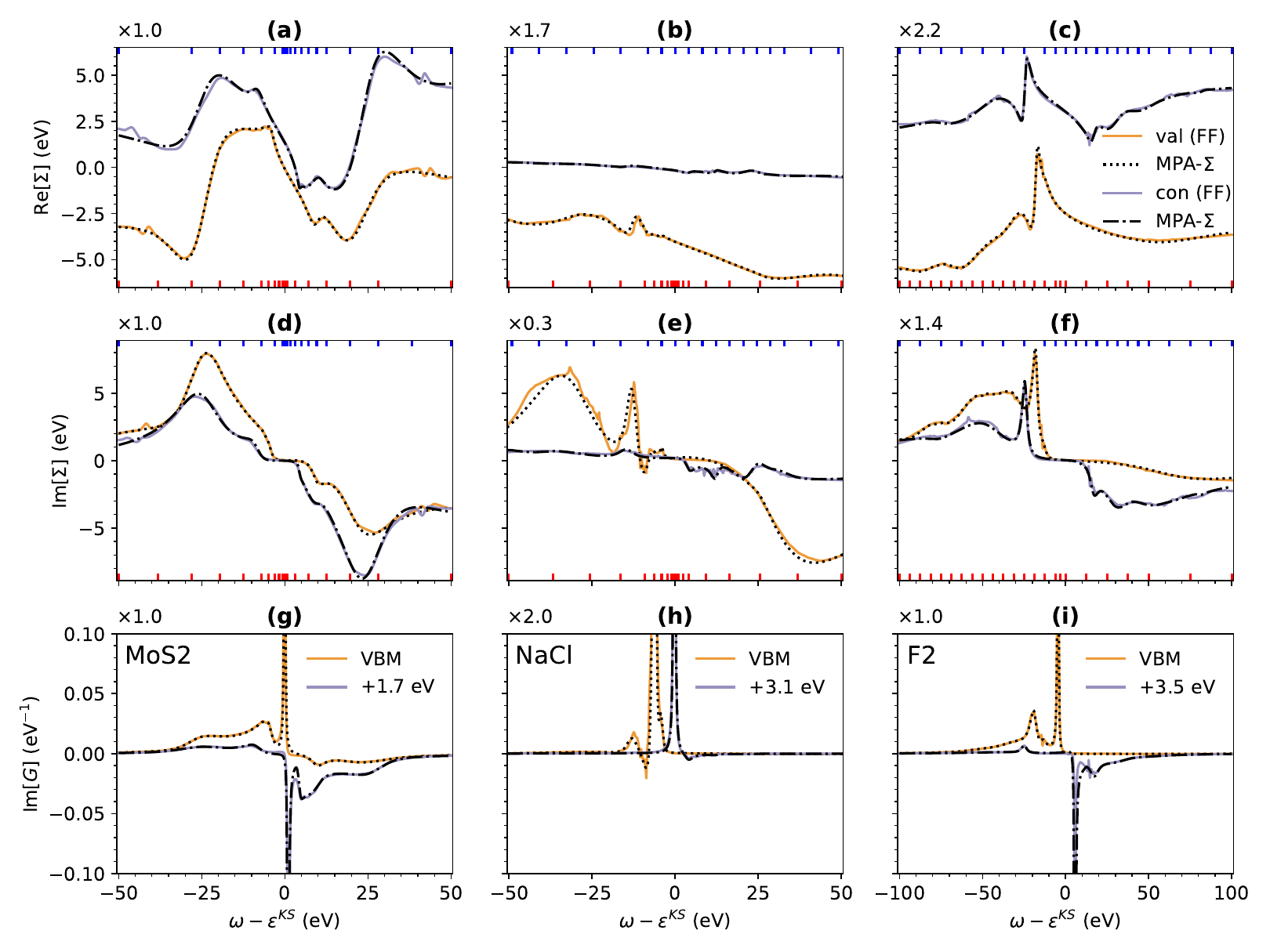}
    \caption{MPA-$\Sigma$ and MPA-$G$ results compared with the corresponding full-frequency (FF) $\Sigma$ and interacting $G$, for materials with reduced dimensionality. The plots show the spectra of a selected valence (yellow) and a conduction (purple) state of the monolayer MoS$_2$ (left), the NaCl ion-pair (middle), and the F$_2$ (right) molecule, with KS energies relative to the valence band minimum (VBM), as specified in their respective panels. 
    As in Fig.~\ref{fig:Si-Na-Cu_mpaS}, the MPA-$\Sigma$ and MPA-$G$ results are plotted as black dotted (valence) and dashed-dotted (conduction) lines, while the red (blue) ticks along the horizontal panel borders indicate the distribution of sampling points along the real frequency axis used as input in the MPA-$\Sigma$ interpolation of the given valence (conduction) state.
    }
    \label{fig:MoS2-F2_mpaS}
\end{figure*}

\section{Results}
\label{sec:results}

\subsection{Self-energy and Green's function of prototypical materials}
\label{sec:benchmarks}
Figure~\ref{fig:Si-Na-Cu_mpaS} shows the real and imaginary parts of the $G_0W_0$ self-energy, i.e. $\Re ~\Sigma$ (a)-(c) and $\Im ~\Sigma$ (d)-(f), and the imaginary part of the Green's function, $\Im ~G$ (g)-(i), for a selected valence and a conduction state of Si, Na, and Cu.
The solid lines give the results computed with a full-frequency evaluation of $\Sigma$ and $G$ (FF), serving as a benchmark, and the (dashed) dotted lines, the MPA-$\Sigma$ approach. 
The FF approach is evaluated on a homogeneous grid of 2000 frequency points. For MPA-$\Sigma$, 18 frequencies are used for Si and 22 for Na and Cu, corresponding to $n_{\Sigma}=9$ and $11$, respectively. 
The sampling frequencies used in the interpolation are indicated with red (valence) and blue (conduction) ticks along the horizontal panel edges.

 \begin{figure*}
    \centering
\includegraphics[width=0.995\textwidth]{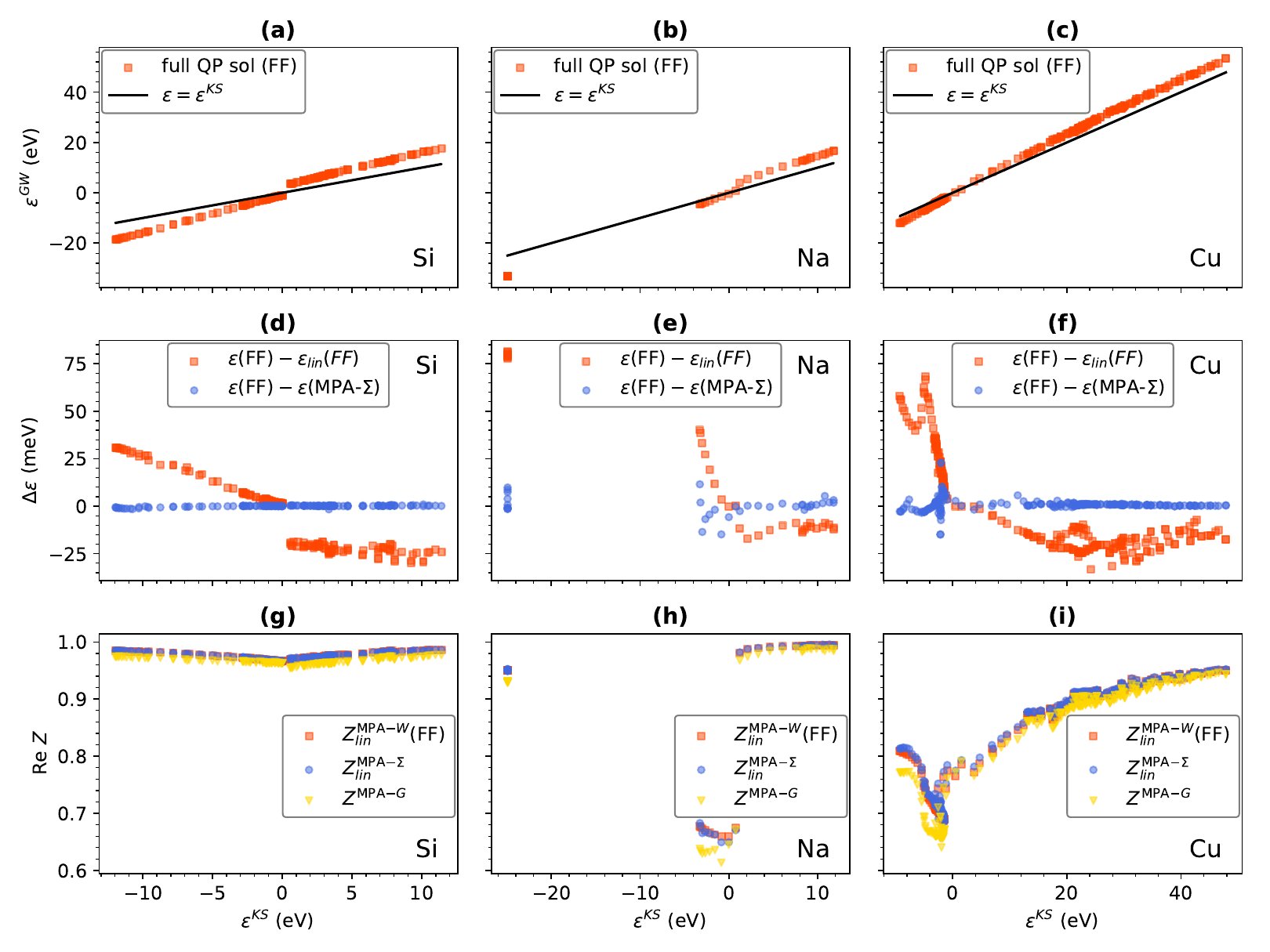}
    \caption{ QP energies and renormalization factors of Si [left: (a), (d), (g)], Na [middle: (b), (e), (h)] and Cu [right: (c), (f), (i)] as a function of their KS energies. (a)-(c) Show the QP solution obtained from the numerical full-frequency (FF) self-energy using a recursive method. (d)-(f) Show the difference between the FF QP equation and the FF linearized results (red squares) and the analytical MPA-$\Sigma$ solution (blue circles). (g)-(i) Correspond to 
    the linearized $\Re [Z^{\text{lin}}]$ obtained with the FF  MPA-$W$ approach of Eq.~\eqref{eq:ZmpaW} (red squares) and the MPA-$\Sigma$ representation of Eq.~\eqref{eq:ZmpaS} (blue circles), and the non-linearized $\Re [Z]$ computed as the residue of the QP pole in the MPA-$G$ representation of Eq.~\eqref{eq:GmpaResidues} (yellow triangles).
    }
    \label{fig:Na-Si-Cu_mpa_QP}
\end{figure*}

The self-energies of Si and Na have the typical two-pole structure characteristic of systems with a screening potential dominated by a single plasmon pole. As seen in the denominators of Eq.~\eqref{eq:Sc}, the two $\Sigma$ poles are the result of the plasmon convoluted, respectively, with valence and conduction states. Cu presents a similar picture, but with several plasmon-like poles in $W_0$, as can be seen in Ref.~\cite{Leon2023PRB}, coupled with the single-particle poles of $G_0$, resulting in a richer structure of $\Sigma$.

We also tested the MPA-$\Sigma$ description for the MoS$_2$ monolayer, the NaCl ion-pair, and the F$_2$ molecule, i.e., materials with lower dimensionality. The results are shown in Fig.~\ref{fig:MoS2-F2_mpaS}.
We used an MPA-$\Sigma$ representation with 10, 11 and up to 14 poles for MoS$_2$, NaCl and F$_2$, respectively. Both Figs.~\ref{fig:Si-Na-Cu_mpaS} and ~\ref{fig:MoS2-F2_mpaS} demonstrate an excellent agreement of MPA-$\Sigma$ and MPA-$G$ with the FF results.

 \begin{figure*}
    \centering
\includegraphics[width=0.995\textwidth]{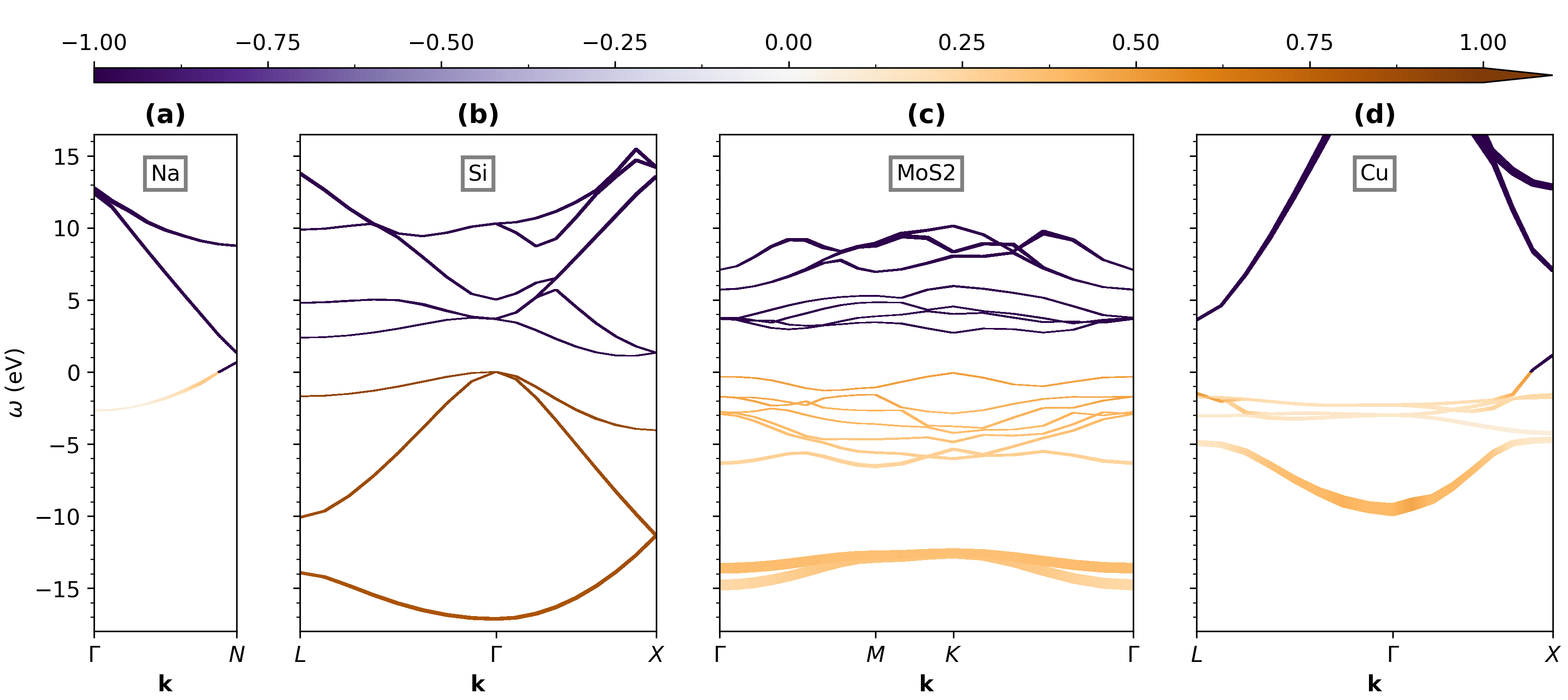}
    \caption{Band structure of Na, Si, monolayer MoS$_2$, and Cu, with a variable width representing the imaginary part of the quasiparticles $\Im [\epsilon_{n \kk}]$. The color map is given by $- \text{sign} \left ( \Re [\epsilon_{n\kk}] \right ) \Re [Z_{n\kk}]$, representing the renormalization factor for valence (orange shades) and conduction (purple shades) states.}
    \label{fig:bs_NaSiMoS2Cu}
\end{figure*}

In order to obtain an accurate MPA-$\Sigma$ representation while preserving the physical meaning of the main poles, in the above calculations, the number of sampling points of $\Sigma$ at each side of the frequency axis was adapted to each particular system and state, using the FF calculations as a reference. 
It is however convenient to establish a general sampling scheme that does not require 
a FF reference and can be applied to all the states of each system. For this purpose, we select a frequency distribution and number of poles similar to the one used for MPA-$W$. 
As mentioned in Sec.~\ref{sec:MPA-S}, in the case of $\Sigma$, the distribution is centered on each KS energy, with a small asymmetry on the frequency sampling, i.e., 1 or 2 frequency points more on the negative (positive) side for valence (conduction) states. The total number of frequency points is given by $2 n_{\Sigma}$, with $n_{\Sigma}=9$ for Si and Na, and $n_{\Sigma}=11$ for Cu.

Figure~\ref{fig:Na-Si-Cu_mpa_QP} shows the QP energies and $\Re ~Z$ of a set of valence and conduction bands of Si, Na, and Cu, in a wide range of energies and momenta. Figures~\ref{fig:Na-Si-Cu_mpa_QP}(a)-\ref{fig:Na-Si-Cu_mpa_QP}(c) show the $G_0W_0$ energies found by recursively solving the QP equation in Eq.~\eqref{eq:QPeq} with the FF $\Sigma$ representation, used here as a reference. 
They show the expected $GW$ stretching of the KS bands and, for Si, a band gap opening. 
Figures~\ref{fig:Na-Si-Cu_mpa_QP}(d)-\ref{fig:Na-Si-Cu_mpa_QP}(f) compare the difference between the results of the non-linearized FF and the linearized FF QP equation, $\Delta \varepsilon^{\text{lin-FF}} = \varepsilon^{\text{FF}} -\varepsilon^{\text{lin-FF}}$, (red squares), and the difference between the non-linearized FF and the analytical MPA-$\Sigma$ results, $\Delta \varepsilon^{\text{MPA-}\Sigma} = \varepsilon^{\text{FF}} -\varepsilon^{\text{MPA-}\Sigma}$, (blue circles).

For valence states $\Delta \varepsilon^{\text{lin-FF}}$ increases with the distance from the Fermi level, consistent with the general trends for materials (see, e.g., Ref.~\cite{Rasmussen2021NPJComputMater}). 
In contrast, $\Delta \varepsilon^{\text{MPA-}\Sigma}$ is almost zero for most of the valence and conduction states, with the exception of a few quasiparticles with more structure around $\Sigma (\omega = \varepsilon^{\text{KS}})$. 
This good agreement between MPA-$\Sigma$ and FF demonstrates the accuracy of the MPA-$\Sigma$ method, even with a simple frequency sampling scheme. 
As mentioned in Sec.~\ref{sec:MPA-S}, solving the linearized QP equation requires $\Sigma$ to be computed for one or two frequency points, whereas MPA-$\Sigma$ requires about 20. 
While the additional sampling points do increase computational costs, this comes with a significant improvement in accuracy. Critically, MPA-$\Sigma$ also allows for a straightforward evaluation of $\Sigma$ in its full-frequency range and gives access to an analytical representation of $G$.

Figures~\ref{fig:Na-Si-Cu_mpa_QP}(g)-\ref{fig:Na-Si-Cu_mpa_QP}(i)  show the linearized renormalization factor of Eq.~\eqref{eq:qp_z}, $Z^{\text{lin}}$, corresponding to the reference FF data evaluated with Eq.~\eqref{eq:ZmpaW} (red squares) and the MPA-$\Sigma$ representation of Eq.~\eqref{eq:ZmpaS} (blue circles). 
They differ by less than 0.003, 0.010, and 0.025 for all the quasiparticles of Si, Na, and Cu, respectively. 
The results labeled MPA-$G$ (yellow triangles) correspond to the non-linearized $Z$, obtained as the residue $Z_{p}$ of the QP pole in Eqs.~\eqref{eq:GmpaResidues} and \eqref{eq:ZmpaG}. Therefore, comparing the MPA-$\Sigma$ and MPA-$G$ results corresponds to comparing $Z^{\text{lin}}$ with $Z$.  
For most of the quasiparticles, the linearization is a good approximation and $Z^{\text{MPA-}\Sigma}$ is quite similar to $Z^{\text{MPA-}G}$; however, for the Na and Cu states with more intense satellites ($\Re ~Z \lesssim 0.8$), the deviation can increase to 0.045 and 0.070, respectively. 
A detailed analysis of 
the limitations of Eq.~\eqref{eq:qp_z}  
as an approximation of the spectral weight of the QP pole in different scenarios is presented 
in Appendix~\ref{sec:toy_models}. 
In particular, aside from the QP correction to the single-particle energies, it is shown that the deviation can increase with the QP broadening.

Figure~\ref{fig:bs_NaSiMoS2Cu} shows the QP band structure of Na (a), Si (b), monolayer MoS$_2$ (c), and Cu (d), interpolated in $\kk$-space, as described in Appendix~\ref{sec:spline_interpolation}. The width of the lines gives 
the imaginary part of the QP poles $\Im [\epsilon_{n \kk}]$, while the color shade indicates the value of the renormalization factors $\Re [Z_{n \kk}]$ for valence (orange shades) and conduction (purple shades) bands. 
In general, we find that both $\Im [\epsilon_{n \kk}]$ and $\Re [Z_{n \kk}]$ increase further away from the Fermi level, as a result of the QP pole and the satellites broadening and merging in a single peak. However, in the energy range in which multiple bands of different character cross, the picture becomes more complex showing a non-monotonic behavior. 
The metallic band of Na is an exception, with $\Im [\epsilon_{n \kk}]$ and $\Re [Z_{n \kk}]$ decreasing without crossing any other band. This is in line with the increased weight of the satellites, as discussed in Sec.~\ref{sec:spectral_bs}.

 \begin{figure*}
    \centering
\includegraphics[width=0.995\textwidth]{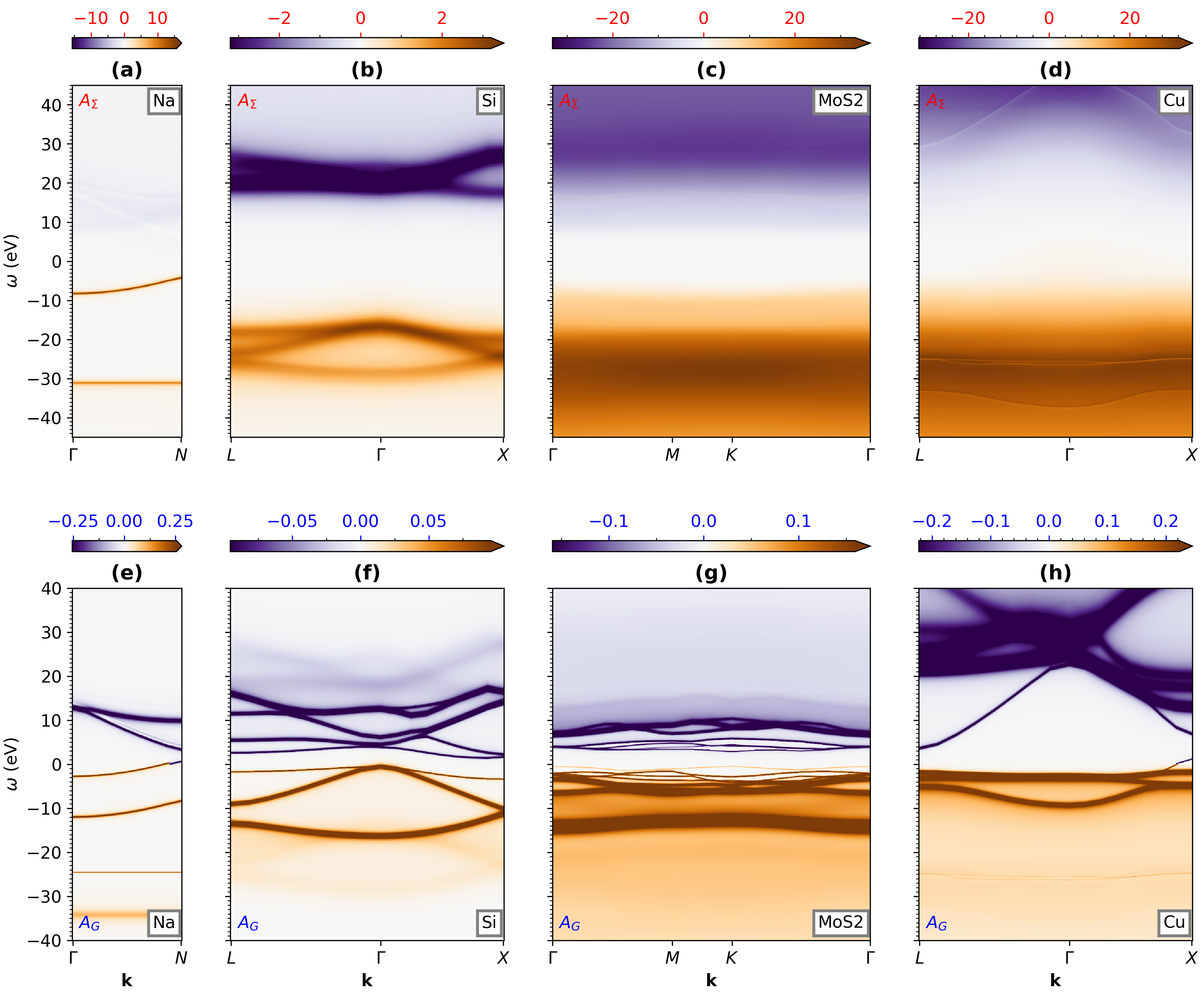}
    \caption{Spectral band structures $A_{\Sigma} (\kk, \omega)$ (a)-(d) and $A_{G} (\kk, \omega)$ (e)-(h) of Na (a), (e); Si (b), (f); MoS$_2$ (c), (g); and Cu (d), (h); computed with MPA-$\Sigma$ and MPA-$G$. 
    We considered the metallic band and the next valence and 2 conduction bands of Na along the $\Gamma N$ path, 4 valence and 6 conduction bands of Si along the $L \Gamma X$ path, 9 valence and 7 conduction bands of MoS$_2$ along the $\Gamma M K \Gamma$ path, and 6 valence and 9 conduction bands of Cu along the $L \Gamma X$ path.
    }
    \label{fig:bandstructureQP}
\end{figure*}

\subsection{Spectral band structures}
\label{sec:spectral_bs}
%
The spectral functions probed in photoemission and inverse photoemission experiments 
account for the contributions of each state, according to the polarization of the incoming light~\cite{hufner2013photoelectron}. In a typical {\it ab initio} calculation, the total $\Sigma$ and $G$ spectral functions, $A_{\Sigma} (\kk, \omega) \equiv 1/\pi \sum_n \Im [\Sigma_{n \kk} (\omega)]$ and $A_{G} (\kk, \omega) \equiv 1/\pi \sum_n \Im [G_{n \kk} (\omega)]$, are evaluated with a finite number of bands. Since we use time-ordered operators, the spectra have opposite signs for valence and conduction states, which makes the background intensity sensitive to the number of bands included. In Sec.~III of \suppinfo{} we provide detailed descriptions on how we plot spectral functions, especially when few bands are included.

Fig.~\ref{fig:bandstructureQP} shows the computed $A_{\Sigma} (\kk, \omega)$ (top panels) and $A_{G} (\kk, \omega)$ (bottom panels) spectral band structures of Na [Figs.~\ref{fig:bandstructureQP}(a) and \ref{fig:bandstructureQP}(e)], Si [Figs.~\ref{fig:bandstructureQP}(b) and \ref{fig:bandstructureQP}(f)], monolayer MoS$_2$ [Figs.~\ref{fig:bandstructureQP}(c) and \ref{fig:bandstructureQP}(g)] and Cu [Figs.~\ref{fig:bandstructureQP}(d) and \ref{fig:bandstructureQP}(h)], obtained with MPA-$\Sigma$ and MPA-$G$. Their accuracy is comparable with the spectral functions obtained with a FF evaluation (see Sec.~III of \suppinfo). As in Fig.~\ref{fig:bs_NaSiMoS2Cu}, to build such spectral band structures we use the spline interpolation in $\kk$-space detailed in Appendix~\ref{sec:spline_interpolation}.
Due to the time ordering, each state has positive (orange shades) and negative (purple shades) $1/\pi \Im [\Sigma_{n \kk} (\omega)]$ components, even if positive (negative) intensities are predominant for valence (conduction) states. 
Therefore, the spectral contribution of each state to the total spectral function $A_{\Sigma} (\kk, \omega)$ is not always trivial, as in the case of Na (see Fig.~S2 of \suppinfo).

$A_{\Sigma} (\kk, \omega)$ exhibits bands that arise from the coupling of the plasmon with the single KS states, as discussed in the previous section.
From now on we will call them $\Sigma$ bands.
Essentially, the position of the $\Sigma$ valence (conduction) bands corresponds to the independent-particle bands of $G_0$ shifted down (up) by the plasmon energy, $\zeta_{n \kk} \sim \varepsilon^{\text{KS}}_{n \kk} \pm \omega_{\text{pl}}$, where 
$\omega_{\text{pl}}$ has a value of $5.8$, $16.6$, $10.98$, and $26.5$~eV for Na, Si, MoS$_2$, and Cu respectively. Therefore, in semiconducting materials like Si, the $\Sigma$ bands present a gap given by the KS gap plus twice the plasmon energy ($\varepsilon_{\text{gap}}^{\Sigma} = \varepsilon_{\text{gap}}^{\text{KS}} + 2 \omega_{\text{pl}}$).

$A_{G} (\kk, \omega)$ (panels (e-h)) also exhibits bands that we will call $G$ bands. There are two types, those coming from the QP pole, $\varepsilon^{\text{QP}}_{n \kk}$, and those formed from the satellites, $\varepsilon^{\text{sat}}_{n \kk}$.
The QP bands are shifted with respect to the $G_0$ bands by the $G_0W_0$ correction, according to Eq.~\eqref{eq:QPeq}. The satellite bands (called sidebands in Ref.~\cite{martin2016book}) correspond to replicas of the QP bands located at larger energies.
As already seen in Figs.~\ref{fig:Si-Na-Cu_mpaS}(g)-\ref{fig:Si-Na-Cu_mpaS}(i), the QP peak typically dominates the spectral function.
To better discern the satellite structures, we impose thresholds to the $A_{G} (\kk, \omega)$ color maps, corresponding to $\pm 0.25$, $\pm 0.10$, $\pm 0.16$, and $\pm 0.22$~eV$^{-1}$ for Na, Si, MoS$_2$, and Cu, respectively.
In the case of the metallic band of Na, the intensity of the satellite around the $\Gamma$ point is similar in magnitude to the QP peak, which is interpreted as a plasmaron~\cite{Guzzo2011PRL, martin2016book, Caruso2016EPJB, Reining2018wcms}.
Since the satellites emerge from the plasmonic structures in $\Sigma$~\cite{Reining2018wcms}, the satellite bands are shifted with respect to the QP bands by roughly the energy of the plasmon~\cite{Gumhalter2016PRB}, $\varepsilon^{\text{sat}}_{n \kk} \sim \varepsilon^{\text{QP}}_{n \kk} \pm \omega_{\text{pl}}$, and, in turn, are shifted with respect to the $\Sigma$ bands by the $G_0W_0$ correction, $\varepsilon^{\text{sat}}_{n \kk}-\zeta_{n \kk} \sim \varepsilon^{\text{QP}}_{n \kk}-\varepsilon^{\text{KS}}_{n \kk}$.
One of the advantages of the MPA-$G$ representation is the possibility to analytically separate the spectral contributions of the QP pole and the satellites, as done in Fig.~S4 of \suppinfo.

Figure~\ref{fig:bandstructureQP} exhibits a qualitative difference between the $\Sigma$ bands of Na [Fig.~\ref{fig:bandstructureQP}(a)] and Si [Fig.~\ref{fig:bandstructureQP}(b)], which are easy to isolate, compared to those of MoS$_2$ [Fig.~\ref{fig:bandstructureQP}(c)] and Cu [Fig.~\ref{fig:bandstructureQP}(d)], which are generally broader and overlap more with each other. This is in line with both the more complex $\Sigma$ structure of MoS$_2$ and Cu, and the fact that the energy separation of the bands is smaller than the width of the main plasmon. 
As a result, MoS$_2$ exhibits rather flat and broadened valence and conduction $\Sigma$ bands, with an apparent gap given by secondary peaks at energies smaller than the main plasmon. The $\Sigma$ bands of Cu also shows a main flat and broadened dispersion, even if some secondary bands are still visible. As a consequence, the satellites bands of MoS$_2$ and Cu are also broadened, resulting in the diffuse background of Fig.~\ref{fig:bandstructureQP}~(g, h).

The results in Fig.~\ref{fig:bandstructureQP} illustrate 
the connection between the poles of $\Sigma$ and $G$, whose analyses allow to identify the QP pole from the satellites, as also illustrated with the toy models introduced in Appendix~\ref{sec:toy_models}.
Due to this connection, the accuracy of both the $\Sigma$ and $G$ bands requires a good description of the screening, which can be improved with vertex corrections~\cite{Guzzo2011PRL,Gumhalter2016PRB} or cumulant expansions~\cite{Gumhalter2016PRB, Caruso2016EPJB, Zhou2018PRB, Zhou2020pnas}. 
At finite temperature, electron-phonon interactions are expected to further renormalize and broaden the QP peak~\cite{Nery2018PRB,Zhou2020pnas}, as shown for example in Refs.~\cite{Gerlach2001PRB, Marini2002PRB, Tamai2013PRB} for the case of Cu.

\section{Conclusions}
\label{sec:comclusions}
We have presented MPA-$\Sigma$, a robust method to efficiently approximate the frequency dependence of the $GW$ self-energy as a multipole-Pad\'e representation, typically with around 10 poles. Such representation, similar to the multipole approximation for the screening interaction, MPA-$W$, is built from numerical data evaluated on around 20 frequency points in the complex plane, thus avoiding explicit evaluations of the self-energy on dense frequency grids (of the order of 1000 frequencies in our cases). 
MPA-$\Sigma$ allows also to solve the QP equation analytically and obtain an MPA-$G$ representation of the interacting Green's function, from which all the spectral properties can be easily extracted, including the positions, broadenings, and the spectral weights of the QP pole and its satellites.

Combining MPA-$W$ and MPA-$\Sigma$ is a computationally powerful approach, reducing the number of frequency evaluations by a factor $\sim 50^2$, compared to full-frequency approaches, while providing spectra with comparable numerical accuracy. 
The excellent accuracy of this method has been verified for several materials: bulk Si, Na, and Cu, monolayer MoS$_2$, the NaCl ion pair, and the F$_2$ molecule. 
The efficiency of the MPA method allows us to compute $\Sigma$ and $G$ spectra in a wide energy range. In particular, our results for NaCl and F$_2$ exhibit features beyond the typical energy range of the $GW$ spectra found in the literature for such molecular species. We have also presented a method for interpolating the spectra of higher-dimensional systems in momentum space, and construct $\Sigma$ and $G$ spectral band structures. For Na, Si, monolayer MoS$_2$, and Cu, we report full $\Sigma$ and $G$ spectral band structures, while isolating the contributions from the QP pole and the satellites.

 The spectral weights of both the QP pole and the satellites computed with MPA-$G$ comply with the sum rule for the number of particles and holes, providing an accurate way to evaluate the renormalization factor beyond the linearized QP equation. Therefore, such analytical representations can be useful in understanding the physical nature of the QP picture and the renormalization factor. 
 In the following Appendixes, we present toy models that capture the most typical situations when solving the QP equation in prototypical materials, exposing the limitations of its linearization in different regimes from weak to strong correlation.

\section*{Acknowledgments}
%
This work was funded by the Research Council of Norway through the MORTY project (315330). Access to high performance computing resources was provided by UNINETT Sigma2 (NN9711K) in Norway, and by EuroHPC Joint Undertaking through the project EHPC-EXT-2022E01-022 that grants access to Leonardo-Booster@Cineca, Italy.
We acknowledge Andrea Ferretti for insightful discussions and comments on the manuscript, and Kristian S. Thygesen and Mikael Kuisma for stimulating discussions.

\section{Data Availability Statements}
The data generated in this article are openly available~\cite{MPA-Sigma_data}.

\begin{appendix}
%
\section{The quasiparticle picture in the MPA representation}
\label{sec:toy_models}
%
As discussed in Sec.~\ref{sec:methods}, it is well known that the linearized renormalization factor 
in Eq.~\eqref{eq:qp_z} is not always a good approximation for the QP spectral weight, 
while there is no established method to compute the spectral weight of satellites. Equation~\eqref{eq:qp_z} is expected to work when the QP correction is small, although it is still widely used in {\it ab initio} calculations even in the case of large corrections, especially in high-throughput studies (see, e.g., Ref.~\cite{Rasmussen2021NPJComputMater}). In the following sections, we expose in detail some of the limitations of such linearization, while highlighting the advantages of the MPA-$G$ representation, consistent with Eqs.~\eqref{eq:ZmpaG} and~\eqref{eq:sumrule_G}, to analyze the QP picture in the limit of weak and strong correlation.
To that aim, we use toy MPA-$\Sigma$ models in which the number of poles is limited to one, $\Sigma^{\text{s1}}$, and two, $\Sigma^{\text{s2}}$, and their corresponding MPA-$G$ solutions, $G^{\text{s1}}$ and $G^{\text{s2}}$.
Such models, although simple, still capture the most typical situations when solving the QP equation in prototypical materials.

In the next sections we show that (1) the linearized QP equation can incorrectly predict values of $Z^{\text{lin}} < 0.5$ even when the QP picture holds ($Z > 0.5$). (2) The error introduced by the linearization, not only increases with the magnitude of the QP correction, but also with the QP broadening. (3) The static term in the self-energy plays an important role in the QP picture, affecting the distribution of the spectral weight between the QP pole and the satellites. For certain conditions, the satellites can have a larger spectral weight than the QP pole. (4) In the case of a self-energy with multiple poles, the linearized $Z^{\text{lin}}$ can deviate from the exact value $Z$, even in the limit 
$Z^{\text{lin}} \to 1$, which does not necessarily correspond to a vanishing linearized QP correction.

\subsection{Toy MPA-$\Sigma$ model with a single pole}
\label{sec:toy1}
%
 \begin{figure*}
    \centering
\includegraphics[width=0.995\textwidth]{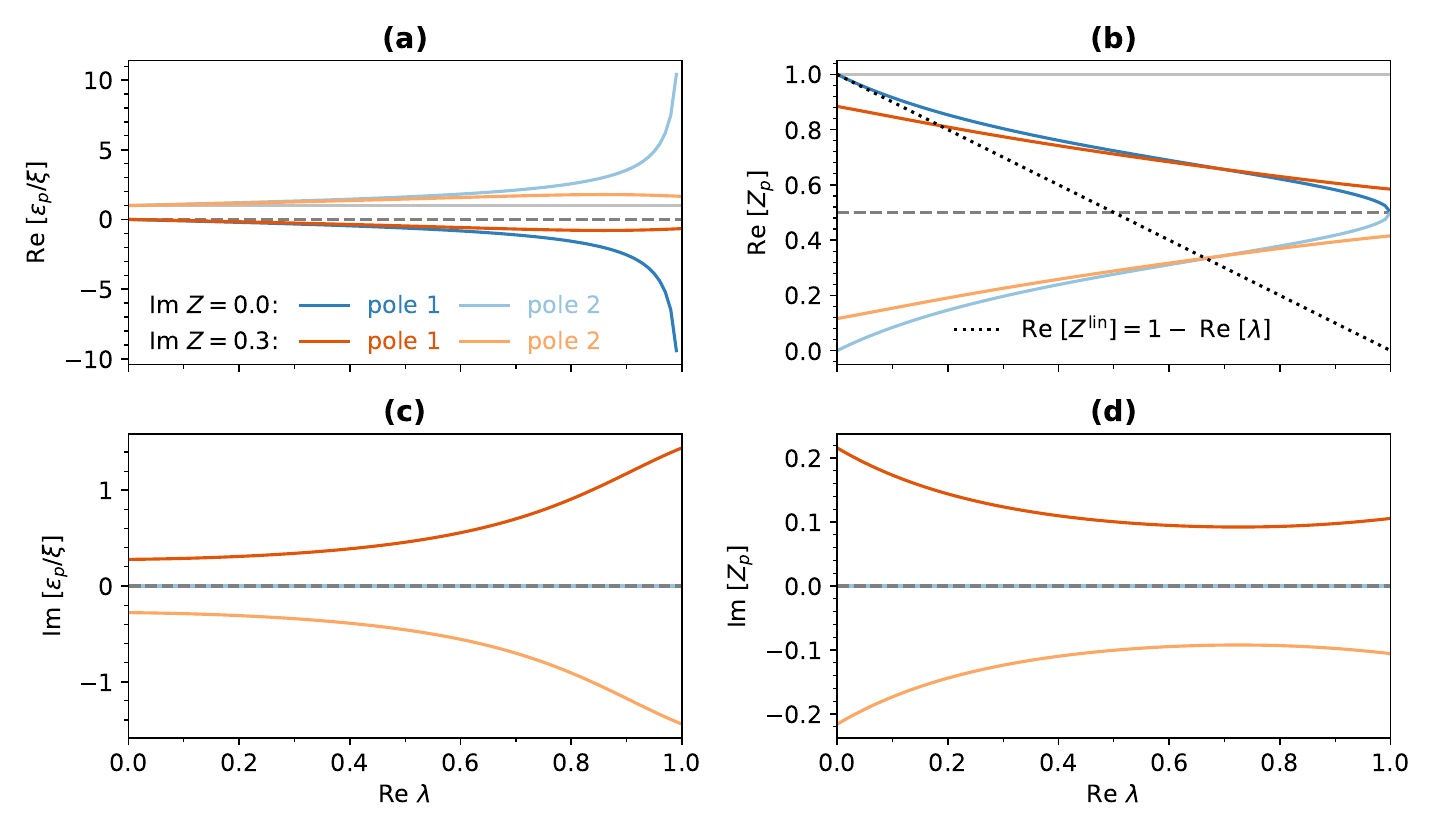}
    \caption{Real and imaginary parts of the scaled poles (a), (c) and residues (b), (d) of the Green's function model of Eq.~\eqref{eq:G1s} as a function of the $\lambda$ parameter, where we have fixed $x=0$. The solid horizontal gray line in (a) corresponds to $\Re [\epsilon_p/ \xi] = 1$, while a dashed line is drawn at zero. In (b), the gray dashed horizontal line at $\Re [Z_p] = 0.5$ represents the limit of the QP picture, while the gray solid line at 1 indicates the sum rule of Eq.~\eqref{eq:sumrule_G} ($Z_1 + Z_2 = 1$). The dotted black line corresponds to $\Re [Z^{\text{lin}}] = 1-\Re ~\lambda$, contrasted 
    with $\Re [Z^{\text{MPA-}G}_{\text{s1}}] = \Re [Z_1]$. 
    }
    \label{fig:qp_picture1}
\end{figure*}
%

%
%
We consider a self-energy model with one pole:  
\begin{equation}
    \Sigma^{\text{s1}} (\omega) = x\xi + \frac{\lambda}{1-\lambda} \frac{\xi^2}{\omega-\xi},
    \label{eq:S1s}
\end{equation}
where $\omega$ is centered on the given QP energy, $x\xi$ accounts for static contributions such as the exchange interaction and vertex corrections, and the second term accounts for correlation with a single pole at the plasmon energy $\xi$. The residue of the pole, 
$S = \xi^2 \lambda/(1-\lambda)$, is defined in terms of a parameter $\lambda$ so that the linearized renormalization factor in Eq.~\eqref{eq:qp_z}, independently of the other two parameters, is given by:
\begin{equation}
  Z^{\text{lin}}_{\text{s1}} \equiv  \left[ 1- \frac{\partial\Sigma^{\text{s1}} (\omega)}{\partial \omega}\bigg|_{\omega=0} \right]^{-1} = 1-\lambda.
    \label{eq:qp_z_toy}
\end{equation}
The interacting Green's function corresponding to Eq.~\eqref{eq:S1s} is obtained by inverting the Dyson equation, $G(\omega) = [\omega - \Sigma(\omega)]^{-1}$, resulting in:
\begin{equation}
    G^{\text{s1}} (\omega) = \sum_{p=1}^2 \frac{Z^{\text{s1}}_p(x, \lambda)}{\omega - \epsilon^{\text{s1}}_p(x, \lambda, \xi)},
    \label{eq:G1s}
\end{equation}
which has two poles, $\epsilon^{\text{s1}}_p$, with residues $Z^{\text{s1}}_p$, given by
\begin{eqnarray}
\begin{aligned}
    \epsilon^{\text{s1}}_{1,2} &= \frac{\xi}{2} \left (1+x \mp \sqrt{D} \right )
    \label{eq:epsilon_s1}, \\
    Z^{\text{s1}}_{1,2} &= \frac{1}{2} \left (1 \pm \frac{1-x}{\sqrt{D}} \right ),
    \label{eq:P_s1}
\end{aligned}
\end{eqnarray}
where
\begin{equation}
    D \equiv (1-x)^2 +4\lambda/(1-\lambda).
\end{equation}
The poles are proportional to $\xi$ and can be rescaled to obtain dimensionless units.
For all the considered values of the parameters, $|\Re [\epsilon^{\text{s1}}_{1}]| < |\Re [\epsilon^{\text{s1}}_{2}]|$. The pole labeled as 1 is identified as the QP pole and the one labeled as 2 is the satellite, thus, $Z^{\text{MPA-}G}_{\text{s1}} = Z^{\text{s1}}_{1}$.

 \begin{figure*}
    \centering
\includegraphics[width=0.995\textwidth]{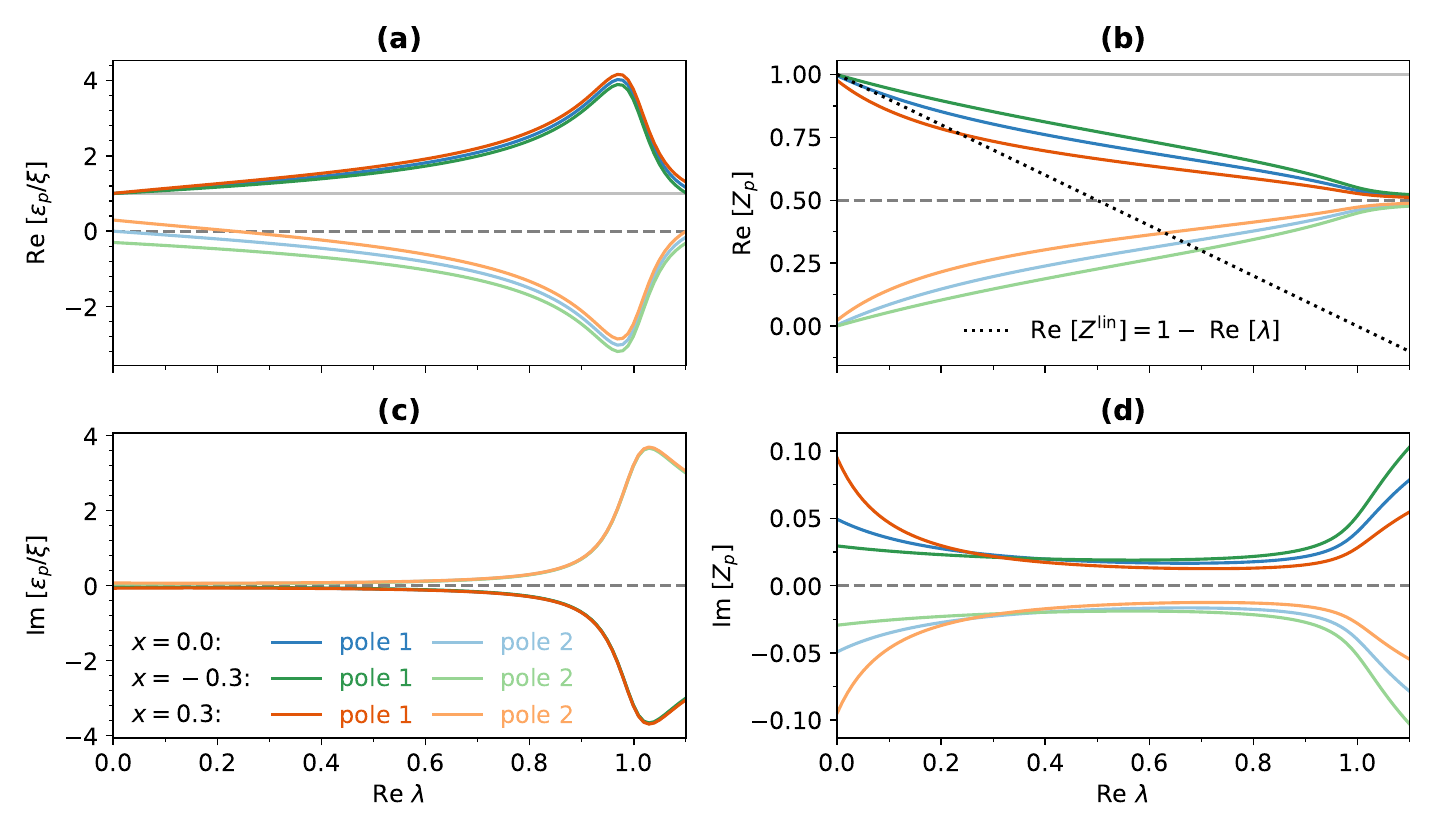}
    \caption{Real and imaginary parts of the scaled poles (a), (c) and residues (b), (d) of the Green's function model of Eq.~\eqref{eq:G1s} as a function of the $\lambda$ parameter, where we have fixed its imaginary part to a small value of $\Im ~\lambda = 0.01$. The solid and dashed horizontal gray lines and the black dotted line in (b) are analogous to Fig.~\ref{fig:qp_picture1}.}
    \label{fig:qp_picture4}
\end{figure*}

We first focus on the correlation effects by setting $x=0$. Notice that $\lambda=0$ corresponds to $S=0$ (zero correlation), while $\lambda \to 1$ corresponds to the limit of infinite correlation.
Figure~\ref{fig:qp_picture1} shows the real and imaginary parts of the scaled poles $\epsilon^{\text{s1}}_p/\xi$ [Figs.~\ref{fig:qp_picture1}(a) and \ref{fig:qp_picture1}(c)], and the residues $Z_p^{\text{s1}}$ [Figs.~\ref{fig:qp_picture1}(b) and \ref{fig:qp_picture1}(d)] of $G^{\text{s1}}$ as functions of $\Re ~\lambda$, for $\Im ~\lambda = 0$ (blue curves) and $\Im ~\lambda = 0.3$ (orange curves).
Since $x=0$, $\Re ~\lambda = 0$ and $\Im ~\lambda = 0$ also correspond to the independent-particle limit, in which the QP state has a zero self-energy correction ($\epsilon^{\text{s1}}_1 = 0$) and carries the whole spectral weight ($Z^{\text{MPA-}G}_{\text{s1}} = 1$). As a consequence, the satellite $\epsilon^{\text{s1}}_2 = \xi$ vanishes ($Z^{\text{s1}}_2 = 0$).
 
In both the cases of $\Im ~\lambda=0$ and $0.3$, as $\Re ~\lambda$ increases, 
$\Re [\epsilon^{\text{s1}}_1]$ goes to negative values with decreasing spectral weight $\Re [Z^{\text{s1}}_1]$, while $\Re [\epsilon^{\text{s1}}_2]$ increases from $\xi$ (solid gray line in panel (a)), with increasing $\Re [Z^{\text{s1}}_2]$. 
The finite $\Im ~\lambda$ induces a finite imaginary part in the poles and residues. As $\Re ~\lambda$ increases, both $\Im [\epsilon^{\text{s1}}_p]$ increase in modulus, while $\Im [Z^{\text{s1}}_p]$ decrease despite $\Im ~\lambda$ being constant. The broadening of the poles avoids the infinite correlation limit, affecting the curvature of $\Re [\epsilon^{\text{s1}}_p]$ and $\Re [Z^{\text{s1}}_p]$ (orange vs. blue curves in panels (a, b)).

As illustrated by the dotted black line in Fig.~\ref{fig:qp_picture1}(b), when we move from the independent-particle limit, the renormalization factor {$Z_{\text{s1}}^{\text{lin}}$} starts deviating from $Z_{\text{s1}}^{\text{MPA-}G}$ (dark blue curve), while for $\Im ~\lambda = 0.3$ (dark orange curve), both definitions already differ at $\Re ~\lambda = 0$ and only their real parts coincide around $\Re ~\lambda = 0.2$. 
The plot shows that, $Z^{\text{lin}}$ cannot be taken as an indicator of the validity of the QP picture beyond the regime of weak correlation ($\lambda \approx 0$), 
since $\Re [Z_{\text{s1}}^{\text{MPA-}G}] > 0.5$ in the whole interval, even when $\Re [Z_{\text{s1}}^{\text{lin}}] < 0.5$. 
In the large correlation limit ($\lambda \to 1$), the energy position of the poles diverges ($\Re [\epsilon^{\text{s1}}_{1,2}] \to \mp \infty$) with similar spectral weight ($\Re [Z^{\text{s1}}_{1,2}] \to 0.5$). 

In Fig.~\ref{fig:qp_picture4} we analyze the effects of the static term. The plots are analogous to Fig.~\ref{fig:qp_picture1}. Notice that we have extended the range of the plots to $\Re ~\lambda = 1.1$, to test the QP picture even for extreme values. In this case, we have fixed $\Im ~\lambda = 0.01$ and considered three values of $x$, resulting in a picture similar to the one described in Fig.~\ref{fig:qp_picture1}, where $\Re [Z^{\text{s1}}_1]$ is always larger than $0.5$. The value $x = 0$ corresponds to the previously discussed case of a self-energy with only the correlation term. The main effect of a finite value, illustrated with $x = \pm 0.3$, is to change the concavity of the residues according to its sign (orange/green vs. blue curves in panel (b)). At variance with $x \le 0$, for positive values $\Re [Z_{\text{s1}}^{\text{MPA-}G}] 
< \Re [Z_{\text{s1}}^{\text{lin}}]$ close to $\Re ~\lambda \to 0$.
In the case of $x > 1$, the spectral weight of the satellite is larger than the QP pole, as $ \Re [Z^{\text{s1}}_1] < \Re [Z^{\text{s1}}_2]$.

\begin{figure*}
    \centering
\includegraphics[width=0.995\textwidth]{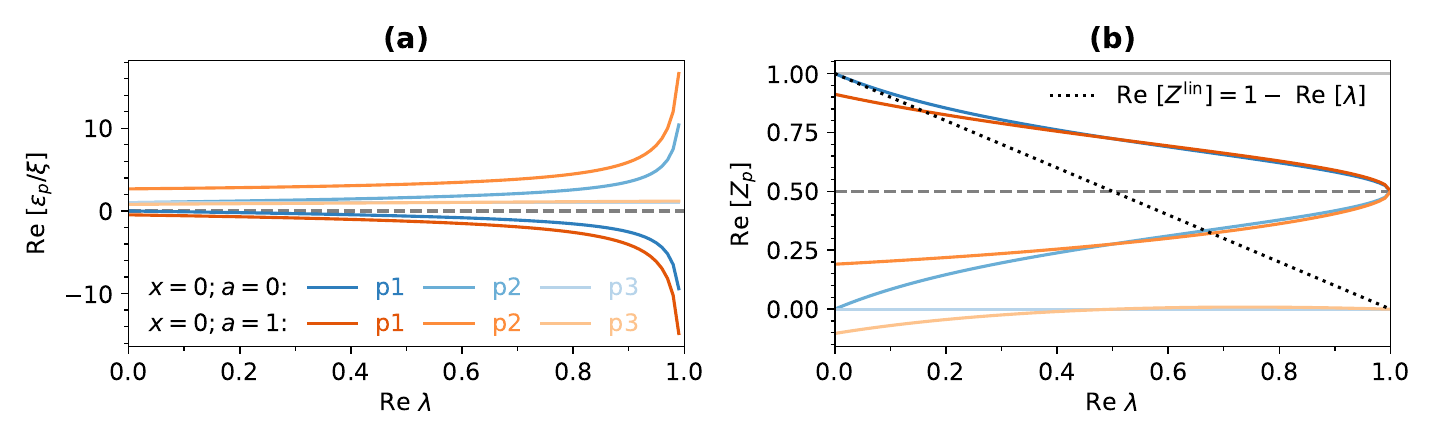}
    \caption{Real part of the poles (a) and residues (b) of the Green's function model of Eq.~\eqref{eq:G2s} as a function of $\Re ~\lambda$, where we have fixed $\Im ~\lambda = 0$. 
    }
    \label{fig:qp_picture3}
\end{figure*}

\subsection{Toy MPA-$\Sigma$ model with two poles}
\label{sec:toy2}
Similarly to the $\Sigma^{\text{s1}}$ model discussed in Sec.~\ref{sec:toy_models}, here we introduce a toy MPA-$\Sigma$ model with two poles, one at $\epsilon_1=\xi$ and the other at a larger energy $\epsilon_2=(1+a)\xi$:
%
%
\begin{multline}
    \Sigma^{\text{s2}} (\omega) = x\xi + \\ \frac{2 \lambda-1}{2(1 - \lambda)} \frac{\xi^2}{\omega-\xi} +  
    \frac{(1+a)^2}{2 (1-\lambda)} \frac{\xi^2}{\omega - (1+a) \xi}.
    \label{eq:S2s}
\end{multline}
where the residues, $S_1 = \xi^2 (2 \lambda -1)/[2 (1-\lambda)]$ and $S_2 = \xi^2 (1+a)^2/[2 (1-\lambda)]$ are constrained so that the following expression remains invariant, as for $\Sigma^{\text{s1}}$:
\begin{equation}
   Z^{\text{lin}}_{\text{s2}} \equiv \left[ 1- \frac{\partial\Sigma^{\text{s2}} (\omega)}{\partial \omega}\bigg|_{\omega=0} \right]^{-1} = 1-\lambda.
    \label{eq:qp_z_toy}
\end{equation}
The corresponding MPA-$G$ representation has three poles and a similar form, only with the additional parameter $a$:
\begin{equation}
    G^{\text{s2}}(\omega) = \sum_{p=1}^3 \frac{Z^{\text{s2}}_p(x, \lambda, a)}{\omega - \epsilon^{\text{s2}}_p(x, \lambda, a, \xi)}.
    \label{eq:G2s}
\end{equation}

Notice that for $a = 0$, $\Sigma^{\text{s2}}$ simplifies to $\Sigma^{\text{s1}}$, and therefore this parameter can be used to turn on the second pole. We then fix $x = 0$ and $\Im ~\lambda = 0$, and compare the two models. Figure~\ref{fig:qp_picture3} 
shows the real part of the scaled poles, $\epsilon^{\text{s2}}_p/\xi$ [Fig.~\ref{fig:qp_picture3}(a)], and the residues, $Z_p^{\text{s2}}$ [Fig.~\ref{fig:qp_picture3}(b)], of $G^{\text{s2}}$ as functions of $\Re ~\lambda$, for $a = 0$ (blue curves) and $a = 1$ (orange curves). 
The overall picture discussed in Sec.~\ref{sec:toy_models} is similar for the two models, 
the first one having a satellite increasing from $\epsilon^{\text{s1}}_2 = \xi$, while the second has two, one increasing from $\epsilon^{\text{s1}}_2 = (1+a)\xi$ and the second remaining around $\epsilon^{\text{s2}}_3 = \xi$ with a vanishing residue $Z^{\text{s2}}_3$. 
However, a finite $a$ induces a finite $Z^{\text{s2}}_3$, with its larger modulus at $\Re ~\lambda = 0$, which introduces a deviation between $Z^{\text{lin}}_{\text{s2}}$ and $Z^{\text{MPA-}G}_{\text{s2}} 
$ around $\lambda = 0$. 
Notice that even if $Z^{\text{lin}}_{\text{s2}}=1$ for $\lambda=0$, the linearized QP correction does not vanish for finite $a$, as $\Sigma^{\text{s2}} (\omega=0)/\xi = -(2l+a)/[2(1-l)]$. The weak correlation limit is found for both $\lambda \to 0$ and $a \to 0$ simultaneously.

\section{Spline interpolation in $\kk$-space}
\label{sec:spline_interpolation}
To obtain smooth plots of the $\Sigma$ and $G$ spectral band structures, similar to the band structures of the response function computed in Ref.~\cite{Leon2024PRB}, we need to perform an interpolation on the $\kk$ space for each frequency $z_i$. Such an interpolation can be cumbersome due to the dispersion of the different bands.  
The MPA-$\Sigma$ and MPA-$G$ representations can simplify this interpolation, since it is sufficient to interpolate the poles and the residues of each multipole-Pad\'e model.
The interpolation of numerical data can also be simplified by considering an auxiliary set of frequencies centered on the KS energies $z'_i \equiv z_i - \epsilon^{\text{KS}}_{n \kk}$. Since $z'_i$ carries the dispersion of the bands, $\Sigma (z'_i, \kk)$ and $G (z'_i, \kk)$ are much smoother than the original $\Sigma (z_i, \kk)$ and $G (z_i, \kk)$.

We start by interpolating each KS band $n$:
\begin{equation}
    \forall n : \\{\epsilon^{\text{KS}}_{n} ( \kk\text{-grid}) } \to f^{\epsilon}_n(\kk),
\end{equation}
where $f^{\epsilon}_n$ is the interpolating function. Similarly, we interpolate $\Sigma$ and $G$ for all frequencies along the direction of each band:
\begin{equation}
\begin{aligned}
    \forall n, i: &&& {\Sigma_n (z'_i, \kk\text{-grid}) } \to f^{\Sigma}_n (z'_i, \kk) \\
   &&& {G_n(z'_i, \kk\text{-grid}) } \to f^{G}_n(z'_i, \kk),
\end{aligned}
\end{equation}
where $f^{\Sigma}_n$ and $f^{G}_n$ are the interpolating functions of $\Sigma$ and $G$, respectively. 
The interpolation for the $\{z_i\}$ frequencies can then be obtained as
\begin{equation}
    f^{\Sigma/G}_n(z_i, \kk) = f^{\Sigma/G}_n[z'_i + f^{\epsilon}_n(\kk), \kk],
\end{equation}
which can be used to evaluate the final spectra on a much denser $\kk$-grid. 
We have used first-order splines as interpolating functions, which is sufficient to obtain smooth spectra for our calculations, although this approach can be applied with other interpolating functions as well. 

\end{appendix}

%
%
\bibliography{biblio}

\begin{thebibliography}{86}%
\makeatletter
\providecommand \@ifxundefined [1]{%
 \@ifx{#1\undefined}
}%
\providecommand \@ifnum [1]{%
 \ifnum #1\expandafter \@firstoftwo
 \else \expandafter \@secondoftwo
 \fi
}%
\providecommand \@ifx [1]{%
 \ifx #1\expandafter \@firstoftwo
 \else \expandafter \@secondoftwo
 \fi
}%
\providecommand \natexlab [1]{#1}%
\providecommand \enquote  [1]{``#1''}%
\providecommand \bibnamefont  [1]{#1}%
\providecommand \bibfnamefont [1]{#1}%
\providecommand \citenamefont [1]{#1}%
\providecommand \href@noop [0]{\@secondoftwo}%
\providecommand \href [0]{\begingroup \@sanitize@url \@href}%
\providecommand \@href[1]{\@@startlink{#1}\@@href}%
\providecommand \@@href[1]{\endgroup#1\@@endlink}%
\providecommand \@sanitize@url [0]{\catcode `\\12\catcode `\$12\catcode `\&12\catcode `\#12\catcode `\^12\catcode `\_12\catcode `\%12\relax}%
\providecommand \@@startlink[1]{}%
\providecommand \@@endlink[0]{}%
\providecommand \url  [0]{\begingroup\@sanitize@url \@url }%
\providecommand \@url [1]{\endgroup\@href {#1}{\urlprefix }}%
\providecommand \urlprefix  [0]{URL }%
\providecommand \Eprint [0]{\href }%
\providecommand \doibase [0]{https://doi.org/}%
\providecommand \selectlanguage [0]{\@gobble}%
\providecommand \bibinfo  [0]{\@secondoftwo}%
\providecommand \bibfield  [0]{\@secondoftwo}%
\providecommand \translation [1]{[#1]}%
\providecommand \BibitemOpen [0]{}%
\providecommand \bibitemStop [0]{}%
\providecommand \bibitemNoStop [0]{.\EOS\space}%
\providecommand \EOS [0]{\spacefactor3000\relax}%
\providecommand \BibitemShut  [1]{\csname bibitem#1\endcsname}%
\let\auto@bib@innerbib\@empty
\bibitem [{\citenamefont {Aryasetiawan}\ and\ \citenamefont {Gunnarsson}(1998)}]{Aryasetiawan1998RPP}%
  \BibitemOpen
  \bibfield  {author} {\bibinfo {author} {\bibfnamefont {F.}~\bibnamefont {Aryasetiawan}}\ and\ \bibinfo {author} {\bibfnamefont {O.}~\bibnamefont {Gunnarsson}},\ }\bibfield  {title} {\bibinfo {title} {The {GW} method},\ }\href {https://doi.org/https://doi.org/10.1088/0034-4885/61/3/002} {\bibfield  {journal} {\bibinfo  {journal} {Rep. Prog. Phys.}\ }\textbf {\bibinfo {volume} {61}},\ \bibinfo {pages} {237} (\bibinfo {year} {1998})}\BibitemShut {NoStop}%
\bibitem [{\citenamefont {Hedin}(1999)}]{Hedin1999JPCM}%
  \BibitemOpen
  \bibfield  {author} {\bibinfo {author} {\bibfnamefont {L.}~\bibnamefont {Hedin}},\ }\bibfield  {title} {\bibinfo {title} {On correlation effects in electron spectroscopies and the gw approximation},\ }\href {https://doi.org/10.1088/0953-8984/11/42/201} {\bibfield  {journal} {\bibinfo  {journal} {Journal of Physics: Condensed Matter}\ }\textbf {\bibinfo {volume} {11}},\ \bibinfo {pages} {R489} (\bibinfo {year} {1999})}\BibitemShut {NoStop}%
\bibitem [{\citenamefont {Onida}\ \emph {et~al.}(2002)\citenamefont {Onida}, \citenamefont {Reining},\ and\ \citenamefont {Rubio}}]{Onida2002RMP}%
  \BibitemOpen
  \bibfield  {author} {\bibinfo {author} {\bibfnamefont {G.}~\bibnamefont {Onida}}, \bibinfo {author} {\bibfnamefont {L.}~\bibnamefont {Reining}},\ and\ \bibinfo {author} {\bibfnamefont {A.}~\bibnamefont {Rubio}},\ }\bibfield  {title} {\bibinfo {title} {Electronic excitations: density-functional versus many-body green’s-function approaches},\ }\href {https://doi.org/https://doi.org/10.1103/RevModPhys.74.601} {\bibfield  {journal} {\bibinfo  {journal} {Rev. Mod. Phys.}\ }\textbf {\bibinfo {volume} {74}},\ \bibinfo {pages} {601} (\bibinfo {year} {2002})}\BibitemShut {NoStop}%
\bibitem [{\citenamefont {Martin}\ \emph {et~al.}(2016)\citenamefont {Martin}, \citenamefont {Reining},\ and\ \citenamefont {Ceperley}}]{martin2016book}%
  \BibitemOpen
  \bibfield  {author} {\bibinfo {author} {\bibfnamefont {R.~M.}\ \bibnamefont {Martin}}, \bibinfo {author} {\bibfnamefont {L.}~\bibnamefont {Reining}},\ and\ \bibinfo {author} {\bibfnamefont {D.~M.}\ \bibnamefont {Ceperley}},\ }\href {https://doi.org/https://doi.org/10.1017/CBO9781139050807} {\emph {\bibinfo {title} {Interacting Electrons}}}\ (\bibinfo  {publisher} {Cambridge University Press},\ \bibinfo {address} {Cambridge},\ \bibinfo {year} {2016})\BibitemShut {NoStop}%
\bibitem [{\citenamefont {Reining}(2018)}]{Reining2018wcms}%
  \BibitemOpen
  \bibfield  {author} {\bibinfo {author} {\bibfnamefont {L.}~\bibnamefont {Reining}},\ }\bibfield  {title} {\bibinfo {title} {The gw approximation: content, successes and limitations},\ }\href {https://doi.org/https://doi.org/10.1002/wcms.1344} {\bibfield  {journal} {\bibinfo  {journal} {WIREs Computational Molecular Science}\ }\textbf {\bibinfo {volume} {8}},\ \bibinfo {pages} {e1344} (\bibinfo {year} {2018})}\BibitemShut {NoStop}%
\bibitem [{\citenamefont {Marzari}\ \emph {et~al.}(2021)\citenamefont {Marzari}, \citenamefont {Ferretti},\ and\ \citenamefont {Wolverton}}]{Marzari2021NatureMat}%
  \BibitemOpen
  \bibfield  {author} {\bibinfo {author} {\bibfnamefont {N.}~\bibnamefont {Marzari}}, \bibinfo {author} {\bibfnamefont {A.}~\bibnamefont {Ferretti}},\ and\ \bibinfo {author} {\bibfnamefont {C.}~\bibnamefont {Wolverton}},\ }\bibfield  {title} {\bibinfo {title} {Electronic-structure methods for materials design},\ }\href {https://doi.org/https://doi.org/10.1038/s41563-021-01013-3} {\bibfield  {journal} {\bibinfo  {journal} {Nature Materials}\ }\textbf {\bibinfo {volume} {20}},\ \bibinfo {pages} {736–749} (\bibinfo {year} {2021})}\BibitemShut {NoStop}%
\bibitem [{\citenamefont {Damascelli}(2004)}]{Damascelli2004PS}%
  \BibitemOpen
  \bibfield  {author} {\bibinfo {author} {\bibfnamefont {A.}~\bibnamefont {Damascelli}},\ }\bibfield  {title} {\bibinfo {title} {Probing the electronic structure of complex systems by arpes},\ }\href {https://doi.org/10.1238/Physica.Topical.109a00061} {\bibfield  {journal} {\bibinfo  {journal} {Physica Scripta}\ }\textbf {\bibinfo {volume} {2004}},\ \bibinfo {pages} {61} (\bibinfo {year} {2004})}\BibitemShut {NoStop}%
\bibitem [{\citenamefont {H{\"u}fner}(2013)}]{hufner2013photoelectron}%
  \BibitemOpen
  \bibfield  {author} {\bibinfo {author} {\bibfnamefont {S.}~\bibnamefont {H{\"u}fner}},\ }\href@noop {} {\emph {\bibinfo {title} {Photoelectron spectroscopy: principles and applications}}}\ (\bibinfo  {publisher} {Springer Science \& Business Media},\ \bibinfo {year} {2013})\BibitemShut {NoStop}%
\bibitem [{\citenamefont {Strinati}\ \emph {et~al.}(1980)\citenamefont {Strinati}, \citenamefont {Mattausch},\ and\ \citenamefont {Hanke}}]{Strinati1980PRL}%
  \BibitemOpen
  \bibfield  {author} {\bibinfo {author} {\bibfnamefont {G.}~\bibnamefont {Strinati}}, \bibinfo {author} {\bibfnamefont {H.~J.}\ \bibnamefont {Mattausch}},\ and\ \bibinfo {author} {\bibfnamefont {W.}~\bibnamefont {Hanke}},\ }\bibfield  {title} {\bibinfo {title} {Dynamical correlation effects on the quasiparticle bloch states of a covalent crystal},\ }\href {https://doi.org/10.1103/PhysRevLett.45.290} {\bibfield  {journal} {\bibinfo  {journal} {Phys. Rev. Lett.}\ }\textbf {\bibinfo {volume} {45}},\ \bibinfo {pages} {290} (\bibinfo {year} {1980})}\BibitemShut {NoStop}%
\bibitem [{\citenamefont {Strinati}\ \emph {et~al.}(1982)\citenamefont {Strinati}, \citenamefont {Mattausch},\ and\ \citenamefont {Hanke}}]{Strinati1982PRB}%
  \BibitemOpen
  \bibfield  {author} {\bibinfo {author} {\bibfnamefont {G.}~\bibnamefont {Strinati}}, \bibinfo {author} {\bibfnamefont {H.~J.}\ \bibnamefont {Mattausch}},\ and\ \bibinfo {author} {\bibfnamefont {W.}~\bibnamefont {Hanke}},\ }\bibfield  {title} {\bibinfo {title} {Dynamical aspects of correlation corrections in a covalent crystal},\ }\href {https://doi.org/10.1103/PhysRevB.25.2867} {\bibfield  {journal} {\bibinfo  {journal} {Phys. Rev. B}\ }\textbf {\bibinfo {volume} {25}},\ \bibinfo {pages} {2867} (\bibinfo {year} {1982})}\BibitemShut {NoStop}%
\bibitem [{\citenamefont {Dolado}\ \emph {et~al.}(2001)\citenamefont {Dolado}, \citenamefont {Silkin}, \citenamefont {Cazalilla}, \citenamefont {Rubio},\ and\ \citenamefont {Echenique}}]{Dolado2001PRB}%
  \BibitemOpen
  \bibfield  {author} {\bibinfo {author} {\bibfnamefont {J.~S.}\ \bibnamefont {Dolado}}, \bibinfo {author} {\bibfnamefont {V.~M.}\ \bibnamefont {Silkin}}, \bibinfo {author} {\bibfnamefont {M.~A.}\ \bibnamefont {Cazalilla}}, \bibinfo {author} {\bibfnamefont {A.}~\bibnamefont {Rubio}},\ and\ \bibinfo {author} {\bibfnamefont {P.~M.}\ \bibnamefont {Echenique}},\ }\bibfield  {title} {\bibinfo {title} {Lifetimes and mean-free paths of hot electrons in the alkali metals},\ }\href {https://doi.org/10.1103/PhysRevB.64.195128} {\bibfield  {journal} {\bibinfo  {journal} {Phys. Rev. B}\ }\textbf {\bibinfo {volume} {64}},\ \bibinfo {pages} {195128} (\bibinfo {year} {2001})}\BibitemShut {NoStop}%
\bibitem [{\citenamefont {Marini}\ \emph {et~al.}(2002{\natexlab{a}})\citenamefont {Marini}, \citenamefont {Del~Sole}, \citenamefont {Rubio},\ and\ \citenamefont {Onida}}]{Marini2002PRB}%
  \BibitemOpen
  \bibfield  {author} {\bibinfo {author} {\bibfnamefont {A.}~\bibnamefont {Marini}}, \bibinfo {author} {\bibfnamefont {R.}~\bibnamefont {Del~Sole}}, \bibinfo {author} {\bibfnamefont {A.}~\bibnamefont {Rubio}},\ and\ \bibinfo {author} {\bibfnamefont {G.}~\bibnamefont {Onida}},\ }\bibfield  {title} {\bibinfo {title} {Quasiparticle band-structure effects on the d hole lifetimes of copper within the gw approximation},\ }\href {https://doi.org/10.1103/PhysRevB.66.161104} {\bibfield  {journal} {\bibinfo  {journal} {Phys. Rev. B}\ }\textbf {\bibinfo {volume} {66}},\ \bibinfo {pages} {161104} (\bibinfo {year} {2002}{\natexlab{a}})}\BibitemShut {NoStop}%
\bibitem [{\citenamefont {Kheifets}\ \emph {et~al.}(2003)\citenamefont {Kheifets}, \citenamefont {Sashin}, \citenamefont {Vos}, \citenamefont {Weigold},\ and\ \citenamefont {Aryasetiawan}}]{Kheifets2003PRB}%
  \BibitemOpen
  \bibfield  {author} {\bibinfo {author} {\bibfnamefont {A.~S.}\ \bibnamefont {Kheifets}}, \bibinfo {author} {\bibfnamefont {V.~A.}\ \bibnamefont {Sashin}}, \bibinfo {author} {\bibfnamefont {M.}~\bibnamefont {Vos}}, \bibinfo {author} {\bibfnamefont {E.}~\bibnamefont {Weigold}},\ and\ \bibinfo {author} {\bibfnamefont {F.}~\bibnamefont {Aryasetiawan}},\ }\bibfield  {title} {\bibinfo {title} {Spectral properties of quasiparticles in silicon: A test of many-body theory},\ }\href {https://doi.org/10.1103/PhysRevB.68.233205} {\bibfield  {journal} {\bibinfo  {journal} {Phys. Rev. B}\ }\textbf {\bibinfo {volume} {68}},\ \bibinfo {pages} {233205} (\bibinfo {year} {2003})}\BibitemShut {NoStop}%
\bibitem [{\citenamefont {Arnaud}\ \emph {et~al.}(2005)\citenamefont {Arnaud}, \citenamefont {Leb\`egue},\ and\ \citenamefont {Alouani}}]{Arnaud2005PRB}%
  \BibitemOpen
  \bibfield  {author} {\bibinfo {author} {\bibfnamefont {B.}~\bibnamefont {Arnaud}}, \bibinfo {author} {\bibfnamefont {S.}~\bibnamefont {Leb\`egue}},\ and\ \bibinfo {author} {\bibfnamefont {M.}~\bibnamefont {Alouani}},\ }\bibfield  {title} {\bibinfo {title} {Excitonic and quasiparticle lifetime effects on silicon electron energy loss spectra from first principles},\ }\href {https://doi.org/10.1103/PhysRevB.71.035308} {\bibfield  {journal} {\bibinfo  {journal} {Phys. Rev. B}\ }\textbf {\bibinfo {volume} {71}},\ \bibinfo {pages} {035308} (\bibinfo {year} {2005})}\BibitemShut {NoStop}%
\bibitem [{\citenamefont {Cazzaniga}(2012)}]{Cazzaniga2012PRB}%
  \BibitemOpen
  \bibfield  {author} {\bibinfo {author} {\bibfnamefont {M.}~\bibnamefont {Cazzaniga}},\ }\bibfield  {title} {\bibinfo {title} {$gw$ and beyond approaches to quasiparticle properties in metals},\ }\href {https://doi.org/https://doi.org/10.1103/PhysRevB.86.035120} {\bibfield  {journal} {\bibinfo  {journal} {Phys. Rev. B}\ }\textbf {\bibinfo {volume} {86}},\ \bibinfo {pages} {035120} (\bibinfo {year} {2012})}\BibitemShut {NoStop}%
\bibitem [{\citenamefont {Zhou}\ \emph {et~al.}(2020)\citenamefont {Zhou}, \citenamefont {Reining}, \citenamefont {Nicolaou}, \citenamefont {Bendounan}, \citenamefont {Ruotsalainen}, \citenamefont {Vanzini}, \citenamefont {Kas}, \citenamefont {Rehr}, \citenamefont {Muntwiler}, \citenamefont {Strocov}, \citenamefont {Sirotti},\ and\ \citenamefont {Gatti}}]{Zhou2020pnas}%
  \BibitemOpen
  \bibfield  {author} {\bibinfo {author} {\bibfnamefont {J.~S.}\ \bibnamefont {Zhou}}, \bibinfo {author} {\bibfnamefont {L.}~\bibnamefont {Reining}}, \bibinfo {author} {\bibfnamefont {A.}~\bibnamefont {Nicolaou}}, \bibinfo {author} {\bibfnamefont {A.}~\bibnamefont {Bendounan}}, \bibinfo {author} {\bibfnamefont {K.}~\bibnamefont {Ruotsalainen}}, \bibinfo {author} {\bibfnamefont {M.}~\bibnamefont {Vanzini}}, \bibinfo {author} {\bibfnamefont {J.~J.}\ \bibnamefont {Kas}}, \bibinfo {author} {\bibfnamefont {J.~J.}\ \bibnamefont {Rehr}}, \bibinfo {author} {\bibfnamefont {M.}~\bibnamefont {Muntwiler}}, \bibinfo {author} {\bibfnamefont {V.~N.}\ \bibnamefont {Strocov}}, \bibinfo {author} {\bibfnamefont {F.}~\bibnamefont {Sirotti}},\ and\ \bibinfo {author} {\bibfnamefont {M.}~\bibnamefont {Gatti}},\ }\bibfield  {title} {\bibinfo {title} {Unraveling intrinsic correlation effects with angle-resolved photoemission spectroscopy},\ }\href {https://doi.org/10.1073/pnas.2012625117} {\bibfield  {journal} {\bibinfo  {journal}
  {Proceedings of the National Academy of Sciences}\ }\textbf {\bibinfo {volume} {117}},\ \bibinfo {pages} {28596} (\bibinfo {year} {2020})}\BibitemShut {NoStop}%
\bibitem [{\citenamefont {Hybertsen}\ and\ \citenamefont {Louie}(1985)}]{Hybertsen1985PRL}%
  \BibitemOpen
  \bibfield  {author} {\bibinfo {author} {\bibfnamefont {M.~S.}\ \bibnamefont {Hybertsen}}\ and\ \bibinfo {author} {\bibfnamefont {S.~G.}\ \bibnamefont {Louie}},\ }\bibfield  {title} {\bibinfo {title} {First-principles theory of quasiparticles: Calculation of band gaps in semiconductors and insulators},\ }\href {https://doi.org/10.1103/PhysRevLett.55.1418} {\bibfield  {journal} {\bibinfo  {journal} {Phys. Rev. Lett.}\ }\textbf {\bibinfo {volume} {55}},\ \bibinfo {pages} {1418} (\bibinfo {year} {1985})}\BibitemShut {NoStop}%
\bibitem [{\citenamefont {van Schilfgaarde}\ \emph {et~al.}(2006)\citenamefont {van Schilfgaarde}, \citenamefont {Kotani},\ and\ \citenamefont {Faleev}}]{vanSchilfgaarde2006PRL}%
  \BibitemOpen
  \bibfield  {author} {\bibinfo {author} {\bibfnamefont {M.}~\bibnamefont {van Schilfgaarde}}, \bibinfo {author} {\bibfnamefont {T.}~\bibnamefont {Kotani}},\ and\ \bibinfo {author} {\bibfnamefont {S.}~\bibnamefont {Faleev}},\ }\bibfield  {title} {\bibinfo {title} {Quasiparticle self-consistent {GW} theory},\ }\href {https://doi.org/https://doi.org/10.1103/PhysRevLett.96.226402} {\bibfield  {journal} {\bibinfo  {journal} {Phys. Rev. Lett.}\ }\textbf {\bibinfo {volume} {96}},\ \bibinfo {pages} {226402} (\bibinfo {year} {2006})}\BibitemShut {NoStop}%
\bibitem [{\citenamefont {H\"user}\ \emph {et~al.}(2013)\citenamefont {H\"user}, \citenamefont {Olsen},\ and\ \citenamefont {Thygesen}}]{Huser2013PRB}%
  \BibitemOpen
  \bibfield  {author} {\bibinfo {author} {\bibfnamefont {F.}~\bibnamefont {H\"user}}, \bibinfo {author} {\bibfnamefont {T.}~\bibnamefont {Olsen}},\ and\ \bibinfo {author} {\bibfnamefont {K.~S.}\ \bibnamefont {Thygesen}},\ }\bibfield  {title} {\bibinfo {title} {Quasiparticle gw calculations for solids, molecules, and two-dimensional materials},\ }\href {https://doi.org/https://doi.org/10.1103/PhysRevB.87.235132} {\bibfield  {journal} {\bibinfo  {journal} {Phys. Rev. B}\ }\textbf {\bibinfo {volume} {87}},\ \bibinfo {pages} {235132} (\bibinfo {year} {2013})}\BibitemShut {NoStop}%
\bibitem [{\citenamefont {Golze}\ \emph {et~al.}(2019)\citenamefont {Golze}, \citenamefont {Dvorak},\ and\ \citenamefont {Rinke}}]{Golze2019FrontChem}%
  \BibitemOpen
  \bibfield  {author} {\bibinfo {author} {\bibfnamefont {D.}~\bibnamefont {Golze}}, \bibinfo {author} {\bibfnamefont {M.}~\bibnamefont {Dvorak}},\ and\ \bibinfo {author} {\bibfnamefont {P.}~\bibnamefont {Rinke}},\ }\bibfield  {title} {\bibinfo {title} {The {GW} {C}ompendium: {A} {P}ractical {G}uide to {T}heoretical {P}hotoemission {S}pectroscopy},\ }\href {https://doi.org/https://doi.org/10.3389/fchem.2019.00377} {\bibfield  {journal} {\bibinfo  {journal} {Front. Chem.}\ }\textbf {\bibinfo {volume} {7}},\ \bibinfo {pages} {377} (\bibinfo {year} {2019})}\BibitemShut {NoStop}%
\bibitem [{\citenamefont {Gumhalter}\ \emph {et~al.}(2016)\citenamefont {Gumhalter}, \citenamefont {Kova\ifmmode~\check{c}\else \v{c}\fi{}}, \citenamefont {Caruso}, \citenamefont {Lambert},\ and\ \citenamefont {Giustino}}]{Gumhalter2016PRB}%
  \BibitemOpen
  \bibfield  {author} {\bibinfo {author} {\bibfnamefont {B.}~\bibnamefont {Gumhalter}}, \bibinfo {author} {\bibfnamefont {V.}~\bibnamefont {Kova\ifmmode~\check{c}\else \v{c}\fi{}}}, \bibinfo {author} {\bibfnamefont {F.}~\bibnamefont {Caruso}}, \bibinfo {author} {\bibfnamefont {H.}~\bibnamefont {Lambert}},\ and\ \bibinfo {author} {\bibfnamefont {F.}~\bibnamefont {Giustino}},\ }\bibfield  {title} {\bibinfo {title} {On the combined use of gw approximation and cumulant expansion in the calculations of quasiparticle spectra: The paradigm of si valence bands},\ }\href {https://doi.org/10.1103/PhysRevB.94.035103} {\bibfield  {journal} {\bibinfo  {journal} {Phys. Rev. B}\ }\textbf {\bibinfo {volume} {94}},\ \bibinfo {pages} {035103} (\bibinfo {year} {2016})}\BibitemShut {NoStop}%
\bibitem [{\citenamefont {Zhou}\ \emph {et~al.}(2018)\citenamefont {Zhou}, \citenamefont {Gatti}, \citenamefont {Kas}, \citenamefont {Rehr},\ and\ \citenamefont {Reining}}]{Zhou2018PRB}%
  \BibitemOpen
  \bibfield  {author} {\bibinfo {author} {\bibfnamefont {J.~S.}\ \bibnamefont {Zhou}}, \bibinfo {author} {\bibfnamefont {M.}~\bibnamefont {Gatti}}, \bibinfo {author} {\bibfnamefont {J.~J.}\ \bibnamefont {Kas}}, \bibinfo {author} {\bibfnamefont {J.~J.}\ \bibnamefont {Rehr}},\ and\ \bibinfo {author} {\bibfnamefont {L.}~\bibnamefont {Reining}},\ }\bibfield  {title} {\bibinfo {title} {Cumulant green's function calculations of plasmon satellites in bulk sodium: Influence of screening and the crystal environment},\ }\href {https://doi.org/10.1103/PhysRevB.97.035137} {\bibfield  {journal} {\bibinfo  {journal} {Phys. Rev. B}\ }\textbf {\bibinfo {volume} {97}},\ \bibinfo {pages} {035137} (\bibinfo {year} {2018})}\BibitemShut {NoStop}%
\bibitem [{\citenamefont {Nery}\ \emph {et~al.}(2018)\citenamefont {Nery}, \citenamefont {Allen}, \citenamefont {Antonius}, \citenamefont {Reining}, \citenamefont {Miglio},\ and\ \citenamefont {Gonze}}]{Nery2018PRB}%
  \BibitemOpen
  \bibfield  {author} {\bibinfo {author} {\bibfnamefont {J.~P.}\ \bibnamefont {Nery}}, \bibinfo {author} {\bibfnamefont {P.~B.}\ \bibnamefont {Allen}}, \bibinfo {author} {\bibfnamefont {G.}~\bibnamefont {Antonius}}, \bibinfo {author} {\bibfnamefont {L.}~\bibnamefont {Reining}}, \bibinfo {author} {\bibfnamefont {A.}~\bibnamefont {Miglio}},\ and\ \bibinfo {author} {\bibfnamefont {X.}~\bibnamefont {Gonze}},\ }\bibfield  {title} {\bibinfo {title} {Quasiparticles and phonon satellites in spectral functions of semiconductors and insulators: Cumulants applied to the full first-principles theory and the fr\"ohlich polaron},\ }\href {https://doi.org/10.1103/PhysRevB.97.115145} {\bibfield  {journal} {\bibinfo  {journal} {Phys. Rev. B}\ }\textbf {\bibinfo {volume} {97}},\ \bibinfo {pages} {115145} (\bibinfo {year} {2018})}\BibitemShut {NoStop}%
\bibitem [{\citenamefont {Marini}\ \emph {et~al.}(2002{\natexlab{b}})\citenamefont {Marini}, \citenamefont {Onida},\ and\ \citenamefont {Sole}}]{Marini2002PRL}%
  \BibitemOpen
  \bibfield  {author} {\bibinfo {author} {\bibfnamefont {A.}~\bibnamefont {Marini}}, \bibinfo {author} {\bibfnamefont {G.}~\bibnamefont {Onida}},\ and\ \bibinfo {author} {\bibfnamefont {R.~D.}\ \bibnamefont {Sole}},\ }\bibfield  {title} {\bibinfo {title} {Quasiparticle electronic structure of copper in the gw approximation},\ }\href {https://doi.org/https://doi.org/10.1103/physrevlett.88.016403} {\bibfield  {journal} {\bibinfo  {journal} {Phys. Rev. Lett.}\ }\textbf {\bibinfo {volume} {88}},\ \bibinfo {pages} {016403} (\bibinfo {year} {2002}{\natexlab{b}})}\BibitemShut {NoStop}%
\bibitem [{\citenamefont {Shishkin}\ and\ \citenamefont {Kresse}(2006)}]{Shishkin2006PRB}%
  \BibitemOpen
  \bibfield  {author} {\bibinfo {author} {\bibfnamefont {M.}~\bibnamefont {Shishkin}}\ and\ \bibinfo {author} {\bibfnamefont {G.}~\bibnamefont {Kresse}},\ }\bibfield  {title} {\bibinfo {title} {Implementation and performance of the frequency-dependent $gw$ method within the paw framework},\ }\href {https://doi.org/https://doi.org/10.1103/PhysRevB.74.035101} {\bibfield  {journal} {\bibinfo  {journal} {Phys. Rev. B}\ }\textbf {\bibinfo {volume} {74}},\ \bibinfo {pages} {035101} (\bibinfo {year} {2006})}\BibitemShut {NoStop}%
\bibitem [{\citenamefont {Liu}\ \emph {et~al.}(2015)\citenamefont {Liu}, \citenamefont {Lin}, \citenamefont {Vigil-Fowler}, \citenamefont {Lischner}, \citenamefont {Kemper}, \citenamefont {Sharifzadeh}, \citenamefont {{da Jornada}}, \citenamefont {Deslippe}, \citenamefont {Yang}, \citenamefont {Neaton},\ and\ \citenamefont {Louie}}]{Liu2015JComputPhys}%
  \BibitemOpen
  \bibfield  {author} {\bibinfo {author} {\bibfnamefont {F.}~\bibnamefont {Liu}}, \bibinfo {author} {\bibfnamefont {L.}~\bibnamefont {Lin}}, \bibinfo {author} {\bibfnamefont {D.}~\bibnamefont {Vigil-Fowler}}, \bibinfo {author} {\bibfnamefont {J.}~\bibnamefont {Lischner}}, \bibinfo {author} {\bibfnamefont {A.~F.}\ \bibnamefont {Kemper}}, \bibinfo {author} {\bibfnamefont {S.}~\bibnamefont {Sharifzadeh}}, \bibinfo {author} {\bibfnamefont {F.~H.}\ \bibnamefont {{da Jornada}}}, \bibinfo {author} {\bibfnamefont {J.}~\bibnamefont {Deslippe}}, \bibinfo {author} {\bibfnamefont {C.}~\bibnamefont {Yang}}, \bibinfo {author} {\bibfnamefont {J.~B.}\ \bibnamefont {Neaton}},\ and\ \bibinfo {author} {\bibfnamefont {S.~G.}\ \bibnamefont {Louie}},\ }\bibfield  {title} {\bibinfo {title} {Numerical integration for ab initio many-electron self energy calculations within the {GW} approximation},\ }\href {https://doi.org/https://doi.org/10.1016/j.jcp.2015.01.023} {\bibfield  {journal} {\bibinfo  {journal} {J. Comput. Phys.}\
  }\textbf {\bibinfo {volume} {286}},\ \bibinfo {pages} {1} (\bibinfo {year} {2015})}\BibitemShut {NoStop}%
\bibitem [{\citenamefont {Godby}\ \emph {et~al.}(1988)\citenamefont {Godby}, \citenamefont {Schl\"uter},\ and\ \citenamefont {Sham}}]{Godby1988PRB}%
  \BibitemOpen
  \bibfield  {author} {\bibinfo {author} {\bibfnamefont {R.~W.}\ \bibnamefont {Godby}}, \bibinfo {author} {\bibfnamefont {M.}~\bibnamefont {Schl\"uter}},\ and\ \bibinfo {author} {\bibfnamefont {L.~J.}\ \bibnamefont {Sham}},\ }\bibfield  {title} {\bibinfo {title} {Self-energy operators and exchange-correlation potentials in semiconductors},\ }\href {https://doi.org/10.1103/PhysRevB.37.10159} {\bibfield  {journal} {\bibinfo  {journal} {Phys. Rev. B}\ }\textbf {\bibinfo {volume} {37}},\ \bibinfo {pages} {10159} (\bibinfo {year} {1988})}\BibitemShut {NoStop}%
\bibitem [{\citenamefont {{F. Aryasetiawan in}}(2000)}]{book_Anisimov2000}%
  \BibitemOpen
  \bibfield  {author} {\bibinfo {author} {\bibnamefont {{F. Aryasetiawan in}}},\ }\bibinfo {title} {Strong coulomb correlations in electronic structure calculations}\ (\bibinfo  {publisher} {CRC Press.},\ \bibinfo {address} {London},\ \bibinfo {year} {2000})\ p.~\bibinfo {pages} {96},\ \bibinfo {edition} {1st}\ ed.,\ \bibinfo {note} {\url{https://doi.org/10.1201/9781482296877}}\BibitemShut {NoStop}%
\bibitem [{\citenamefont {Kotani}\ \emph {et~al.}(2007)\citenamefont {Kotani}, \citenamefont {van Schilfgaarde},\ and\ \citenamefont {Faleev}}]{Kotani2007PRB}%
  \BibitemOpen
  \bibfield  {author} {\bibinfo {author} {\bibfnamefont {T.}~\bibnamefont {Kotani}}, \bibinfo {author} {\bibfnamefont {M.}~\bibnamefont {van Schilfgaarde}},\ and\ \bibinfo {author} {\bibfnamefont {S.~V.}\ \bibnamefont {Faleev}},\ }\bibfield  {title} {\bibinfo {title} {Quasiparticle self-consistent $gw$ method: A basis for the independent-particle approximation},\ }\href {https://doi.org/https://doi.org/10.1103/PhysRevB.76.165106} {\bibfield  {journal} {\bibinfo  {journal} {Phys. Rev. B}\ }\textbf {\bibinfo {volume} {76}},\ \bibinfo {pages} {165106} (\bibinfo {year} {2007})}\BibitemShut {NoStop}%
\bibitem [{\citenamefont {Daling}\ \emph {et~al.}(1991)\citenamefont {Daling}, \citenamefont {van Haeringen},\ and\ \citenamefont {Farid}}]{DalingPRB1991}%
  \BibitemOpen
  \bibfield  {author} {\bibinfo {author} {\bibfnamefont {R.}~\bibnamefont {Daling}}, \bibinfo {author} {\bibfnamefont {W.}~\bibnamefont {van Haeringen}},\ and\ \bibinfo {author} {\bibfnamefont {B.}~\bibnamefont {Farid}},\ }\bibfield  {title} {\bibinfo {title} {Plasmon dispersion in silicon obtained by analytic continuation of the random-phase-approximation dielectric matrix},\ }\href {https://doi.org/10.1103/PhysRevB.44.2952} {\bibfield  {journal} {\bibinfo  {journal} {Phys. Rev. B}\ }\textbf {\bibinfo {volume} {44}},\ \bibinfo {pages} {2952} (\bibinfo {year} {1991})}\BibitemShut {NoStop}%
\bibitem [{\citenamefont {Engel}\ \emph {et~al.}(1991)\citenamefont {Engel}, \citenamefont {Farid}, \citenamefont {Nex},\ and\ \citenamefont {March}}]{Engel1991PRB}%
  \BibitemOpen
  \bibfield  {author} {\bibinfo {author} {\bibfnamefont {G.~E.}\ \bibnamefont {Engel}}, \bibinfo {author} {\bibfnamefont {B.}~\bibnamefont {Farid}}, \bibinfo {author} {\bibfnamefont {C.~M.~M.}\ \bibnamefont {Nex}},\ and\ \bibinfo {author} {\bibfnamefont {N.~H.}\ \bibnamefont {March}},\ }\bibfield  {title} {\bibinfo {title} {Calculation of the gw self-energy in semiconducting crystals},\ }\href {https://doi.org/10.1103/PhysRevB.44.13356} {\bibfield  {journal} {\bibinfo  {journal} {Phys. Rev. B}\ }\textbf {\bibinfo {volume} {44}},\ \bibinfo {pages} {13356} (\bibinfo {year} {1991})}\BibitemShut {NoStop}%
\bibitem [{\citenamefont {Duchemin}\ and\ \citenamefont {Blase}(2020)}]{Duchemin2020JCTC}%
  \BibitemOpen
  \bibfield  {author} {\bibinfo {author} {\bibfnamefont {I.}~\bibnamefont {Duchemin}}\ and\ \bibinfo {author} {\bibfnamefont {X.}~\bibnamefont {Blase}},\ }\bibfield  {title} {\bibinfo {title} {Robust {A}nalytic-{C}ontinuation {A}pproach to {M}any-{B}ody {GW} {C}alculations},\ }\href {https://doi.org/https://doi.org/10.1021/acs.jctc.9b01235} {\bibfield  {journal} {\bibinfo  {journal} {J. Chem. Theory Comput.}\ }\textbf {\bibinfo {volume} {16}},\ \bibinfo {pages} {1742} (\bibinfo {year} {2020})}\BibitemShut {NoStop}%
\bibitem [{\citenamefont {Hybertsen}\ and\ \citenamefont {Louie}(1986)}]{Hybertsen1986PRB}%
  \BibitemOpen
  \bibfield  {author} {\bibinfo {author} {\bibfnamefont {M.~S.}\ \bibnamefont {Hybertsen}}\ and\ \bibinfo {author} {\bibfnamefont {S.~G.}\ \bibnamefont {Louie}},\ }\bibfield  {title} {\bibinfo {title} {Electron correlation in semiconductors and insulators: Band gaps and quasiparticle energies},\ }\href {https://doi.org/https://doi.org/10.1103/PhysRevB.34.5390} {\bibfield  {journal} {\bibinfo  {journal} {Phys. Rev. B}\ }\textbf {\bibinfo {volume} {34}},\ \bibinfo {pages} {5390} (\bibinfo {year} {1986})}\BibitemShut {NoStop}%
\bibitem [{\citenamefont {Zhang}\ \emph {et~al.}(1989)\citenamefont {Zhang}, \citenamefont {Tom\'anek}, \citenamefont {Cohen}, \citenamefont {Louie},\ and\ \citenamefont {Hybertsen}}]{Zhang1989PRB}%
  \BibitemOpen
  \bibfield  {author} {\bibinfo {author} {\bibfnamefont {S.~B.}\ \bibnamefont {Zhang}}, \bibinfo {author} {\bibfnamefont {D.}~\bibnamefont {Tom\'anek}}, \bibinfo {author} {\bibfnamefont {M.~L.}\ \bibnamefont {Cohen}}, \bibinfo {author} {\bibfnamefont {S.~G.}\ \bibnamefont {Louie}},\ and\ \bibinfo {author} {\bibfnamefont {M.~S.}\ \bibnamefont {Hybertsen}},\ }\bibfield  {title} {\bibinfo {title} {Evaluation of quasiparticle energies for semiconductors without inversion symmetry},\ }\href {https://doi.org/https://doi.org/10.1103/PhysRevB.40.3162} {\bibfield  {journal} {\bibinfo  {journal} {Phys. Rev. B}\ }\textbf {\bibinfo {volume} {40}},\ \bibinfo {pages} {3162} (\bibinfo {year} {1989})}\BibitemShut {NoStop}%
\bibitem [{\citenamefont {Godby}\ and\ \citenamefont {Needs}(1989)}]{Godby1989PRL}%
  \BibitemOpen
  \bibfield  {author} {\bibinfo {author} {\bibfnamefont {R.~W.}\ \bibnamefont {Godby}}\ and\ \bibinfo {author} {\bibfnamefont {R.~J.}\ \bibnamefont {Needs}},\ }\bibfield  {title} {\bibinfo {title} {Metal-insulator transition in kohn-sham theory and quasiparticle theory},\ }\href {https://doi.org/https://doi.org/10.1103/PhysRevLett.62.1169} {\bibfield  {journal} {\bibinfo  {journal} {Phys. Rev. Lett.}\ }\textbf {\bibinfo {volume} {62}},\ \bibinfo {pages} {1169} (\bibinfo {year} {1989})}\BibitemShut {NoStop}%
\bibitem [{\citenamefont {{von der Linden}}\ and\ \citenamefont {Horsch}(1988)}]{vonderLinden1988PRB}%
  \BibitemOpen
  \bibfield  {author} {\bibinfo {author} {\bibfnamefont {W.}~\bibnamefont {{von der Linden}}}\ and\ \bibinfo {author} {\bibfnamefont {P.}~\bibnamefont {Horsch}},\ }\bibfield  {title} {\bibinfo {title} {Precise quasiparticle energies and hartree-fock bands of semiconductors and insulators},\ }\href {https://doi.org/https://doi.org/10.1103/PhysRevB.37.8351} {\bibfield  {journal} {\bibinfo  {journal} {Phys. Rev. B}\ }\textbf {\bibinfo {volume} {37}},\ \bibinfo {pages} {8351} (\bibinfo {year} {1988})}\BibitemShut {NoStop}%
\bibitem [{\citenamefont {Engel}\ and\ \citenamefont {Farid}(1993)}]{Engel1993PRB}%
  \BibitemOpen
  \bibfield  {author} {\bibinfo {author} {\bibfnamefont {G.~E.}\ \bibnamefont {Engel}}\ and\ \bibinfo {author} {\bibfnamefont {B.}~\bibnamefont {Farid}},\ }\bibfield  {title} {\bibinfo {title} {Generalized plasmon-pole model and plasmon band structures of crystals},\ }\href {https://doi.org/https://doi.org/10.1103/PhysRevB.47.15931} {\bibfield  {journal} {\bibinfo  {journal} {Phys. Rev. B}\ }\textbf {\bibinfo {volume} {47}},\ \bibinfo {pages} {15931} (\bibinfo {year} {1993})}\BibitemShut {NoStop}%
\bibitem [{\citenamefont {Farid}\ \emph {et~al.}(1991)\citenamefont {Farid}, \citenamefont {Engel}, \citenamefont {Daling},\ and\ \citenamefont {van Haeringen}}]{FaridPRB1991}%
  \BibitemOpen
  \bibfield  {author} {\bibinfo {author} {\bibfnamefont {B.}~\bibnamefont {Farid}}, \bibinfo {author} {\bibfnamefont {G.~E.}\ \bibnamefont {Engel}}, \bibinfo {author} {\bibfnamefont {R.}~\bibnamefont {Daling}},\ and\ \bibinfo {author} {\bibfnamefont {W.}~\bibnamefont {van Haeringen}},\ }\bibfield  {title} {\bibinfo {title} {Plasmon excitations in crystals},\ }\href {https://doi.org/10.1103/PhysRevB.44.13349} {\bibfield  {journal} {\bibinfo  {journal} {Phys. Rev. B}\ }\textbf {\bibinfo {volume} {44}},\ \bibinfo {pages} {13349} (\bibinfo {year} {1991})}\BibitemShut {NoStop}%
\bibitem [{\citenamefont {Lee}\ and\ \citenamefont {Chang}(1994)}]{Lee1994PRB}%
  \BibitemOpen
  \bibfield  {author} {\bibinfo {author} {\bibfnamefont {K.-H.}\ \bibnamefont {Lee}}\ and\ \bibinfo {author} {\bibfnamefont {K.~J.}\ \bibnamefont {Chang}},\ }\bibfield  {title} {\bibinfo {title} {First-principles study of the optical properties and the dielectric response of al},\ }\href {https://doi.org/https://doi.org/10.1103/PhysRevB.49.2362} {\bibfield  {journal} {\bibinfo  {journal} {Phys. Rev. B}\ }\textbf {\bibinfo {volume} {49}},\ \bibinfo {pages} {2362} (\bibinfo {year} {1994})}\BibitemShut {NoStop}%
\bibitem [{\citenamefont {Soininen}\ \emph {et~al.}(2003)\citenamefont {Soininen}, \citenamefont {Rehr},\ and\ \citenamefont {Shirley}}]{Soininen2003JPCM}%
  \BibitemOpen
  \bibfield  {author} {\bibinfo {author} {\bibfnamefont {J.~A.}\ \bibnamefont {Soininen}}, \bibinfo {author} {\bibfnamefont {J.~J.}\ \bibnamefont {Rehr}},\ and\ \bibinfo {author} {\bibfnamefont {E.~L.}\ \bibnamefont {Shirley}},\ }\bibfield  {title} {\bibinfo {title} {Electron self-energy calculation using a general multi-pole approximation},\ }\href {https://doi.org/https://doi.org/10.1088/0953-8984/15/17/312} {\bibfield  {journal} {\bibinfo  {journal} {J. Phys.: Condens. Matter}\ }\textbf {\bibinfo {volume} {15}},\ \bibinfo {pages} {2573} (\bibinfo {year} {2003})}\BibitemShut {NoStop}%
\bibitem [{\citenamefont {Leon}\ \emph {et~al.}(2021)\citenamefont {Leon}, \citenamefont {Cardoso}, \citenamefont {Chiarotti}, \citenamefont {Varsano}, \citenamefont {Molinari},\ and\ \citenamefont {Ferretti}}]{Leon2021PRB}%
  \BibitemOpen
  \bibfield  {author} {\bibinfo {author} {\bibfnamefont {D.~A.}\ \bibnamefont {Leon}}, \bibinfo {author} {\bibfnamefont {C.}~\bibnamefont {Cardoso}}, \bibinfo {author} {\bibfnamefont {T.}~\bibnamefont {Chiarotti}}, \bibinfo {author} {\bibfnamefont {D.}~\bibnamefont {Varsano}}, \bibinfo {author} {\bibfnamefont {E.}~\bibnamefont {Molinari}},\ and\ \bibinfo {author} {\bibfnamefont {A.}~\bibnamefont {Ferretti}},\ }\bibfield  {title} {\bibinfo {title} {Frequency dependence in $gw$ made simple using a multipole approximation},\ }\href {https://doi.org/10.1103/PhysRevB.104.115157} {\bibfield  {journal} {\bibinfo  {journal} {Phys. Rev. B}\ }\textbf {\bibinfo {volume} {104}},\ \bibinfo {pages} {115157} (\bibinfo {year} {2021})}\BibitemShut {NoStop}%
\bibitem [{\citenamefont {Leon}\ \emph {et~al.}(2023)\citenamefont {Leon}, \citenamefont {Ferretti}, \citenamefont {Varsano}, \citenamefont {Molinari},\ and\ \citenamefont {Cardoso}}]{Leon2023PRB}%
  \BibitemOpen
  \bibfield  {author} {\bibinfo {author} {\bibfnamefont {D.~A.}\ \bibnamefont {Leon}}, \bibinfo {author} {\bibfnamefont {A.}~\bibnamefont {Ferretti}}, \bibinfo {author} {\bibfnamefont {D.}~\bibnamefont {Varsano}}, \bibinfo {author} {\bibfnamefont {E.}~\bibnamefont {Molinari}},\ and\ \bibinfo {author} {\bibfnamefont {C.}~\bibnamefont {Cardoso}},\ }\bibfield  {title} {\bibinfo {title} {Efficient full frequency gw for metals using a multipole approach for the dielectric screening},\ }\href {https://doi.org/10.1103/PhysRevB.107.155130} {\bibfield  {journal} {\bibinfo  {journal} {Phys. Rev. B}\ }\textbf {\bibinfo {volume} {107}},\ \bibinfo {pages} {155130} (\bibinfo {year} {2023})}\BibitemShut {NoStop}%
\bibitem [{\citenamefont {Guandalini}\ \emph {et~al.}(2024)\citenamefont {Guandalini}, \citenamefont {Leon}, \citenamefont {D'Amico}, \citenamefont {Cardoso}, \citenamefont {Ferretti}, \citenamefont {Rontani},\ and\ \citenamefont {Varsano}}]{Guandalini2024PRB}%
  \BibitemOpen
  \bibfield  {author} {\bibinfo {author} {\bibfnamefont {A.}~\bibnamefont {Guandalini}}, \bibinfo {author} {\bibfnamefont {D.~A.}\ \bibnamefont {Leon}}, \bibinfo {author} {\bibfnamefont {P.}~\bibnamefont {D'Amico}}, \bibinfo {author} {\bibfnamefont {C.}~\bibnamefont {Cardoso}}, \bibinfo {author} {\bibfnamefont {A.}~\bibnamefont {Ferretti}}, \bibinfo {author} {\bibfnamefont {M.}~\bibnamefont {Rontani}},\ and\ \bibinfo {author} {\bibfnamefont {D.}~\bibnamefont {Varsano}},\ }\bibfield  {title} {\bibinfo {title} {Efficient {GW} calculations via interpolation of the screened interaction in momentum and frequency space: The case of graphene},\ }\href {https://doi.org/10.1103/PhysRevB.109.075120} {\bibfield  {journal} {\bibinfo  {journal} {Phys. Rev. B}\ }\textbf {\bibinfo {volume} {109}},\ \bibinfo {pages} {075120} (\bibinfo {year} {2024})}\BibitemShut {NoStop}%
\bibitem [{\citenamefont {Raether}(1980)}]{book_Raether1980}%
  \BibitemOpen
  \bibfield  {author} {\bibinfo {author} {\bibfnamefont {H.}~\bibnamefont {Raether}},\ }\href {https://doi.org/https://doi.org/10.1007/BFb0045951} {\emph {\bibinfo {title} {Excitation of Plasmons and Interband Transitions by Electrons}}},\ \bibinfo {edition} {1st}\ ed.,\ \bibinfo {series} {Springer Tracts in Modern Physics 88}, Vol.~\bibinfo {volume} {88}\ (\bibinfo  {publisher} {Springer, Berlin, Heidelberg},\ \bibinfo {year} {1980})\BibitemShut {NoStop}%
\bibitem [{\citenamefont {Giuliani}\ and\ \citenamefont {Vignale}(2005)}]{Giuliani_Vignale_2005}%
  \BibitemOpen
  \bibfield  {author} {\bibinfo {author} {\bibfnamefont {G.}~\bibnamefont {Giuliani}}\ and\ \bibinfo {author} {\bibfnamefont {G.}~\bibnamefont {Vignale}},\ }\href@noop {} {\emph {\bibinfo {title} {Quantum Theory of the Electron Liquid}}}\ (\bibinfo  {publisher} {Cambridge University Press},\ \bibinfo {year} {2005})\BibitemShut {NoStop}%
\bibitem [{\citenamefont {Pines}(2018)}]{Pines2018book}%
  \BibitemOpen
  \bibfield  {author} {\bibinfo {author} {\bibfnamefont {D.}~\bibnamefont {Pines}},\ }\href@noop {} {\emph {\bibinfo {title} {Theory of quantum liquids}}}\ (\bibinfo  {publisher} {CRC Press},\ \bibinfo {year} {2018})\BibitemShut {NoStop}%
\bibitem [{\citenamefont {Allen}\ and\ \citenamefont {Mikkelsen}(1977)}]{Allen1977PRB}%
  \BibitemOpen
  \bibfield  {author} {\bibinfo {author} {\bibfnamefont {J.~W.}\ \bibnamefont {Allen}}\ and\ \bibinfo {author} {\bibfnamefont {J.~C.}\ \bibnamefont {Mikkelsen}},\ }\bibfield  {title} {\bibinfo {title} {Optical properties of crsb, mnsb, nisb, and nias},\ }\href {https://doi.org/https://doi.org/10.1103/PhysRevB.15.2952} {\bibfield  {journal} {\bibinfo  {journal} {Phys. Rev. B}\ }\textbf {\bibinfo {volume} {15}},\ \bibinfo {pages} {2952} (\bibinfo {year} {1977})}\BibitemShut {NoStop}%
\bibitem [{\citenamefont {Smith}\ and\ \citenamefont {Segall}(1986)}]{Smith1986PRB}%
  \BibitemOpen
  \bibfield  {author} {\bibinfo {author} {\bibfnamefont {D.~Y.}\ \bibnamefont {Smith}}\ and\ \bibinfo {author} {\bibfnamefont {B.}~\bibnamefont {Segall}},\ }\bibfield  {title} {\bibinfo {title} {Intraband and interband processes in the infrared spectrum of metallic aluminum},\ }\href {https://doi.org/https://doi.org/10.1103/PhysRevB.34.5191} {\bibfield  {journal} {\bibinfo  {journal} {Phys. Rev. B}\ }\textbf {\bibinfo {volume} {34}},\ \bibinfo {pages} {5191} (\bibinfo {year} {1986})}\BibitemShut {NoStop}%
\bibitem [{\citenamefont {Lee}\ and\ \citenamefont {Chang}(1996)}]{Lee1996PRB}%
  \BibitemOpen
  \bibfield  {author} {\bibinfo {author} {\bibfnamefont {K.-H.}\ \bibnamefont {Lee}}\ and\ \bibinfo {author} {\bibfnamefont {K.~J.}\ \bibnamefont {Chang}},\ }\bibfield  {title} {\bibinfo {title} {Analytic continuation of the dynamic response function using an {N}-point {P}ad\'e approximant},\ }\href {https://doi.org/https://doi.org/10.1103/physrevb.54.r8285} {\bibfield  {journal} {\bibinfo  {journal} {Phys. Rev. B}\ }\textbf {\bibinfo {volume} {54}},\ \bibinfo {pages} {R8285} (\bibinfo {year} {1996})}\BibitemShut {NoStop}%
\bibitem [{\citenamefont {Jin}\ and\ \citenamefont {Chang}(1999)}]{Jin1999PRB}%
  \BibitemOpen
  \bibfield  {author} {\bibinfo {author} {\bibfnamefont {Y.-G.}\ \bibnamefont {Jin}}\ and\ \bibinfo {author} {\bibfnamefont {K.~J.}\ \bibnamefont {Chang}},\ }\bibfield  {title} {\bibinfo {title} {Dynamic response function and energy-loss spectrum for {L}i using an {N}-point {P}ad\'e approximant},\ }\href {https://doi.org/https://doi.org/10.1103/PhysRevB.59.14841} {\bibfield  {journal} {\bibinfo  {journal} {Phys. Rev. B}\ }\textbf {\bibinfo {volume} {59}},\ \bibinfo {pages} {R8285} (\bibinfo {year} {1999})}\BibitemShut {NoStop}%
\bibitem [{\citenamefont {Kas}\ \emph {et~al.}(2007)\citenamefont {Kas}, \citenamefont {Sorini}, \citenamefont {Prange}, \citenamefont {Cambell}, \citenamefont {Soininen},\ and\ \citenamefont {Rehr}}]{Kas2007PRB}%
  \BibitemOpen
  \bibfield  {author} {\bibinfo {author} {\bibfnamefont {J.~J.}\ \bibnamefont {Kas}}, \bibinfo {author} {\bibfnamefont {A.~P.}\ \bibnamefont {Sorini}}, \bibinfo {author} {\bibfnamefont {M.~P.}\ \bibnamefont {Prange}}, \bibinfo {author} {\bibfnamefont {L.~W.}\ \bibnamefont {Cambell}}, \bibinfo {author} {\bibfnamefont {J.~A.}\ \bibnamefont {Soininen}},\ and\ \bibinfo {author} {\bibfnamefont {J.~J.}\ \bibnamefont {Rehr}},\ }\bibfield  {title} {\bibinfo {title} {Many-pole model of inelastic losses in x-ray absorption spectra},\ }\href {https://doi.org/http://dx.doi.org/10.1103/PhysRevB.76.195116} {\bibfield  {journal} {\bibinfo  {journal} {Phys. Rev. B}\ }\textbf {\bibinfo {volume} {76}},\ \bibinfo {pages} {195116} (\bibinfo {year} {2007})}\BibitemShut {NoStop}%
\bibitem [{\citenamefont {Kas}\ \emph {et~al.}(2009)\citenamefont {Kas}, \citenamefont {Vinson}, \citenamefont {Trcera}, \citenamefont {Cabaret}, \citenamefont {Shirley},\ and\ \citenamefont {Rehr}}]{Kas2009JPCS}%
  \BibitemOpen
  \bibfield  {author} {\bibinfo {author} {\bibfnamefont {J.~J.}\ \bibnamefont {Kas}}, \bibinfo {author} {\bibfnamefont {J.}~\bibnamefont {Vinson}}, \bibinfo {author} {\bibfnamefont {N.}~\bibnamefont {Trcera}}, \bibinfo {author} {\bibfnamefont {D.}~\bibnamefont {Cabaret}}, \bibinfo {author} {\bibfnamefont {E.~L.}\ \bibnamefont {Shirley}},\ and\ \bibinfo {author} {\bibfnamefont {J.~J.}\ \bibnamefont {Rehr}},\ }\bibfield  {title} {\bibinfo {title} {{M}any-{P}ole {M}odel of {I}nelastic {L}osses {A}pplied to {C}alculations of {XANES}},\ }\href {https://doi.org/10.1088/1742-6596/190/1/012009} {\bibfield  {journal} {\bibinfo  {journal} {J. Phys. Conf. Ser.}\ }\textbf {\bibinfo {volume} {190}},\ \bibinfo {pages} {012009} (\bibinfo {year} {2009})}\BibitemShut {NoStop}%
\bibitem [{\citenamefont {Liang}\ and\ \citenamefont {Yang}(2015)}]{Liang2015PRL}%
  \BibitemOpen
  \bibfield  {author} {\bibinfo {author} {\bibfnamefont {Y.}~\bibnamefont {Liang}}\ and\ \bibinfo {author} {\bibfnamefont {L.}~\bibnamefont {Yang}},\ }\bibfield  {title} {\bibinfo {title} {Carrier plasmon induced nonlinear band gap renormalization in two-dimensional semiconductors},\ }\href {https://doi.org/10.1103/PhysRevLett.114.063001} {\bibfield  {journal} {\bibinfo  {journal} {Phys. Rev. Lett.}\ }\textbf {\bibinfo {volume} {114}},\ \bibinfo {pages} {063001} (\bibinfo {year} {2015})}\BibitemShut {NoStop}%
\bibitem [{\citenamefont {Champagne}\ \emph {et~al.}(2023)\citenamefont {Champagne}, \citenamefont {Haber}, \citenamefont {Pokawanvit}, \citenamefont {Qiu}, \citenamefont {Biswas}, \citenamefont {Atwater}, \citenamefont {da~Jornada},\ and\ \citenamefont {Neaton}}]{Champagne2023nanolett}%
  \BibitemOpen
  \bibfield  {author} {\bibinfo {author} {\bibfnamefont {A.}~\bibnamefont {Champagne}}, \bibinfo {author} {\bibfnamefont {J.~B.}\ \bibnamefont {Haber}}, \bibinfo {author} {\bibfnamefont {S.}~\bibnamefont {Pokawanvit}}, \bibinfo {author} {\bibfnamefont {D.~Y.}\ \bibnamefont {Qiu}}, \bibinfo {author} {\bibfnamefont {S.}~\bibnamefont {Biswas}}, \bibinfo {author} {\bibfnamefont {H.~A.}\ \bibnamefont {Atwater}}, \bibinfo {author} {\bibfnamefont {F.~H.}\ \bibnamefont {da~Jornada}},\ and\ \bibinfo {author} {\bibfnamefont {J.~B.}\ \bibnamefont {Neaton}},\ }\bibfield  {title} {\bibinfo {title} {Quasiparticle and optical properties of carrier-doped monolayer mote2 from first principles},\ }\href {https://doi.org/10.1021/acs.nanolett.3c00386} {\bibfield  {journal} {\bibinfo  {journal} {Nano Letters}\ }\textbf {\bibinfo {volume} {23}},\ \bibinfo {pages} {4274} (\bibinfo {year} {2023})},\ \bibinfo {note} {pMID: 37159934}\BibitemShut {NoStop}%
\bibitem [{\citenamefont {Lihm}\ and\ \citenamefont {Park}(2024)}]{Lihm2024PRL}%
  \BibitemOpen
  \bibfield  {author} {\bibinfo {author} {\bibfnamefont {J.-M.}\ \bibnamefont {Lihm}}\ and\ \bibinfo {author} {\bibfnamefont {C.-H.}\ \bibnamefont {Park}},\ }\bibfield  {title} {\bibinfo {title} {Plasmon-phonon hybridization in doped semiconductors from first principles},\ }\href {https://doi.org/10.1103/PhysRevLett.133.116402} {\bibfield  {journal} {\bibinfo  {journal} {Phys. Rev. Lett.}\ }\textbf {\bibinfo {volume} {133}},\ \bibinfo {pages} {116402} (\bibinfo {year} {2024})}\BibitemShut {NoStop}%
\bibitem [{\citenamefont {Riegera}\ \emph {et~al.}(1999)\citenamefont {Riegera}, \citenamefont {Steinbeck}, \citenamefont {White}, \citenamefont {Rojas},\ and\ \citenamefont {Godby}}]{Riegera1999CPC}%
  \BibitemOpen
  \bibfield  {author} {\bibinfo {author} {\bibfnamefont {M.~M.}\ \bibnamefont {Riegera}}, \bibinfo {author} {\bibfnamefont {L.}~\bibnamefont {Steinbeck}}, \bibinfo {author} {\bibfnamefont {I.~D.}\ \bibnamefont {White}}, \bibinfo {author} {\bibfnamefont {H.~N.}\ \bibnamefont {Rojas}},\ and\ \bibinfo {author} {\bibfnamefont {R.~W.}\ \bibnamefont {Godby}},\ }\bibfield  {title} {\bibinfo {title} {The gw space-time method for the self-energy of large systems},\ }\href {https://doi.org/https://doi.org/10.1016/S0010-4655(98)00174-X} {\bibfield  {journal} {\bibinfo  {journal} {Comput. Phys. Commun.}\ }\textbf {\bibinfo {volume} {117}},\ \bibinfo {pages} {211} (\bibinfo {year} {1999})}\BibitemShut {NoStop}%
\bibitem [{\citenamefont {Soininen}\ \emph {et~al.}(2005)\citenamefont {Soininen}, \citenamefont {Rehr},\ and\ \citenamefont {Shirley}}]{Soininen2005PS}%
  \BibitemOpen
  \bibfield  {author} {\bibinfo {author} {\bibfnamefont {J.~A.}\ \bibnamefont {Soininen}}, \bibinfo {author} {\bibfnamefont {J.~J.}\ \bibnamefont {Rehr}},\ and\ \bibinfo {author} {\bibfnamefont {E.~L.}\ \bibnamefont {Shirley}},\ }\bibfield  {title} {\bibinfo {title} {Multipole representation of the dielectric matrix},\ }\href {https://doi.org/https://doi.org/10.1238/physica.topical.115a00243} {\bibfield  {journal} {\bibinfo  {journal} {Phys. Scripta}\ }\textbf {\bibinfo {volume} {2005}},\ \bibinfo {pages} {243} (\bibinfo {year} {2005})}\BibitemShut {NoStop}%
\bibitem [{\citenamefont {van Setten}\ \emph {et~al.}(2015)\citenamefont {van Setten}, \citenamefont {Caruso}, \citenamefont {Sharifzadeh}, \citenamefont {Ren}, \citenamefont {Scheffler}, \citenamefont {Liu}, \citenamefont {Lischner}, \citenamefont {Lin}, \citenamefont {Deslippe}, \citenamefont {Yang}, \citenamefont {Weigend}, \citenamefont {Neaton}, \citenamefont {Evers},\ and\ \citenamefont {Rinke}}]{vanSetten2015JCTC}%
  \BibitemOpen
  \bibfield  {author} {\bibinfo {author} {\bibfnamefont {M.~J.}\ \bibnamefont {van Setten}}, \bibinfo {author} {\bibfnamefont {F.}~\bibnamefont {Caruso}}, \bibinfo {author} {\bibfnamefont {S.}~\bibnamefont {Sharifzadeh}}, \bibinfo {author} {\bibfnamefont {X.}~\bibnamefont {Ren}}, \bibinfo {author} {\bibfnamefont {M.}~\bibnamefont {Scheffler}}, \bibinfo {author} {\bibfnamefont {F.}~\bibnamefont {Liu}}, \bibinfo {author} {\bibfnamefont {J.}~\bibnamefont {Lischner}}, \bibinfo {author} {\bibfnamefont {L.}~\bibnamefont {Lin}}, \bibinfo {author} {\bibfnamefont {J.~R.}\ \bibnamefont {Deslippe}}, \bibinfo {author} {\bibfnamefont {S.~G. L.~C.}\ \bibnamefont {Yang}}, \bibinfo {author} {\bibfnamefont {F.}~\bibnamefont {Weigend}}, \bibinfo {author} {\bibfnamefont {J.~B.}\ \bibnamefont {Neaton}}, \bibinfo {author} {\bibfnamefont {F.}~\bibnamefont {Evers}},\ and\ \bibinfo {author} {\bibfnamefont {P.}~\bibnamefont {Rinke}},\ }\bibfield  {title} {\bibinfo {title} {{GW}100: {B}enchmarking {G0W0} for {M}olecular {S}ystems},\
  }\href {https://doi.org/https://doi.org/10.1021/acs.jctc.5b00453} {\bibfield  {journal} {\bibinfo  {journal} {J. Chem. Theory Comput.}\ }\textbf {\bibinfo {volume} {11}},\ \bibinfo {pages} {5665} (\bibinfo {year} {2015})}\BibitemShut {NoStop}%
\bibitem [{\citenamefont {Chiarotti}\ \emph {et~al.}(2022)\citenamefont {Chiarotti}, \citenamefont {Marzari},\ and\ \citenamefont {Ferretti}}]{Chiarotti2022PRR}%
  \BibitemOpen
  \bibfield  {author} {\bibinfo {author} {\bibfnamefont {T.}~\bibnamefont {Chiarotti}}, \bibinfo {author} {\bibfnamefont {N.}~\bibnamefont {Marzari}},\ and\ \bibinfo {author} {\bibfnamefont {A.}~\bibnamefont {Ferretti}},\ }\bibfield  {title} {\bibinfo {title} {Unified green's function approach for spectral and thermodynamic properties from algorithmic inversion of dynamical potentials},\ }\href {https://doi.org/10.1103/PhysRevResearch.4.013242} {\bibfield  {journal} {\bibinfo  {journal} {Phys. Rev. Res.}\ }\textbf {\bibinfo {volume} {4}},\ \bibinfo {pages} {013242} (\bibinfo {year} {2022})}\BibitemShut {NoStop}%
\bibitem [{\citenamefont {Chiarotti}\ \emph {et~al.}(2024)\citenamefont {Chiarotti}, \citenamefont {Ferretti},\ and\ \citenamefont {Marzari}}]{Chiarotti2024PRR}%
  \BibitemOpen
  \bibfield  {author} {\bibinfo {author} {\bibfnamefont {T.}~\bibnamefont {Chiarotti}}, \bibinfo {author} {\bibfnamefont {A.}~\bibnamefont {Ferretti}},\ and\ \bibinfo {author} {\bibfnamefont {N.}~\bibnamefont {Marzari}},\ }\bibfield  {title} {\bibinfo {title} {Energies and spectra of solids from the algorithmic inversion of dynamical hubbard functionals},\ }\href {https://doi.org/10.1103/PhysRevResearch.6.L032023} {\bibfield  {journal} {\bibinfo  {journal} {Phys. Rev. Res.}\ }\textbf {\bibinfo {volume} {6}},\ \bibinfo {pages} {L032023} (\bibinfo {year} {2024})}\BibitemShut {NoStop}%
\bibitem [{\citenamefont {Ferretti}\ \emph {et~al.}(2024)\citenamefont {Ferretti}, \citenamefont {Chiarotti},\ and\ \citenamefont {Marzari}}]{Ferretti2024PRB}%
  \BibitemOpen
  \bibfield  {author} {\bibinfo {author} {\bibfnamefont {A.}~\bibnamefont {Ferretti}}, \bibinfo {author} {\bibfnamefont {T.}~\bibnamefont {Chiarotti}},\ and\ \bibinfo {author} {\bibfnamefont {N.}~\bibnamefont {Marzari}},\ }\bibfield  {title} {\bibinfo {title} {Green's function embedding using sum-over-pole representations},\ }\href {https://doi.org/10.1103/PhysRevB.110.045149} {\bibfield  {journal} {\bibinfo  {journal} {Phys. Rev. B}\ }\textbf {\bibinfo {volume} {110}},\ \bibinfo {pages} {045149} (\bibinfo {year} {2024})}\BibitemShut {NoStop}%
\bibitem [{\citenamefont {Quinzi}\ \emph {et~al.}(2025)\citenamefont {Quinzi}, \citenamefont {Chiarotti}, \citenamefont {Gibertini},\ and\ \citenamefont {Ferretti}}]{Quinzi2025PRB}%
  \BibitemOpen
  \bibfield  {author} {\bibinfo {author} {\bibfnamefont {M.}~\bibnamefont {Quinzi}}, \bibinfo {author} {\bibfnamefont {T.}~\bibnamefont {Chiarotti}}, \bibinfo {author} {\bibfnamefont {M.}~\bibnamefont {Gibertini}},\ and\ \bibinfo {author} {\bibfnamefont {A.}~\bibnamefont {Ferretti}},\ }\bibfield  {title} {\bibinfo {title} {Broken symmetry solutions in one-dimensional lattice models via many-body perturbation theory},\ }\href {https://doi.org/10.1103/PhysRevB.111.125148} {\bibfield  {journal} {\bibinfo  {journal} {Phys. Rev. B}\ }\textbf {\bibinfo {volume} {111}},\ \bibinfo {pages} {125148} (\bibinfo {year} {2025})}\BibitemShut {NoStop}%
\bibitem [{\citenamefont {Ismail-Beigi}(2010)}]{Ismail-Beigi2010PRB}%
  \BibitemOpen
  \bibfield  {author} {\bibinfo {author} {\bibfnamefont {S.}~\bibnamefont {Ismail-Beigi}},\ }\bibfield  {title} {\bibinfo {title} {Correlation energy functional within the {GW-RPA}: {E}xact forms, approximate forms, and challenges},\ }\href {https://doi.org/https://doi.org/10.1103/PhysRevB.81.195126} {\bibfield  {journal} {\bibinfo  {journal} {Phys. Rev. B}\ }\textbf {\bibinfo {volume} {81}},\ \bibinfo {pages} {195126} (\bibinfo {year} {2010})}\BibitemShut {NoStop}%
\bibitem [{\citenamefont {Guo}\ and\ \citenamefont {Liu}(2024)}]{Guo2024Arxiv}%
  \BibitemOpen
  \bibfield  {author} {\bibinfo {author} {\bibfnamefont {Z.}~\bibnamefont {Guo}}\ and\ \bibinfo {author} {\bibfnamefont {J.}~\bibnamefont {Liu}},\ }\href@noop {} {\bibinfo {title} {Beyond-mean-field studies of wigner crystal transitions in various interacting two-dimensional systems}},\ \bibinfo {howpublished} {\url{https://arxiv.org/abs/2409.14658v2}} (\bibinfo {year} {2024})\BibitemShut {NoStop}%
\bibitem [{\citenamefont {Lehmann}\ and\ \citenamefont {Taut}(1972)}]{Lehmann1972}%
  \BibitemOpen
  \bibfield  {author} {\bibinfo {author} {\bibfnamefont {G.}~\bibnamefont {Lehmann}}\ and\ \bibinfo {author} {\bibfnamefont {M.}~\bibnamefont {Taut}},\ }\bibfield  {title} {\bibinfo {title} {On the numerical calculation of the density of states and related properties},\ }\href {https://doi.org/https://doi.org/10.1002/pssb.2220540211} {\bibfield  {journal} {\bibinfo  {journal} {Phys. Status Solidi (b)}\ }\textbf {\bibinfo {volume} {54}},\ \bibinfo {pages} {469} (\bibinfo {year} {1972})}\BibitemShut {NoStop}%
\bibitem [{\citenamefont {Farid}(2002)}]{Farid2002PMB}%
  \BibitemOpen
  \bibfield  {author} {\bibinfo {author} {\bibfnamefont {B.}~\bibnamefont {Farid}},\ }\bibfield  {title} {\bibinfo {title} {Dynamical correlation functions expressed in terms of many-particle ground-state wavefunction; the dynamical self-energy operator},\ }\href {https://doi.org/10.1080/13642810208222682} {\bibfield  {journal} {\bibinfo  {journal} {Philosophical Magazine B}\ }\textbf {\bibinfo {volume} {82}},\ \bibinfo {pages} {1413} (\bibinfo {year} {2002})}\BibitemShut {NoStop}%
\bibitem [{\citenamefont {Rasmussen}\ \emph {et~al.}(2021)\citenamefont {Rasmussen}, \citenamefont {Deilmann},\ and\ \citenamefont {Thygesen}}]{Rasmussen2021NPJComputMater}%
  \BibitemOpen
  \bibfield  {author} {\bibinfo {author} {\bibfnamefont {A.}~\bibnamefont {Rasmussen}}, \bibinfo {author} {\bibfnamefont {T.}~\bibnamefont {Deilmann}},\ and\ \bibinfo {author} {\bibfnamefont {K.~S.}\ \bibnamefont {Thygesen}},\ }\bibfield  {title} {\bibinfo {title} {Towards fully automatized {GW} band structure calculations: {W}hat we can learn from 60.000 self-energy evaluations},\ }\href {https://doi.org/https://doi.org/10.1038/s41524-020-00480-7} {\bibfield  {journal} {\bibinfo  {journal} {NPJ Comput. Mater.}\ }\textbf {\bibinfo {volume} {7}} (\bibinfo {year} {2021})}\BibitemShut {NoStop}%
\bibitem [{\citenamefont {Georges}\ \emph {et~al.}(1996)\citenamefont {Georges}, \citenamefont {Kotliar}, \citenamefont {Krauth},\ and\ \citenamefont {Rozenberg}}]{Georges1996RevModPhys}%
  \BibitemOpen
  \bibfield  {author} {\bibinfo {author} {\bibfnamefont {A.}~\bibnamefont {Georges}}, \bibinfo {author} {\bibfnamefont {G.}~\bibnamefont {Kotliar}}, \bibinfo {author} {\bibfnamefont {W.}~\bibnamefont {Krauth}},\ and\ \bibinfo {author} {\bibfnamefont {M.~J.}\ \bibnamefont {Rozenberg}},\ }\bibfield  {title} {\bibinfo {title} {Dynamical mean-field theory of strongly correlated fermion systems and the limit of infinite dimensions},\ }\href {https://doi.org/10.1103/RevModPhys.68.13} {\bibfield  {journal} {\bibinfo  {journal} {Rev. Mod. Phys.}\ }\textbf {\bibinfo {volume} {68}},\ \bibinfo {pages} {13} (\bibinfo {year} {1996})}\BibitemShut {NoStop}%
\bibitem [{\citenamefont {von Barth}\ and\ \citenamefont {Holm}(1996)}]{vonBarth1996PRB}%
  \BibitemOpen
  \bibfield  {author} {\bibinfo {author} {\bibfnamefont {U.}~\bibnamefont {von Barth}}\ and\ \bibinfo {author} {\bibfnamefont {B.}~\bibnamefont {Holm}},\ }\bibfield  {title} {\bibinfo {title} {Self-consistent ${\mathit{gw}}_{0}$ results for the electron gas: Fixed screened potential ${\mathit{w}}_{0}$ within the random-phase approximation},\ }\href {https://doi.org/10.1103/PhysRevB.54.8411} {\bibfield  {journal} {\bibinfo  {journal} {Phys. Rev. B}\ }\textbf {\bibinfo {volume} {54}},\ \bibinfo {pages} {8411} (\bibinfo {year} {1996})}\BibitemShut {NoStop}%
\bibitem [{\citenamefont {Marini}\ \emph {et~al.}(2009)\citenamefont {Marini}, \citenamefont {Hogan}, \citenamefont {Gr{\" u}ning},\ and\ \citenamefont {Varsano}}]{Marini2009CPC}%
  \BibitemOpen
  \bibfield  {author} {\bibinfo {author} {\bibfnamefont {A.}~\bibnamefont {Marini}}, \bibinfo {author} {\bibfnamefont {C.}~\bibnamefont {Hogan}}, \bibinfo {author} {\bibfnamefont {M.}~\bibnamefont {Gr{\" u}ning}},\ and\ \bibinfo {author} {\bibfnamefont {D.}~\bibnamefont {Varsano}},\ }\bibfield  {title} {\bibinfo {title} {yambo: An ab initio tool for excited state calculations},\ }\href {https://doi.org/https://doi.org/10.1016/j.cpc.2009.02.003} {\bibfield  {journal} {\bibinfo  {journal} {Comput. Phys. Commun.}\ }\textbf {\bibinfo {volume} {180}},\ \bibinfo {pages} {1392} (\bibinfo {year} {2009})}\BibitemShut {NoStop}%
\bibitem [{\citenamefont {Sangalli}\ \emph {et~al.}(2019)\citenamefont {Sangalli}, \citenamefont {Ferretti}, \citenamefont {Miranda}, \citenamefont {Attaccalite}, \citenamefont {Marri}, \citenamefont {Cannuccia}, \citenamefont {Melo}, \citenamefont {Marsili}, \citenamefont {Paleari}, \citenamefont {Marrazzo}, \citenamefont {Prandini}, \citenamefont {Bonf{\`{a}}}, \citenamefont {Atambo}, \citenamefont {Affinito}, \citenamefont {Palummo}, \citenamefont {Molina-S{\'{a}}nchez}, \citenamefont {Hogan}, \citenamefont {Gr{\" u}ning}, \citenamefont {Varsano},\ and\ \citenamefont {Marini}}]{Sangalli2019JPCM}%
  \BibitemOpen
  \bibfield  {author} {\bibinfo {author} {\bibfnamefont {D.}~\bibnamefont {Sangalli}}, \bibinfo {author} {\bibfnamefont {A.}~\bibnamefont {Ferretti}}, \bibinfo {author} {\bibfnamefont {H.}~\bibnamefont {Miranda}}, \bibinfo {author} {\bibfnamefont {C.}~\bibnamefont {Attaccalite}}, \bibinfo {author} {\bibfnamefont {I.}~\bibnamefont {Marri}}, \bibinfo {author} {\bibfnamefont {E.}~\bibnamefont {Cannuccia}}, \bibinfo {author} {\bibfnamefont {P.}~\bibnamefont {Melo}}, \bibinfo {author} {\bibfnamefont {M.}~\bibnamefont {Marsili}}, \bibinfo {author} {\bibfnamefont {F.}~\bibnamefont {Paleari}}, \bibinfo {author} {\bibfnamefont {A.}~\bibnamefont {Marrazzo}}, \bibinfo {author} {\bibfnamefont {G.}~\bibnamefont {Prandini}}, \bibinfo {author} {\bibfnamefont {P.}~\bibnamefont {Bonf{\`{a}}}}, \bibinfo {author} {\bibfnamefont {M.~O.}\ \bibnamefont {Atambo}}, \bibinfo {author} {\bibfnamefont {F.}~\bibnamefont {Affinito}}, \bibinfo {author} {\bibfnamefont {M.}~\bibnamefont {Palummo}}, \bibinfo {author} {\bibfnamefont
  {A.}~\bibnamefont {Molina-S{\'{a}}nchez}}, \bibinfo {author} {\bibfnamefont {C.}~\bibnamefont {Hogan}}, \bibinfo {author} {\bibfnamefont {M.}~\bibnamefont {Gr{\" u}ning}}, \bibinfo {author} {\bibfnamefont {D.}~\bibnamefont {Varsano}},\ and\ \bibinfo {author} {\bibfnamefont {A.}~\bibnamefont {Marini}},\ }\bibfield  {title} {\bibinfo {title} {Many-body perturbation theory calculations using the yambo code},\ }\href {https://doi.org/https://doi.org/10.1088/1361-648X/ab15d0} {\bibfield  {journal} {\bibinfo  {journal} {J. Phys.: Condens. Matter}\ }\textbf {\bibinfo {volume} {31}},\ \bibinfo {pages} {325902} (\bibinfo {year} {2019})}\BibitemShut {NoStop}%
\bibitem [{\citenamefont {Mortensen}\ \emph {et~al.}(2024)\citenamefont {Mortensen}, \citenamefont {Larsen}, \citenamefont {Kuisma}, \citenamefont {Ivanov}, \citenamefont {Taghizadeh}, \citenamefont {Peterson}, \citenamefont {Haldar}, \citenamefont {Dohn}, \citenamefont {Sch{\" a}fer}, \citenamefont {Jónsson}, \citenamefont {Hermes}, \citenamefont {Nilsson}, \citenamefont {Kastlunger}, \citenamefont {Levi}, \citenamefont {Jónsson}, \citenamefont {H{\" a}kkinen}, \citenamefont {Fojt}, \citenamefont {Kangsabanik}, \citenamefont {S{\o}dequist}, \citenamefont {Lehtom{\" a}ki}, \citenamefont {Heske}, \citenamefont {Enkovaara}, \citenamefont {Winther}, \citenamefont {Dulak}, \citenamefont {Melander}, \citenamefont {Ovesen}, \citenamefont {Louhivuori}, \citenamefont {Walter}, \citenamefont {Gjerding}, \citenamefont {Lopez-Acevedo}, \citenamefont {Erhart}, \citenamefont {Warmbier}, \citenamefont {W{\" u}rdemann}, \citenamefont {Kaappa}, \citenamefont {Latini}, \citenamefont {Boland}, \citenamefont {Bligaard},
  \citenamefont {Skovhus}, \citenamefont {Susi}, \citenamefont {Maxson}, \citenamefont {Rossi}, \citenamefont {Chen}, \citenamefont {Schmerwitz}, \citenamefont {Schi{\o}tz}, \citenamefont {Olsen}, \citenamefont {Jacobsen},\ and\ \citenamefont {Thygesen}}]{gpaw2024JCP}%
  \BibitemOpen
  \bibfield  {author} {\bibinfo {author} {\bibfnamefont {J.~J.}\ \bibnamefont {Mortensen}}, \bibinfo {author} {\bibfnamefont {A.~H.}\ \bibnamefont {Larsen}}, \bibinfo {author} {\bibfnamefont {M.}~\bibnamefont {Kuisma}}, \bibinfo {author} {\bibfnamefont {A.~V.}\ \bibnamefont {Ivanov}}, \bibinfo {author} {\bibfnamefont {A.}~\bibnamefont {Taghizadeh}}, \bibinfo {author} {\bibfnamefont {A.}~\bibnamefont {Peterson}}, \bibinfo {author} {\bibfnamefont {A.}~\bibnamefont {Haldar}}, \bibinfo {author} {\bibfnamefont {A.~O.}\ \bibnamefont {Dohn}}, \bibinfo {author} {\bibfnamefont {C.}~\bibnamefont {Sch{\" a}fer}}, \bibinfo {author} {\bibfnamefont {E.~{\" O}.}\ \bibnamefont {Jónsson}}, \bibinfo {author} {\bibfnamefont {E.~D.}\ \bibnamefont {Hermes}}, \bibinfo {author} {\bibfnamefont {F.~A.}\ \bibnamefont {Nilsson}}, \bibinfo {author} {\bibfnamefont {G.}~\bibnamefont {Kastlunger}}, \bibinfo {author} {\bibfnamefont {G.}~\bibnamefont {Levi}}, \bibinfo {author} {\bibfnamefont {H.}~\bibnamefont {Jónsson}}, \bibinfo {author}
  {\bibfnamefont {H.}~\bibnamefont {H{\" a}kkinen}}, \bibinfo {author} {\bibfnamefont {J.}~\bibnamefont {Fojt}}, \bibinfo {author} {\bibfnamefont {J.}~\bibnamefont {Kangsabanik}}, \bibinfo {author} {\bibfnamefont {J.}~\bibnamefont {S{\o}dequist}}, \bibinfo {author} {\bibfnamefont {J.}~\bibnamefont {Lehtom{\" a}ki}}, \bibinfo {author} {\bibfnamefont {J.}~\bibnamefont {Heske}}, \bibinfo {author} {\bibfnamefont {J.}~\bibnamefont {Enkovaara}}, \bibinfo {author} {\bibfnamefont {K.~T.}\ \bibnamefont {Winther}}, \bibinfo {author} {\bibfnamefont {M.}~\bibnamefont {Dulak}}, \bibinfo {author} {\bibfnamefont {M.~M.}\ \bibnamefont {Melander}}, \bibinfo {author} {\bibfnamefont {M.}~\bibnamefont {Ovesen}}, \bibinfo {author} {\bibfnamefont {M.}~\bibnamefont {Louhivuori}}, \bibinfo {author} {\bibfnamefont {M.}~\bibnamefont {Walter}}, \bibinfo {author} {\bibfnamefont {M.}~\bibnamefont {Gjerding}}, \bibinfo {author} {\bibfnamefont {O.}~\bibnamefont {Lopez-Acevedo}}, \bibinfo {author} {\bibfnamefont {P.}~\bibnamefont {Erhart}},
  \bibinfo {author} {\bibfnamefont {R.}~\bibnamefont {Warmbier}}, \bibinfo {author} {\bibfnamefont {R.}~\bibnamefont {W{\" u}rdemann}}, \bibinfo {author} {\bibfnamefont {S.}~\bibnamefont {Kaappa}}, \bibinfo {author} {\bibfnamefont {S.}~\bibnamefont {Latini}}, \bibinfo {author} {\bibfnamefont {T.~M.}\ \bibnamefont {Boland}}, \bibinfo {author} {\bibfnamefont {T.}~\bibnamefont {Bligaard}}, \bibinfo {author} {\bibfnamefont {T.}~\bibnamefont {Skovhus}}, \bibinfo {author} {\bibfnamefont {T.}~\bibnamefont {Susi}}, \bibinfo {author} {\bibfnamefont {T.}~\bibnamefont {Maxson}}, \bibinfo {author} {\bibfnamefont {T.}~\bibnamefont {Rossi}}, \bibinfo {author} {\bibfnamefont {X.}~\bibnamefont {Chen}}, \bibinfo {author} {\bibfnamefont {Y.~L.~A.}\ \bibnamefont {Schmerwitz}}, \bibinfo {author} {\bibfnamefont {J.}~\bibnamefont {Schi{\o}tz}}, \bibinfo {author} {\bibfnamefont {T.}~\bibnamefont {Olsen}}, \bibinfo {author} {\bibfnamefont {K.~W.}\ \bibnamefont {Jacobsen}},\ and\ \bibinfo {author} {\bibfnamefont {K.~S.}\ \bibnamefont
  {Thygesen}},\ }\bibfield  {title} {\bibinfo {title} {{GPAW: An open Python package for electronic structure calculations}},\ }\href {https://doi.org/10.1063/5.0182685} {\bibfield  {journal} {\bibinfo  {journal} {The Journal of Chemical Physics}\ }\textbf {\bibinfo {volume} {160}},\ \bibinfo {pages} {092503} (\bibinfo {year} {2024})}\BibitemShut {NoStop}%
\bibitem [{sup()}]{supp-info}%
  \BibitemOpen
  \href@noop {} {}\bibinfo {note} {See Supplemental Materials for a detailed description.}\BibitemShut {Stop}%
\bibitem [{\citenamefont {Gesenhues}\ \emph {et~al.}(2017)\citenamefont {Gesenhues}, \citenamefont {Nabok}, \citenamefont {Rohlfing},\ and\ \citenamefont {Draxl}}]{Gesenhue2017PRB}%
  \BibitemOpen
  \bibfield  {author} {\bibinfo {author} {\bibfnamefont {J.}~\bibnamefont {Gesenhues}}, \bibinfo {author} {\bibfnamefont {D.}~\bibnamefont {Nabok}}, \bibinfo {author} {\bibfnamefont {M.}~\bibnamefont {Rohlfing}},\ and\ \bibinfo {author} {\bibfnamefont {C.}~\bibnamefont {Draxl}},\ }\bibfield  {title} {\bibinfo {title} {Analytical representation of dynamical quantities in $gw$ from a matrix resolvent},\ }\href {https://doi.org/10.1103/PhysRevB.96.245124} {\bibfield  {journal} {\bibinfo  {journal} {Phys. Rev. B}\ }\textbf {\bibinfo {volume} {96}},\ \bibinfo {pages} {245124} (\bibinfo {year} {2017})}\BibitemShut {NoStop}%
\bibitem [{\citenamefont {Güttel}\ and\ \citenamefont {Tisseur}(2017)}]{Guttel2017ActNum}%
  \BibitemOpen
  \bibfield  {author} {\bibinfo {author} {\bibfnamefont {S.}~\bibnamefont {Güttel}}\ and\ \bibinfo {author} {\bibfnamefont {F.}~\bibnamefont {Tisseur}},\ }\bibfield  {title} {\bibinfo {title} {The nonlinear eigenvalue problem},\ }\href {https://doi.org/10.1017/S0962492917000034} {\bibfield  {journal} {\bibinfo  {journal} {Acta Numerica}\ }\textbf {\bibinfo {volume} {26}},\ \bibinfo {pages} {1–94} (\bibinfo {year} {2017})}\BibitemShut {NoStop}%
\bibitem [{\citenamefont {Giannozzi}\ \emph {et~al.}(2009)\citenamefont {Giannozzi}, \citenamefont {Baroni}, \citenamefont {Bonini}, \citenamefont {Calandra}, \citenamefont {Car}, \citenamefont {Cavazzoni}, \citenamefont {Ceresoli}, \citenamefont {Chiarotti}, \citenamefont {Cococcioni}, \citenamefont {Dabo}, \citenamefont {Corso}, \citenamefont {de~Gironcoli}, \citenamefont {Fabris}, \citenamefont {Fratesi}, \citenamefont {Gebauer}, \citenamefont {Gerstmann}, \citenamefont {Gougoussis}, \citenamefont {Kokalj}, \citenamefont {Lazzeri}, \citenamefont {Martin-Samos}, \citenamefont {Marzari}, \citenamefont {Mauri}, \citenamefont {Mazzarello}, \citenamefont {Paolini}, \citenamefont {Pasquarello}, \citenamefont {Paulatto}, \citenamefont {Sbraccia}, \citenamefont {Scandolo}, \citenamefont {Sclauzero}, \citenamefont {Seitsonen}, \citenamefont {Smogunov}, \citenamefont {Umari},\ and\ \citenamefont {Wentzcovitch}}]{QE1}%
  \BibitemOpen
  \bibfield  {author} {\bibinfo {author} {\bibfnamefont {P.}~\bibnamefont {Giannozzi}}, \bibinfo {author} {\bibfnamefont {S.}~\bibnamefont {Baroni}}, \bibinfo {author} {\bibfnamefont {N.}~\bibnamefont {Bonini}}, \bibinfo {author} {\bibfnamefont {M.}~\bibnamefont {Calandra}}, \bibinfo {author} {\bibfnamefont {R.}~\bibnamefont {Car}}, \bibinfo {author} {\bibfnamefont {C.}~\bibnamefont {Cavazzoni}}, \bibinfo {author} {\bibfnamefont {D.}~\bibnamefont {Ceresoli}}, \bibinfo {author} {\bibfnamefont {G.~L.}\ \bibnamefont {Chiarotti}}, \bibinfo {author} {\bibfnamefont {M.}~\bibnamefont {Cococcioni}}, \bibinfo {author} {\bibfnamefont {I.}~\bibnamefont {Dabo}}, \bibinfo {author} {\bibfnamefont {A.~D.}\ \bibnamefont {Corso}}, \bibinfo {author} {\bibfnamefont {S.}~\bibnamefont {de~Gironcoli}}, \bibinfo {author} {\bibfnamefont {S.}~\bibnamefont {Fabris}}, \bibinfo {author} {\bibfnamefont {G.}~\bibnamefont {Fratesi}}, \bibinfo {author} {\bibfnamefont {R.}~\bibnamefont {Gebauer}}, \bibinfo {author} {\bibfnamefont
  {U.}~\bibnamefont {Gerstmann}}, \bibinfo {author} {\bibfnamefont {C.}~\bibnamefont {Gougoussis}}, \bibinfo {author} {\bibfnamefont {A.}~\bibnamefont {Kokalj}}, \bibinfo {author} {\bibfnamefont {M.}~\bibnamefont {Lazzeri}}, \bibinfo {author} {\bibfnamefont {L.}~\bibnamefont {Martin-Samos}}, \bibinfo {author} {\bibfnamefont {N.}~\bibnamefont {Marzari}}, \bibinfo {author} {\bibfnamefont {F.}~\bibnamefont {Mauri}}, \bibinfo {author} {\bibfnamefont {R.}~\bibnamefont {Mazzarello}}, \bibinfo {author} {\bibfnamefont {S.}~\bibnamefont {Paolini}}, \bibinfo {author} {\bibfnamefont {A.}~\bibnamefont {Pasquarello}}, \bibinfo {author} {\bibfnamefont {L.}~\bibnamefont {Paulatto}}, \bibinfo {author} {\bibfnamefont {C.}~\bibnamefont {Sbraccia}}, \bibinfo {author} {\bibfnamefont {S.}~\bibnamefont {Scandolo}}, \bibinfo {author} {\bibfnamefont {G.}~\bibnamefont {Sclauzero}}, \bibinfo {author} {\bibfnamefont {A.~P.}\ \bibnamefont {Seitsonen}}, \bibinfo {author} {\bibfnamefont {A.}~\bibnamefont {Smogunov}}, \bibinfo {author}
  {\bibfnamefont {P.}~\bibnamefont {Umari}},\ and\ \bibinfo {author} {\bibfnamefont {R.~M.}\ \bibnamefont {Wentzcovitch}},\ }\bibfield  {title} {\bibinfo {title} {{QUANTUM} {ESPRESSO}: a modular and open-source software project for quantum simulations of materials},\ }\href {https://doi.org/https://doi.org/10.1088/0953-8984/21/39/395502} {\bibfield  {journal} {\bibinfo  {journal} {J. Phys.: Condens. Matter}\ }\textbf {\bibinfo {volume} {21}},\ \bibinfo {pages} {395502} (\bibinfo {year} {2009})}\BibitemShut {NoStop}%
\bibitem [{\citenamefont {Giannozzi}\ \emph {et~al.}(2017)\citenamefont {Giannozzi}, \citenamefont {Andreussi}, \citenamefont {Brumme}, \citenamefont {Bunau}, \citenamefont {Nardelli}, \citenamefont {Calandra}, \citenamefont {Car}, \citenamefont {Cavazzoni}, \citenamefont {Ceresoli}, \citenamefont {Cococcioni}, \citenamefont {Colonna}, \citenamefont {Carnimeo}, \citenamefont {Corso}, \citenamefont {de~Gironcoli}, \citenamefont {Delugas}, \citenamefont {DiStasio}, \citenamefont {Ferretti}, \citenamefont {Floris}, \citenamefont {Fratesi}, \citenamefont {Fugallo}, \citenamefont {Gebauer}, \citenamefont {Gerstmann}, \citenamefont {Giustino}, \citenamefont {Gorni}, \citenamefont {Jia}, \citenamefont {Kawamura}, \citenamefont {Ko}, \citenamefont {Kokalj}, \citenamefont {K{\" u}{\c{c}}{\" u}kbenli}, \citenamefont {Lazzeri}, \citenamefont {Marsili}, \citenamefont {Marzari}, \citenamefont {Mauri}, \citenamefont {Nguyen}, \citenamefont {Nguyen}, \citenamefont {de-la Roza}, \citenamefont {Paulatto}, \citenamefont
  {Ponc{\'{e}}}, \citenamefont {Rocca}, \citenamefont {Sabatini}, \citenamefont {Santra}, \citenamefont {Schlipf}, \citenamefont {Seitsonen}, \citenamefont {Smogunov}, \citenamefont {Timrov}, \citenamefont {Thonhauser}, \citenamefont {Umari}, \citenamefont {Vast}, \citenamefont {Wu},\ and\ \citenamefont {Baroni}}]{QE2}%
  \BibitemOpen
  \bibfield  {author} {\bibinfo {author} {\bibfnamefont {P.}~\bibnamefont {Giannozzi}}, \bibinfo {author} {\bibfnamefont {O.}~\bibnamefont {Andreussi}}, \bibinfo {author} {\bibfnamefont {T.}~\bibnamefont {Brumme}}, \bibinfo {author} {\bibfnamefont {O.}~\bibnamefont {Bunau}}, \bibinfo {author} {\bibfnamefont {M.~B.}\ \bibnamefont {Nardelli}}, \bibinfo {author} {\bibfnamefont {M.}~\bibnamefont {Calandra}}, \bibinfo {author} {\bibfnamefont {R.}~\bibnamefont {Car}}, \bibinfo {author} {\bibfnamefont {C.}~\bibnamefont {Cavazzoni}}, \bibinfo {author} {\bibfnamefont {D.}~\bibnamefont {Ceresoli}}, \bibinfo {author} {\bibfnamefont {M.}~\bibnamefont {Cococcioni}}, \bibinfo {author} {\bibfnamefont {N.}~\bibnamefont {Colonna}}, \bibinfo {author} {\bibfnamefont {I.}~\bibnamefont {Carnimeo}}, \bibinfo {author} {\bibfnamefont {A.~D.}\ \bibnamefont {Corso}}, \bibinfo {author} {\bibfnamefont {S.}~\bibnamefont {de~Gironcoli}}, \bibinfo {author} {\bibfnamefont {P.}~\bibnamefont {Delugas}}, \bibinfo {author} {\bibfnamefont
  {R.~A.}\ \bibnamefont {DiStasio}}, \bibinfo {author} {\bibfnamefont {A.}~\bibnamefont {Ferretti}}, \bibinfo {author} {\bibfnamefont {A.}~\bibnamefont {Floris}}, \bibinfo {author} {\bibfnamefont {G.}~\bibnamefont {Fratesi}}, \bibinfo {author} {\bibfnamefont {G.}~\bibnamefont {Fugallo}}, \bibinfo {author} {\bibfnamefont {R.}~\bibnamefont {Gebauer}}, \bibinfo {author} {\bibfnamefont {U.}~\bibnamefont {Gerstmann}}, \bibinfo {author} {\bibfnamefont {F.}~\bibnamefont {Giustino}}, \bibinfo {author} {\bibfnamefont {T.}~\bibnamefont {Gorni}}, \bibinfo {author} {\bibfnamefont {J.}~\bibnamefont {Jia}}, \bibinfo {author} {\bibfnamefont {M.}~\bibnamefont {Kawamura}}, \bibinfo {author} {\bibfnamefont {H.-Y.}\ \bibnamefont {Ko}}, \bibinfo {author} {\bibfnamefont {A.}~\bibnamefont {Kokalj}}, \bibinfo {author} {\bibfnamefont {E.}~\bibnamefont {K{\" u}{\c{c}}{\" u}kbenli}}, \bibinfo {author} {\bibfnamefont {M.}~\bibnamefont {Lazzeri}}, \bibinfo {author} {\bibfnamefont {M.}~\bibnamefont {Marsili}}, \bibinfo {author}
  {\bibfnamefont {N.}~\bibnamefont {Marzari}}, \bibinfo {author} {\bibfnamefont {F.}~\bibnamefont {Mauri}}, \bibinfo {author} {\bibfnamefont {N.~L.}\ \bibnamefont {Nguyen}}, \bibinfo {author} {\bibfnamefont {H.-V.}\ \bibnamefont {Nguyen}}, \bibinfo {author} {\bibfnamefont {A.~O.}\ \bibnamefont {de-la Roza}}, \bibinfo {author} {\bibfnamefont {L.}~\bibnamefont {Paulatto}}, \bibinfo {author} {\bibfnamefont {S.}~\bibnamefont {Ponc{\'{e}}}}, \bibinfo {author} {\bibfnamefont {D.}~\bibnamefont {Rocca}}, \bibinfo {author} {\bibfnamefont {R.}~\bibnamefont {Sabatini}}, \bibinfo {author} {\bibfnamefont {B.}~\bibnamefont {Santra}}, \bibinfo {author} {\bibfnamefont {M.}~\bibnamefont {Schlipf}}, \bibinfo {author} {\bibfnamefont {A.~P.}\ \bibnamefont {Seitsonen}}, \bibinfo {author} {\bibfnamefont {A.}~\bibnamefont {Smogunov}}, \bibinfo {author} {\bibfnamefont {I.}~\bibnamefont {Timrov}}, \bibinfo {author} {\bibfnamefont {T.}~\bibnamefont {Thonhauser}}, \bibinfo {author} {\bibfnamefont {P.}~\bibnamefont {Umari}}, \bibinfo
  {author} {\bibfnamefont {N.}~\bibnamefont {Vast}}, \bibinfo {author} {\bibfnamefont {X.}~\bibnamefont {Wu}},\ and\ \bibinfo {author} {\bibfnamefont {S.}~\bibnamefont {Baroni}},\ }\bibfield  {title} {\bibinfo {title} {Advanced capabilities for materials modelling with {Q}uantum {ESPRESSO}},\ }\href {https://doi.org/https://doi.org/10.1088/1361-648x/aa8f79} {\bibfield  {journal} {\bibinfo  {journal} {J. Phys.: Condens. Matter}\ }\textbf {\bibinfo {volume} {29}},\ \bibinfo {pages} {465901} (\bibinfo {year} {2017})}\BibitemShut {NoStop}%
\bibitem [{\citenamefont {Perdew}\ \emph {et~al.}(1996)\citenamefont {Perdew}, \citenamefont {Burke},\ and\ \citenamefont {Ernzerhof}}]{Perdew1996PRL}%
  \BibitemOpen
  \bibfield  {author} {\bibinfo {author} {\bibfnamefont {J.~P.}\ \bibnamefont {Perdew}}, \bibinfo {author} {\bibfnamefont {K.}~\bibnamefont {Burke}},\ and\ \bibinfo {author} {\bibfnamefont {M.}~\bibnamefont {Ernzerhof}},\ }\bibfield  {title} {\bibinfo {title} {Generalized gradient approximation made simple},\ }\href {https://doi.org/https://doi.org/10.1103/PhysRevLett.77.3865} {\bibfield  {journal} {\bibinfo  {journal} {Phys. Rev. Lett.}\ }\textbf {\bibinfo {volume} {77}},\ \bibinfo {pages} {3865} (\bibinfo {year} {1996})}\BibitemShut {NoStop}%
\bibitem [{\citenamefont {Hamann}(2013)}]{Hamann2013PRB}%
  \BibitemOpen
  \bibfield  {author} {\bibinfo {author} {\bibfnamefont {D.~R.}\ \bibnamefont {Hamann}},\ }\bibfield  {title} {\bibinfo {title} {Optimized norm-conserving vanderbilt pseudopotentials},\ }\href {https://doi.org/https://doi.org/10.1103/PhysRevB.88.085117} {\bibfield  {journal} {\bibinfo  {journal} {Phys. Rev. B}\ }\textbf {\bibinfo {volume} {88}},\ \bibinfo {pages} {085117} (\bibinfo {year} {2013})}\BibitemShut {NoStop}%
\bibitem [{\citenamefont {Guandalini}\ \emph {et~al.}(2023)\citenamefont {Guandalini}, \citenamefont {D'Amico}, \citenamefont {Ferretti},\ and\ \citenamefont {Varsano}}]{Guandalini2023npjCM}%
  \BibitemOpen
  \bibfield  {author} {\bibinfo {author} {\bibfnamefont {A.}~\bibnamefont {Guandalini}}, \bibinfo {author} {\bibfnamefont {P.}~\bibnamefont {D'Amico}}, \bibinfo {author} {\bibfnamefont {A.}~\bibnamefont {Ferretti}},\ and\ \bibinfo {author} {\bibfnamefont {D.}~\bibnamefont {Varsano}},\ }\bibfield  {title} {\bibinfo {title} {Efficient {GW} calculations in two dimensional materials through a stochastic integration of the screened potential},\ }\href {https://doi.org/10.1038/s41524-023-00989-7} {\bibfield  {journal} {\bibinfo  {journal} {npj Computational Materials}\ }\textbf {\bibinfo {volume} {9}},\ \bibinfo {pages} {44} (\bibinfo {year} {2023})}\BibitemShut {NoStop}%
\bibitem [{\citenamefont {Guzzo}\ \emph {et~al.}(2011)\citenamefont {Guzzo}, \citenamefont {Lani}, \citenamefont {Sottile}, \citenamefont {Romaniello}, \citenamefont {Gatti}, \citenamefont {Kas}, \citenamefont {Rehr}, \citenamefont {Silly}, \citenamefont {Sirotti},\ and\ \citenamefont {Reining}}]{Guzzo2011PRL}%
  \BibitemOpen
  \bibfield  {author} {\bibinfo {author} {\bibfnamefont {M.}~\bibnamefont {Guzzo}}, \bibinfo {author} {\bibfnamefont {G.}~\bibnamefont {Lani}}, \bibinfo {author} {\bibfnamefont {F.}~\bibnamefont {Sottile}}, \bibinfo {author} {\bibfnamefont {P.}~\bibnamefont {Romaniello}}, \bibinfo {author} {\bibfnamefont {M.}~\bibnamefont {Gatti}}, \bibinfo {author} {\bibfnamefont {J.~J.}\ \bibnamefont {Kas}}, \bibinfo {author} {\bibfnamefont {J.~J.}\ \bibnamefont {Rehr}}, \bibinfo {author} {\bibfnamefont {M.~G.}\ \bibnamefont {Silly}}, \bibinfo {author} {\bibfnamefont {F.}~\bibnamefont {Sirotti}},\ and\ \bibinfo {author} {\bibfnamefont {L.}~\bibnamefont {Reining}},\ }\bibfield  {title} {\bibinfo {title} {Valence electron photoemission spectrum of semiconductors: Ab initio description of multiple satellites},\ }\href {https://doi.org/10.1103/PhysRevLett.107.166401} {\bibfield  {journal} {\bibinfo  {journal} {Phys. Rev. Lett.}\ }\textbf {\bibinfo {volume} {107}},\ \bibinfo {pages} {166401} (\bibinfo {year} {2011})}\BibitemShut
  {NoStop}%
\bibitem [{\citenamefont {Caruso}\ and\ \citenamefont {Giustino}(2016)}]{Caruso2016EPJB}%
  \BibitemOpen
  \bibfield  {author} {\bibinfo {author} {\bibfnamefont {F.}~\bibnamefont {Caruso}}\ and\ \bibinfo {author} {\bibfnamefont {F.}~\bibnamefont {Giustino}},\ }\bibfield  {title} {\bibinfo {title} {The {GW} plus cumulant method and plasmonic polarons: application to the homogeneous electron gas*},\ }\bibfield  {journal} {\bibinfo  {journal} {Eur. Phys. J. B}\ }\textbf {\bibinfo {volume} {89}},\ \href {https://doi.org/https://doi.org/10.1140/epjb/e2016-70028-4} {https://doi.org/10.1140/epjb/e2016-70028-4} (\bibinfo {year} {2016})\BibitemShut {NoStop}%
\bibitem [{\citenamefont {Gerlach}\ \emph {et~al.}(2001)\citenamefont {Gerlach}, \citenamefont {Berge}, \citenamefont {Goldmann}, \citenamefont {Campillo}, \citenamefont {Rubio}, \citenamefont {Pitarke},\ and\ \citenamefont {Echenique}}]{Gerlach2001PRB}%
  \BibitemOpen
  \bibfield  {author} {\bibinfo {author} {\bibfnamefont {A.}~\bibnamefont {Gerlach}}, \bibinfo {author} {\bibfnamefont {K.}~\bibnamefont {Berge}}, \bibinfo {author} {\bibfnamefont {A.}~\bibnamefont {Goldmann}}, \bibinfo {author} {\bibfnamefont {I.}~\bibnamefont {Campillo}}, \bibinfo {author} {\bibfnamefont {A.}~\bibnamefont {Rubio}}, \bibinfo {author} {\bibfnamefont {J.~M.}\ \bibnamefont {Pitarke}},\ and\ \bibinfo {author} {\bibfnamefont {P.~M.}\ \bibnamefont {Echenique}},\ }\bibfield  {title} {\bibinfo {title} {Lifetime of d holes at cu surfaces: Theory and experiment},\ }\href {https://doi.org/10.1103/PhysRevB.64.085423} {\bibfield  {journal} {\bibinfo  {journal} {Phys. Rev. B}\ }\textbf {\bibinfo {volume} {64}},\ \bibinfo {pages} {085423} (\bibinfo {year} {2001})}\BibitemShut {NoStop}%
\bibitem [{\citenamefont {Tamai}\ \emph {et~al.}(2013)\citenamefont {Tamai}, \citenamefont {Meevasana}, \citenamefont {King}, \citenamefont {Nicholson}, \citenamefont {de~la Torre}, \citenamefont {Rozbicki},\ and\ \citenamefont {Baumberger}}]{Tamai2013PRB}%
  \BibitemOpen
  \bibfield  {author} {\bibinfo {author} {\bibfnamefont {A.}~\bibnamefont {Tamai}}, \bibinfo {author} {\bibfnamefont {W.}~\bibnamefont {Meevasana}}, \bibinfo {author} {\bibfnamefont {P.~D.~C.}\ \bibnamefont {King}}, \bibinfo {author} {\bibfnamefont {C.~W.}\ \bibnamefont {Nicholson}}, \bibinfo {author} {\bibfnamefont {A.}~\bibnamefont {de~la Torre}}, \bibinfo {author} {\bibfnamefont {E.}~\bibnamefont {Rozbicki}},\ and\ \bibinfo {author} {\bibfnamefont {F.}~\bibnamefont {Baumberger}},\ }\bibfield  {title} {\bibinfo {title} {Spin-orbit splitting of the shockley surface state on {Cu(111)}},\ }\href {https://doi.org/10.1103/PhysRevB.87.075113} {\bibfield  {journal} {\bibinfo  {journal} {Phys. Rev. B}\ }\textbf {\bibinfo {volume} {87}},\ \bibinfo {pages} {075113} (\bibinfo {year} {2013})}\BibitemShut {NoStop}%
\bibitem [{\citenamefont {Leon}(2025)}]{MPA-Sigma_data}%
  \BibitemOpen
  \bibfield  {author} {\bibinfo {author} {\bibfnamefont {D.~A.}\ \bibnamefont {Leon}},\ }\href {https://doi.org/https://doi.org/10.5281/zenodo.15193879} {\bibinfo {title} {{MPA}-{S}igma{\_}data}} (\bibinfo {year} {2025})\BibitemShut {NoStop}%
\bibitem [{\citenamefont {Leon}\ \emph {et~al.}(2024)\citenamefont {Leon}, \citenamefont {Elgvin}, \citenamefont {Nguyen}, \citenamefont {Prytz}, \citenamefont {Hage},\ and\ \citenamefont {Berland}}]{Leon2024PRB}%
  \BibitemOpen
  \bibfield  {author} {\bibinfo {author} {\bibfnamefont {D.~A.}\ \bibnamefont {Leon}}, \bibinfo {author} {\bibfnamefont {C.}~\bibnamefont {Elgvin}}, \bibinfo {author} {\bibfnamefont {P.~D.}\ \bibnamefont {Nguyen}}, \bibinfo {author} {\bibfnamefont {O.}~\bibnamefont {Prytz}}, \bibinfo {author} {\bibfnamefont {F.~S.}\ \bibnamefont {Hage}},\ and\ \bibinfo {author} {\bibfnamefont {K.}~\bibnamefont {Berland}},\ }\bibfield  {title} {\bibinfo {title} {Unraveling many-body effects in {ZnO}: Combined study using momentum-resolved electron energy-loss spectroscopy and first-principles calculations},\ }\href {https://doi.org/10.1103/PhysRevB.109.115153} {\bibfield  {journal} {\bibinfo  {journal} {Phys. Rev. B}\ }\textbf {\bibinfo {volume} {109}},\ \bibinfo {pages} {115153} (\bibinfo {year} {2024})}\BibitemShut {NoStop}%
\end{thebibliography}%


\begin{thebibliography}{3}%
\makeatletter
\providecommand \@ifxundefined [1]{%
 \@ifx{#1\undefined}
}%
\providecommand \@ifnum [1]{%
 \ifnum #1\expandafter \@firstoftwo
 \else \expandafter \@secondoftwo
 \fi
}%
\providecommand \@ifx [1]{%
 \ifx #1\expandafter \@firstoftwo
 \else \expandafter \@secondoftwo
 \fi
}%
\providecommand \natexlab [1]{#1}%
\providecommand \enquote  [1]{``#1''}%
\providecommand \bibnamefont  [1]{#1}%
\providecommand \bibfnamefont [1]{#1}%
\providecommand \citenamefont [1]{#1}%
\providecommand \href@noop [0]{\@secondoftwo}%
\providecommand \href [0]{\begingroup \@sanitize@url \@href}%
\providecommand \@href[1]{\@@startlink{#1}\@@href}%
\providecommand \@@href[1]{\endgroup#1\@@endlink}%
\providecommand \@sanitize@url [0]{\catcode `\\12\catcode `\$12\catcode `\&12\catcode `\#12\catcode `\^12\catcode `\_12\catcode `\%12\relax}%
\providecommand \@@startlink[1]{}%
\providecommand \@@endlink[0]{}%
\providecommand \url  [0]{\begingroup\@sanitize@url \@url }%
\providecommand \@url [1]{\endgroup\@href {#1}{\urlprefix }}%
\providecommand \urlprefix  [0]{URL }%
\providecommand \Eprint [0]{\href }%
\providecommand \doibase [0]{https://doi.org/}%
\providecommand \selectlanguage [0]{\@gobble}%
\providecommand \bibinfo  [0]{\@secondoftwo}%
\providecommand \bibfield  [0]{\@secondoftwo}%
\providecommand \translation [1]{[#1]}%
\providecommand \BibitemOpen [0]{}%
\providecommand \bibitemStop [0]{}%
\providecommand \bibitemNoStop [0]{.\EOS\space}%
\providecommand \EOS [0]{\spacefactor3000\relax}%
\providecommand \BibitemShut  [1]{\csname bibitem#1\endcsname}%
\let\auto@bib@innerbib\@empty
\bibitem [{\citenamefont {Leon}\ \emph {et~al.}(2021)\citenamefont {Leon}, \citenamefont {Cardoso}, \citenamefont {Chiarotti}, \citenamefont {Varsano}, \citenamefont {Molinari},\ and\ \citenamefont {Ferretti}}]{Leon2021PRB}%
  \BibitemOpen
  \bibfield  {author} {\bibinfo {author} {\bibfnamefont {D.~A.}\ \bibnamefont {Leon}}, \bibinfo {author} {\bibfnamefont {C.}~\bibnamefont {Cardoso}}, \bibinfo {author} {\bibfnamefont {T.}~\bibnamefont {Chiarotti}}, \bibinfo {author} {\bibfnamefont {D.}~\bibnamefont {Varsano}}, \bibinfo {author} {\bibfnamefont {E.}~\bibnamefont {Molinari}},\ and\ \bibinfo {author} {\bibfnamefont {A.}~\bibnamefont {Ferretti}},\ }\bibfield  {title} {\bibinfo {title} {Frequency dependence in $gw$ made simple using a multipole approximation},\ }\href {https://doi.org/10.1103/PhysRevB.104.115157} {\bibfield  {journal} {\bibinfo  {journal} {Phys. Rev. B}\ }\textbf {\bibinfo {volume} {104}},\ \bibinfo {pages} {115157} (\bibinfo {year} {2021})}\BibitemShut {NoStop}%
\bibitem [{\citenamefont {Vidberg}\ and\ \citenamefont {Serene}(1977)}]{Pade1977JLTPhys}%
  \BibitemOpen
  \bibfield  {author} {\bibinfo {author} {\bibfnamefont {H.~J.}\ \bibnamefont {Vidberg}}\ and\ \bibinfo {author} {\bibfnamefont {J.~W.}\ \bibnamefont {Serene}},\ }\bibfield  {title} {\bibinfo {title} {Solving the eliashberg equations by means of n-point pad\'e approximants},\ }\href {https://doi.org/https://doi.org/10.1007/BF00655090} {\bibfield  {journal} {\bibinfo  {journal} {J. Low Temp. Phys.}\ }\textbf {\bibinfo {volume} {29}},\ \bibinfo {pages} {179–192} (\bibinfo {year} {1977})}\BibitemShut {NoStop}%
\bibitem [{\citenamefont {Leon}\ \emph {et~al.}(2023)\citenamefont {Leon}, \citenamefont {Ferretti}, \citenamefont {Varsano}, \citenamefont {Molinari},\ and\ \citenamefont {Cardoso}}]{Leon2023PRB}%
  \BibitemOpen
  \bibfield  {author} {\bibinfo {author} {\bibfnamefont {D.~A.}\ \bibnamefont {Leon}}, \bibinfo {author} {\bibfnamefont {A.}~\bibnamefont {Ferretti}}, \bibinfo {author} {\bibfnamefont {D.}~\bibnamefont {Varsano}}, \bibinfo {author} {\bibfnamefont {E.}~\bibnamefont {Molinari}},\ and\ \bibinfo {author} {\bibfnamefont {C.}~\bibnamefont {Cardoso}},\ }\bibfield  {title} {\bibinfo {title} {Efficient full frequency gw for metals using a multipole approach for the dielectric screening},\ }\href {https://doi.org/10.1103/PhysRevB.107.155130} {\bibfield  {journal} {\bibinfo  {journal} {Phys. Rev. B}\ }\textbf {\bibinfo {volume} {107}},\ \bibinfo {pages} {155130} (\bibinfo {year} {2023})}\BibitemShut {NoStop}%
\end{thebibliography}%

\end{document}


\title{
Spectral properties from an efficient analytical representation of the $GW$ self-energy within a multipole approximation: Supplemental materials
}

\author{Dario A. Leon}
\email{dario.alejandro.leon.valido@nmbu.no}
\affiliation{
 Department of Mechanical Engineering and Technology Management, \\ Norwegian University of Life Sciences, NO-1432 Ås, Norway
}%

\author{Kristian Berland}
\affiliation{
 Department of Mechanical Engineering and Technology Management, \\ Norwegian University of Life Sciences, NO-1432 Ås, Norway
}

\author{Claudia Cardoso}
\affiliation{
 S3 Centre, Istituto Nanoscienze, CNR, 41125 Modena, Italy
}

\maketitle

\section{MPA-$\Sigma$ interpolation} 
\label{section:interpolation}
The procedure we use to obtain the MPA-$\Sigma$ representation is analogous to the one for MPA-$W$ (see Appendix A of Ref.~\cite{Leon2021PRB}).
We need to solve the following non linear system of $2 n_{\Sigma}$ equations and variables:
\begin{eqnarray}
    \Sigma_c^{\text{MPA}}(z_i) &\equiv& \sum_p^{n_{\Sigma}} \frac{S_p}{z_i-\xi_p} = \Sigma_c(z_i)\text{, } i = 1, \ldots, 2 n_{\Sigma}, 
    \label{eq:SysEqs}
\end{eqnarray}
where $n_{\Sigma}$ is the number of poles and $\{z_i, \Sigma_c(z_i) \}$ correspond to the numerical data according to the given sampling (see Sec.~\ref{sec:sampling}).
Notice that finding the residues $S_p$ if the poles $\xi_p$ are known is a simple linear least square problem:
\begin{equation}
   \min\limits_{\mathbf{S_{n_{\Sigma}}}} ||\mathbf{M_{2 n_{\Sigma} n_{\Sigma}}} \cdot \mathbf{S_{n_{\Sigma}}} - \mathbf{\Sigma_{2 n_{\Sigma}}}||,
    \label{eq:Rfit}
\end{equation}
where we have defined the following vectors and matrix:
\begin{eqnarray}
  \mathbf{S_{n_{\Sigma}}}&\equiv& \begin{bmatrix}
   S_1 & S_2 & \ldots & S_{n_{\Sigma}}
   \end{bmatrix}\\[4pt]
  \mathbf{\Sigma_{2 n_{\Sigma}}}&\equiv& \begin{bmatrix}
    \Sigma_c(z_1) & \Sigma_c(z_2) & \ldots & \Sigma_c(z_{2 n_{\Sigma}})
    \end{bmatrix}\\[4pt]
  \mathbf{M_{2 n_{\Sigma} n_{\Sigma}}}&\equiv&  
    \begin{bmatrix}
 \frac{1}{z_1-\xi_1} & \frac{1}{z_1-\xi_2} & \ldots & \frac{1}{z_1-\xi_{n_{\Sigma}}} \\
 \frac{1}{z_2-\xi_1} & \frac{1}{z_2-\xi_2} & \ldots & \frac{1}{z_2-\xi_{n_{\Sigma}}} \\
 \vdots & \vdots &\ddots&\vdots \\
 \frac{1}{z_{2 n_{\Sigma}}-\xi_1} & \frac{1}{z_{2 n_{\Sigma}}-\xi_2} & \ldots & \frac{1}{z_{2 n_{\Sigma}}-\xi_{n_{\Sigma}}}
 \end{bmatrix}
\label{eq:SEmpaPade2}
\end{eqnarray}

To find the poles, we start by writing the MPA self-energy in its Pad\'e form:
\begin{equation}
    \sum_p^{n_{\Sigma}} \frac{S_p}{z-\xi_p} = \frac{A_{n_{\Sigma} -1}(z)}{B_{n_{\Sigma}}(z)} \equiv \frac{\mathbf{a_{n_{\Sigma}}} \cdot \mathbf{z_{n_{\Sigma}}}}{\mathbf{b_{n_{\Sigma}}} \cdot \mathbf{z_{n_{\Sigma}}} + z^{n_{\Sigma}}},
    \label{eq:SEmpaPade}
\end{equation}
where we have defined the following vectors:
\begin{equation}
\begin{aligned}
   \mathbf{a_{n_{\Sigma}}} &\equiv& \begin{bmatrix}
   a_0&a_1&\ldots&a_{n_{\Sigma} -1} \end{bmatrix} \\
    \mathbf{b_{n_{\Sigma}}} &\equiv& \begin{bmatrix}
    b_0&b_1&\ldots&b_{n_{\Sigma} -1}\end{bmatrix} \\
    \mathbf{z_{n_{\Sigma}}}(z) &\equiv& \begin{bmatrix}
    1&z&\ldots&z^{n_{\Sigma} -1}\end{bmatrix}.
\end{aligned}
    \label{eq:SEmpaPade2}
\end{equation}
Notice that finding the poles means to find $\mathbf{b_{n_{\Sigma}}}$ and factorize the $B_{n_{\Sigma}}(z)$ polynomial, which can be done by diagonalizing its corresponding companion matrix:
\begin{equation}
 \mathbf{C_{n_{\Sigma} n_{\Sigma}}}=
 \begin{bmatrix}
 0&0&\ldots&0&-b_0\\
 1&0&\ldots&0&-b_1\\
 0&1&\ldots&0&-b_2\\
 \vdots&\vdots&\ddots&\vdots&\vdots\\
 0&0&\ldots&1&-b_{n_{\Sigma}-1}\\
 \end{bmatrix}
    \label{eq:CM}
\end{equation}
To compute $\mathbf{b_{n_{\Sigma}}}$, we can then 
take one of the two routes in the next sections.

\subsection{Linear solver
\label{section:linear_solver}}
%
With the definitions in Eq.~\eqref{eq:SEmpaPade} we can transform our non linear problem into an equivalent linear one:
\begin{equation}
\Sigma_c(z_i) z_i^{n_{\Sigma}} + \Sigma_c(z_i)\, {\bf z_{n_{\Sigma}}}(z_i)\cdot{\bf b_{n_{\Sigma}}}=
   {\bf z_{n_{\Sigma}}}(z_i)\cdot{\bf a_{n_{\Sigma}}}.
    \label{eq:ab_z}  
\end{equation}
We can split the sampled points into two sets with the same number of elements $n_{\Sigma}$:
\begin{equation}
\begin{aligned}
    \text{set 1}: \text{ } & i = 1, \ldots, n_{\Sigma}, \\
    \text{set 2}: \text{ } & i = n_{\Sigma}+1, \ldots, 2 n_{\Sigma},
\end{aligned}
 \label{eq:split}
\end{equation}
and define the following vectors and matrices with the first set of Eq.~\eqref{eq:split}:
\begin{eqnarray}
 \mathbf{v_{n_{\Sigma}}^1}&=&
 \begin{bmatrix}
 \Sigma_c(z_1)z_1^{n_{\Sigma}}&
 \Sigma_c(z_2)z_2^{n_{\Sigma}}&
 \ldots&
 \Sigma_c(z_{n_{\Sigma}})z_{n_{\Sigma}}^{n_{\Sigma}}
 \end{bmatrix} 
    \label{eq:v1}
\\[4pt]
 \mathbf{z_{n_{\Sigma} n_{\Sigma}}^1}&=&
 \begin{bmatrix}
 1&z_1&\ldots&z_1^{n_{\Sigma}-1}\\
 1&z_2&\ldots&z_2^{n_{\Sigma}-1}\\
 \vdots&\vdots&\ddots& \vdots\\
 1&z_{n_{\Sigma}}&\ldots&z_{n_{\Sigma}}^{n_{\Sigma}-1}\\
 \end{bmatrix}
    \label{eq:Z1}
\\[4pt]
 \mathbf{M_{n_{\Sigma} n_{\Sigma}}^1}&=&
 \begin{bmatrix}
 \Sigma_c(z_1)&\Sigma_c(z_1)z_1&\ldots&\Sigma_c(z_1)z_1^{n_{\Sigma}-1}\\
 \Sigma_c(z_2)&\Sigma_c(z_2)z_2&\ldots&\Sigma_c(z_2)z_2^{n_{\Sigma}-1}\\
 \vdots&\vdots&\ddots&\vdots\\
 \Sigma_c(z_{n_{\Sigma}})& \Sigma_c(z_{n_{\Sigma}})z_{n_{\Sigma}}&\ldots& \Sigma_c(z_{n_{\Sigma}})z_{n_{\Sigma}}^{n_{\Sigma}-1}\\
 \end{bmatrix},
 \,\,\,\,\,\,\,\,\,\,\,\,
    \label{eq:M1}
\end{eqnarray}
%
likewise, we can define $\mathbf{v_{n_{\Sigma}}^2}$, $\mathbf{z_{n_{\Sigma} n_{\Sigma}}^2}$ and $\mathbf{M_{n_{\Sigma} n_{\Sigma}}^2}$ with the second set. The system of equations is then simplified as
\begin{equation}
\begin{aligned}
 {\bf v_{n_{\Sigma}}^1}+{\bf M_{n_{\Sigma} n_{\Sigma}}^1}\cdot{\bf b_{n_{\Sigma}}}=
{\bf z_{n_{\Sigma} n_{\Sigma}}^1}\cdot{\bf a_{n_{\Sigma}}}\\
 {\bf v_{n_{\Sigma}}^2}+{\bf M_{n_{\Sigma} n_{\Sigma}}^2}\cdot{\bf b_{n_{\Sigma}}}=
{\bf z_{n_{\Sigma} n_{\Sigma}}^2}\cdot{\bf a_{n_{\Sigma}}}.
    \label{eq:ab_1}
\end{aligned}
\end{equation}
By combining the two equations in Eq.~\eqref{eq:ab_1} we obtain a linear system for $\bf b_{n_{\Sigma}}$: 
\begin{equation}
     {\bf M_{n_{\Sigma} n_{\Sigma}}} \cdot {\bf b_{n_{\Sigma}}} = {\bf v_{n_{\Sigma}}},
  \label{eq:b}
\end{equation}
where ${\bf M_{n_{\Sigma} n_{\Sigma}}}$ and ${\bf v_{n_{\Sigma}}}$ are defined as
\begin{equation}
\begin{aligned}
 {\bf M_{n_{\Sigma} n_{\Sigma}}} &\equiv& \mathbf{z_{n_{\Sigma} n_{\Sigma}}^2} \cdot (\mathbf{z_{n_{\Sigma} n_{\Sigma}}^1})^{-1} \cdot \mathbf{M_{n_{\Sigma} n_{\Sigma}}^1} - \mathbf{M_{n_{\Sigma} n_{\Sigma}}^2}  \\
  {\bf v_{n_{\Sigma}}} &\equiv&  -\mathbf{z_{n_{\Sigma} n_{\Sigma}}^2}\cdot (\mathbf{z_{n_{\Sigma} n_{\Sigma}}^1})^{-1}\cdot \mathbf{v_{n_{\Sigma}}^1}+\mathbf{v_{n_{\Sigma}}^2}.
 \label{eq:Mv}
\end{aligned}
\end{equation}

\subsection{Pad\'e/Thiele solver}
\label{section:pade}
%
It is also possible to write the Pad\'e representation as a $2 n_{\Sigma}$-point continued fraction of reciprocal differences and use Thiele's interpolation formula~\cite{Pade1977JLTPhys}:
\begin{multline}
   \frac{A_{n_{\Sigma}-1}(z)/b_0}{B_{n_{\Sigma}}(z)/b_0}=\\
   \frac{c_1}{1+}\frac{c_{2}(z-z_1)}{1+}\ldots\frac{c_{2 n_{\Sigma}}(z-z_{2 n_{\Sigma}-1})}{1+(z-z_{2 n_{\Sigma}-1})g_{2 n_{\Sigma}}(z)}, 
    \label{eq:thiele}
\end{multline}
where the coefficients $c_i$ and functions $g_i(z)$ are given by the following recursion relations:
\begin{equation}
\begin{aligned}
  c_i&= g_i(z_i)
  \\[4pt]
  g_i(z) &= \left\{
  \begin{aligned}
  & \Sigma_c(z_i), \qquad \qquad i=1 \\
  & \frac{g_{i-1}(z_{i-1})-g_{i-1}(z)}{(z-z_{i-1})g_{i-1}(z)},  s\geq 2,      
  \end{aligned}
  \right.
  \end{aligned}
    \label{eq:recursion_g}
\end{equation}
where index $i=1,\ldots,2 n_{\Sigma}$ corresponds to both the iteration step and the index of the sampled point.

Since we are primarily interested in the polynomial in the denominator of Eq.~\eqref{eq:thiele}, we can define the following vectors:
\begin{equation}
\begin{aligned}
    \mathbf{d_{n_{\Sigma}+1}} &= \begin{bmatrix}
        1 & \frac{b_1}{b_0} & \ldots & \frac{b_{n_{\Sigma}-1}}{b_0} & \frac{1}{b_0}
    \end{bmatrix} \\
    \mathbf{c_{2 n_{\Sigma}}} &= \begin{bmatrix}
    g(z_1) & g(z_2) & \ldots & g(z_{2 n_{\Sigma}})
    \end{bmatrix},
\end{aligned}
\end{equation}
and recast the recursivity in vectorial form~\cite{Leon2021PRB}:
\begin{multline}
 \begin{aligned}
 {\bf d_{n_{\Sigma}+1}^s} &= \left\{
  \begin{aligned}
   & d^s_j = \delta_{j 1} , \qquad \qquad \qquad \qquad \qquad  s=0,1 \\
   & d^s_j=d^{s-1}_j+c_s z_{s}d^{s-2}_{j+1}-c_s z_{s-1}d^{s-2}_j,  s\geq 2
  \end{aligned}
   \right.
   \\[8pt]
 {\bf c_{2 n_{\Sigma}}^s} &= \left\{
  \begin{aligned}
   c^s_i&=\Sigma_c(z_i) ,  &s&=1 \\
    c^s_i&=\frac{c^{s-1}_{i-1}-c^{s-1}_i}{(z_i-z_{i-1})c^{s-1}_i} , &s&\geq 2,
  \end{aligned}
   \right.
   \end{aligned}
    \label{eq:recursion_vec_d}
\end{multline}
where $j=1,\ldots,n_{\Sigma}+1$, $s=0,\ldots,2 n_{\Sigma}$ is the iteration step, and $\delta_{j 1}$ is a Kronecker delta.

Once $\mathbf{d_{n_{\Sigma}+1}}$ has been computed in the last iteration, we can retrieve the vector $\mathbf{b_{n_{\Sigma}}}$ as 
\begin{equation}
\mathbf{b_{n_{\Sigma}}} = \begin{bmatrix}
        \frac{1}{d_{n_{\Sigma}+1}} & \frac{d_1}{d_{n_{\Sigma}+1}} & \ldots & \frac{d_{n_{\Sigma}}}{d_{n_{\Sigma}+1}}
    \end{bmatrix}    
\end{equation}
%

\section{MPA-$\Sigma$ sampling}
\label{sec:sampling}
As mentioned in the main text, MPA-$\Sigma$ uses a sampling in the complex frequency plane $z' \equiv z - \epsilon_{n \kk}^{\text{KS}}$, that depends on the number of poles $n_{\Sigma}$. The set of frequency points, $\{ z'_i: i=1,..,2 n_{\Sigma}\}$, is divided into two subsets corresponding to the first and third quadrants of the complex frequency plane:
\begin{equation}
   \{z'_i\} : 
    \left\{
    \begin{aligned}
    z^+_n &= \omega_{n} + i \varpi \text{; } &n=1,.., n_{\Sigma}^+ \\
    z^-_n &= -\omega_{n} - i \varpi \text{; } &n=1,.., n_{\Sigma}^-
    \end{aligned}
    \right.  
    \label{eq:sampling}
\end{equation}
where the imaginary part is typically set to $\varpi=0.1~$eV, $n_{\Sigma}^+ + n_{\Sigma}^- = 2 n_{\Sigma}$, and the frequency points are distributed inhomogeneously along the real axis (see also Eq.~(10) of Ref.~\cite{Leon2023PRB}) as
\begin{equation}
   \{\omega_{n}\}_{\alpha}: \left\{
    \begin{aligned}
        \left(0\right) \text{, } n_{\Sigma}^+ = 1  \\
        \left(0,1\right) \times \omega_m \text{, } n_{\Sigma}^+ = 2  \\
        \left(0,\frac{1}{2},1\right) ^\alpha \times \omega_m \text{, } n_{\Sigma}^+ = 3  \\
        \left( 0,\frac{1}{4},\frac{1}{2},1\right) ^\alpha \times \omega_m \text{, } n_{\Sigma}^+ = 4  \\
        \left( 0,\frac{1}{8},\frac{1}{4},\frac{1}{2},1\right) ^\alpha \times \omega_m \text{, } n_{\Sigma}^+ = 5   \\
        \left( 0,\frac{1}{8},\frac{1}{4},\frac{1}{2},\frac{3}{4},1\right) ^\alpha \times \omega_m \text{, } n_{\Sigma}^+ = 6   \\
        \left( 0,\frac{1}{8},\frac{1}{4},\frac{3}{8},\frac{1}{2},\frac{3}{4},1\right) ^\alpha \times \omega_m \text{, } n_{\Sigma}^+ = 7   \\
        ...
    \end{aligned}
    \right.
    \label{eq:w_grid}
\end{equation}

In Eq.~\eqref{eq:w_grid}, the maximum frequency $\omega_m$ defines the desired sampling interval. The exponent $\alpha$  is usually set to $\alpha=1$ or $\alpha=2$ respectively corresponding to a linear or a quadratic semi-homogeneous partition in powers of 2~\cite{Leon2023PRB}. In the main manuscript, it is common practice to set the sampling with a minimum of asymmetry, using $n_{\Sigma}^- = n_{\Sigma}^+ +1$ ($n_{\Sigma}^+ = n_{\Sigma}^- +1$) for valence (conduction) states.

\begin{figure*}
    \centering
    \includegraphics[width=0.98\linewidth]{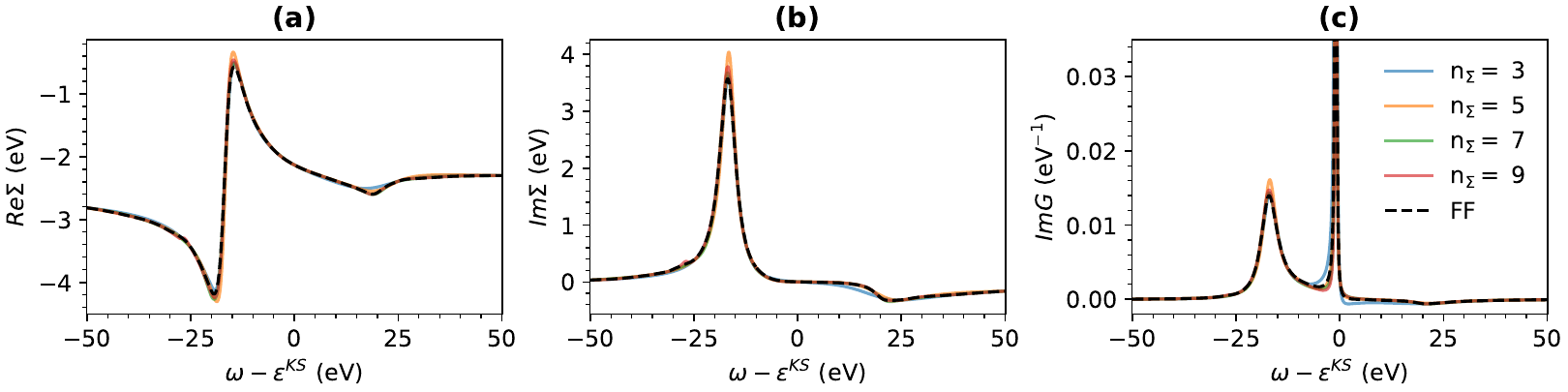}
    \caption{$\Re ~\Sigma$ (a), $\Im ~\Sigma$ (b), and $\Im ~G$ (c) functions corresponding to the top valence state of Si computed with MPA-$\Sigma$ with a variable number of poles, $n_{\Sigma}$, compared to the full-frequency approach (FF).}
    \label{fig:convergence}
\end{figure*}

In Fig.~\ref{fig:convergence} we show the convergence of $\Sigma$ and the interacting $G$ corresponding to the top valence state of Si, with respect to the total number of poles $n_{\Sigma}$. For the MPA-$\Sigma$ sampling, we set $n_{\Sigma}^- = n_{\Sigma}^+ +1$ with a linear partition on both the positive and negative sides.

\section{Insights into spectral band structures}
Here we provide spectral band structures of Na, Si, MoS$_2$, and Cu, computed with a full-frequency evaluation of $\Sigma$ and $G$, to benchmark the analogous ones in Fig.~6 of the main manuscript that correspond to MPA-$\Sigma$. We show the details on how they are built, by isolating the bands of Na as an example. We also make use of the MPA-$G$ representation, to separate the spectral contributions of the QP pole and the satellites from the total MPA-$G$ spectral function computed in the main manuscript.

As mentioned in the main manuscript, the $\Sigma$ and $G$ spectral functions, $A_{\Sigma}$ and $A_{G}$, are computed for a finite number of valence and conduction states, respecting the time ordering. For visualization purposes, we have divided the intensity of the valence (conduction) contribution to $A_{\Sigma}$ and $A_{G}$ by the corresponding number of valence (conduction) bands, to ensure a more uniform background intensity: 
\begin{equation}
    A_{\Sigma} (\kk, \omega) \equiv 
\frac{1}{n_v \pi} \sum_v^{n_v} \Im ~\Sigma_{v \kk} (\omega) + \frac{1}{n_c \pi} \sum_c^{n_c} \Im ~\Sigma_{c \kk} (\omega)
    \label{eq:spectralS}
\end{equation}
\begin{equation}
    A_{G} (\kk, \omega) \equiv \frac{1}{n_v \pi} \sum_v^{n_v} \Im ~G_{v \kk} (\omega) + \frac{1}{n_c \pi} \sum_c^{n_c} \Im ~G_{c \kk} (\omega) ,
    \label{eq:spectralG}
\end{equation}
where $v$ and $c$ run over valence and conduction states respectively and $n_v$ and $n_c$ correspond to the total number of valence and conduction bands considered.
The normalization with respect to $n_v$ and $n_c$ is not necessary if a large enough number of bands is included in Eqs.~\eqref{eq:spectralS} and ~\eqref{eq:spectralG}.

\begin{figure*}
    \centering
\includegraphics[width=0.995\textwidth]{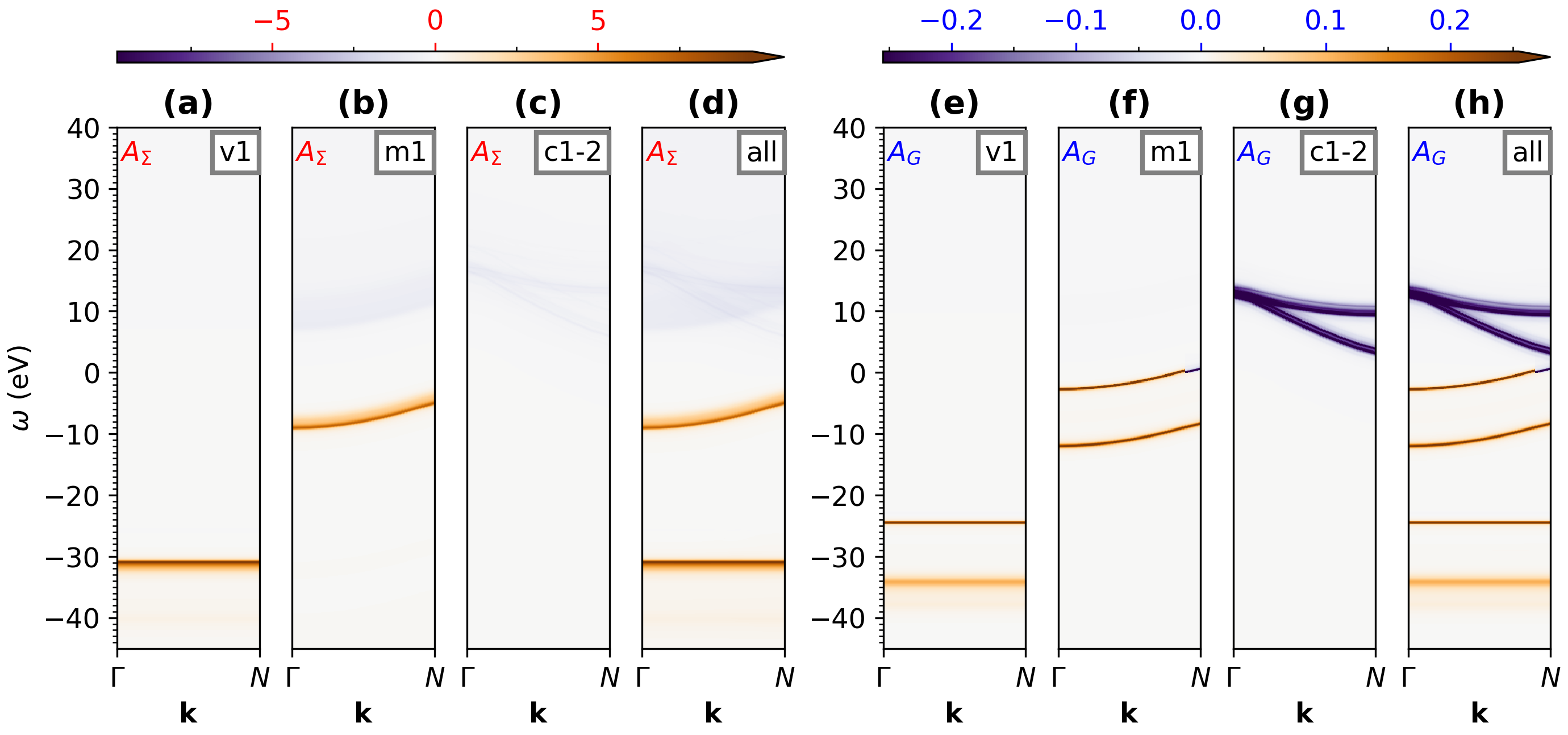}
    \caption{Spectral $\Sigma$ (a-d) and $G$ (e-h) bands of Na, for the first valence state (a, e), the metallic band (b, f), the first two valence bands (c, g) and their combinations (d, h).}
    \label{fig:Na_tot}
\end{figure*}

Fig.~\ref{fig:Na_tot} shows $1/\pi\Im \Sigma_{n \kk}$ (a-d) and $1/\pi\Im G_{n \kk}$ (e-h) of Na along the $\Gamma N$ $\kk$-path, where $n$ corresponds to the highest non-metallic valence band (v1), the metallic band that crosses the Fermi level close to $N$ (m1), the two lowest conduction bands (c1-c2), and the combination of all these 4 bands (all). The last panels (d, h) are analogous to panels (a, e) of Fig.~6 in the main manuscript.
Panel (b) shows discernible intensities in both the positive and negative sides. The same happens for other states at smaller intensities, as a consequence of the time ordering. Therefore, the total spectral function $A_{\Sigma} (\kk, \omega)$~(d) shows a non trivial superposition in the region corresponding to the conduction bands.

\begin{figure*}
    \centering
\includegraphics[width=0.995\textwidth]{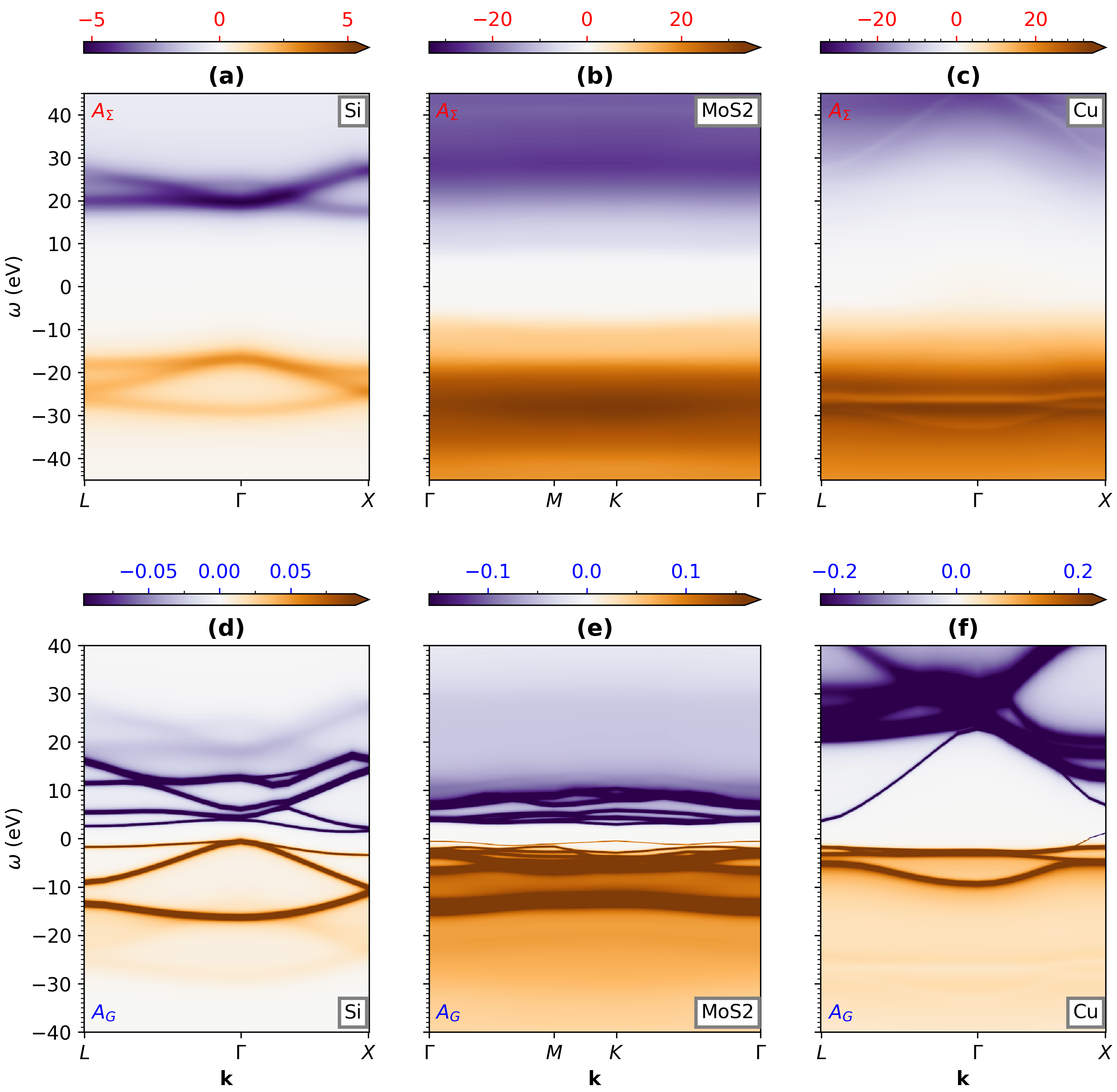}
    \caption{Spectral band structures $A_{\Sigma} (\kk, \omega)$ (a-c) and $A_{G} (\kk, \omega)$ (d-f) of Si (a, d), MoS$_2$ (b, e), and Cu (c, f), computed with a FF $\Sigma$ evaluation.}
    \label{fig:Na-Si-MoS2-Cu_qp}
\end{figure*}

In Fig.~\ref{fig:Na-Si-MoS2-Cu_qp} we provide 
full-frequency (FF) $\Sigma$ and $G$ spectral band structures of Si (a, d), monolayer MoS$_2$ (b, e), and Cu (c, f), analogous to the MPA-$\Sigma$ and MPA-$G$ ones presented in Figs.~6 of the main manuscript.

\begin{figure*}
    \centering
\includegraphics[width=0.995\textwidth]{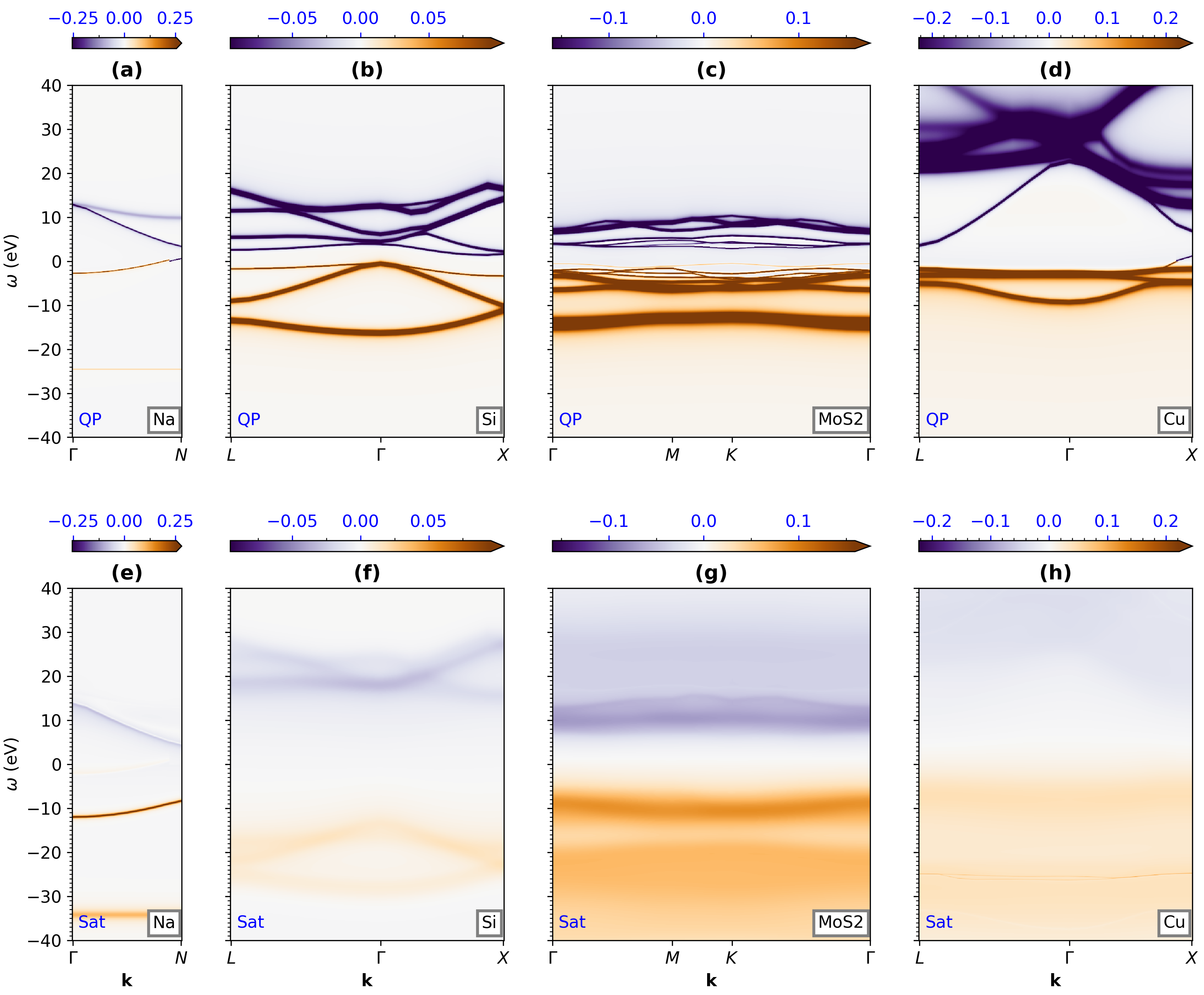}
    \caption{MPA-$G$ spectral band structures of Na (a, e), Si (b, f), MoS$_2$ (c, g), and Cu (d, h), analogous to panels (e-h) of Fig.~6 of the main manuscript, but corresponding to the QP pole only (a-d) and satellites (e-h).}
    \label{fig:Na-Si-MoS2-Cu_sa}
\end{figure*}

In Fig.~\ref{fig:Na-Si-MoS2-Cu_sa} we have separated the spectral contributions of the QP pole (a-d) and the satellites (e-h)  from the total MPA-$G$ spectral function plotted in panels (e-h) of Fig.~6 of the main manuscript.  
By a simple inspection, we can see that the superposition of the QP pole and the satellites spectra gives the total spectral $G$ function. 

%
%
\bibliography{biblio}